%% file: main.tex
\documentclass{article}

\usepackage{arxiv}

\input{extra/before-begin}

\setboolean{useSingleColumn}{true}

\title{\input{content/title}}

\input{extra/authors/arxiv}
\renewcommand{\headeright}{}
\renewcommand{\undertitle}{}
\renewcommand{\shorttitle}{Continual Learning in Medical Image Analysis: A Comprehensive Review}

\hypersetup{
    pdftitle={\input{content/title}},
    pdfsubject={q-bio.NC, q-bio.QM},
    pdfauthor={\input{extra/authors/short-camma}},
    pdfkeywords={\input{extra/keywords/camma}},
}

\begin{document}
    \maketitle
    
    \begin{abstract}
    	\input{content/abstract}

    \end{abstract}

    % keywords can be removed

    % ================= SECTIONS ==================
    
    % Introduction
    \input{content/sections/introduction/index}

    % Background
    \input{content/sections/background/index}\input{content/sections/cl-scenarios/index}

    % Continues learning technique
    \input{content/sections/cl-technique/index}

    % Level of supervision
    \input{content/sections/cl-supervision/index}
    
    % Evaluation
    \input{content/sections/evaluation/index}\input{content/tables/classification}
    \input{content/tables/segmentation}
    \input{content/tables/class-seg-litreature}
    % <<<<<<<<<<<<<<<<<<<<<<<<<<<<<<<<<<<<<<<<<<<
    
    % Discussion
    \input{content/sections/discussion/index}

    % Conclusion
    \input{content/sections/conclusion/index}
    
    %%Harvard
    \bibliographystyle{unsrtnat}
    \bibliography{refs}

\end{document}

%% file: extra/before-begin.tex
\usepackage[utf8]{inputenc} % allow utf-8 input
\usepackage[T1]{fontenc}    % use 8-bit T1 fonts
\usepackage{hyperref}       % hyperlinks
\usepackage{adjustbox}
\usepackage{url}            % simple URL typesetting
\usepackage{booktabs}       % professional-quality tables
\usepackage{framed,multirow}
\usepackage{amsfonts,amssymb,amsmath}       % blackboard math symbols
\usepackage{nicefrac}       % compact symbols for 1/2, etc.
\usepackage{microtype}      % microtypography
\usepackage[bottom]{footmisc}
\usepackage{lipsum}
\usepackage{natbib}
\usepackage{doi}
\usepackage{fancyhdr}       % header
\usepackage{graphicx}       % graphics
\usepackage{float}
\usepackage{wrapfig}
\usepackage{soul}
\usepackage[nameinlink]{cleveref}
\usepackage{latexsym}
\usepackage{colortbl}
\usepackage{letltxmacro}
\usepackage{xcolor}
\usepackage{longtable}
\usepackage{caption}

\hypersetup{
    colorlinks=true,
    linkcolor=teal,
    filecolor=cyan,
    urlcolor=magenta,
    allcolors=blue
}

\definecolor{classinccolor}{rgb}{0.3,0.65,1}
\definecolor{domaininccolor}{rgb}{0.25,1,0.45}
\definecolor{taskinccolor}{rgb}{1,0.45,0.75}
\definecolor{hybridinccolor}{rgb}{1,1,0.2}
\definecolor{datainccolor}{rgb}{1,0.63,0.48}

\definecolor{rehearsalcolor}{rgb}{0.3,0.65,1}
\definecolor{regularizationcolor}{rgb}{0.25,1,0.45}
\definecolor{architecturecolor}{rgb}{1,0.4,0.7}
\definecolor{hybridcolor}{rgb}{1,1,0.2}

\definecolor{classificationcolor}{rgb}{0.3,0.65,1}
\definecolor{segmentationcolor}{rgb}{1,0.5,0.8}
\definecolor{bothcolor}{rgb}{1,1,0.0}

\definecolor{eqmarky}{rgb}{0.9,0.8,0.2}
\definecolor{eqmarkp}{rgb}{0.9,0.5,0.8}

\usepackage[customcolors]{hf-tikz}
\tikzset{offset definition/.style={
		above left offset={-0.01, 0.3},
		below right offset={0.01, -0.12},
	},
	y1/.style={
		offset definition,
		set fill color=eqmarky!90,
		set border color=eqmarky!90,
	},
	p1/.style={
		offset definition,
		set fill color=eqmarkp!80,
		set border color=eqmarkp!80,
	}
}

\newboolean{useSingleColumn}

\newcommand{\tabrefcite}[1]{\citeauthor{#1} (\citeyear{#1})}
\newcommand{\tabdescite}[1]{(\citealt{#1})}

\newcommand{\tikzmarky}[2][a1]{\tikzmarkin[y1]{#1}#2\tikzmarkend{#1}}
\newcommand{\tikzmarkp}[2][a2]{\tikzmarkin[p1]{#1}#2\tikzmarkend{#1}}

\newcommand{\refx}[1]{\Cref{#1}}
\LetLtxMacro\oldcitet\citet
\renewcommand{\citet}[1]{\citeauthor{#1} (\citeyear{#1})}

\LetLtxMacro\oldcite\cite
\renewcommand{\cite}[1]{(\citealt{#1})}

\def\tabcellci{3}
\def\tabcellcis{7}

%% file: content/title.tex
Continual Learning in Medical Image Analysis: A Comprehensive Review of Recent Advancements and Future Prospects

%% file: extra/authors/arxiv.tex
\usepackage{authblk}

\author[1]{Pratibha~Kumari}
\author[2,3]{Joohi~Chauhan}
\author[1]{Afshin~Bozorgpour}
\author[1]{Boqiang~Huang}
\author[4]{Reza~Azad}
\author[1,5]{Dorit~Merhof\hspace{0.75pt}\thanks{Corresponding author: Dorit Merhof. \textit{e-mail}: \texttt{dorit.merhof@ur.de}}}

\affil[1]{Faculty of Informatics and Data Science, University of Regensburg, Regensburg, 93053, Germany}
\affil[2]{University of California, Davis, CA 95616, USA}
\affil[3]{Motilal Nehru National Institute of Technology Allahabad, Prayagraj, 211004, India}
\affil[4]{Faculty of Electrical Engineering and Information Technology, RWTH Aachen University, Aachen, 52062, Germany}
\affil[5]{Fraunhofer Institute for Digital Medicine MEVIS, Bremen 28359, Germany}

%% file: extra/authors/short-camma.tex
Pratibha~Kumari, Joohi~Chauhan, Afshin~Bozorgpour, Boqiang~Huang, Reza~Azad, Dorit~Merhof

%% file: extra/keywords/camma.tex
Continual Learning,
Medical Data Drift,
Domain Shift,
Concept Drift,
Medical Image Analysis,
Histopathology,
Radiology

%% file: content/abstract.tex
Medical image analysis has witnessed remarkable advancements, even surpassing human-level performance in recent years, driven by the rapid development of advanced deep-learning algorithms. 
However, when the inference dataset slightly differs from what the model has seen during one-time training, the model performance is greatly compromised. The situation requires restarting the training process using both the old and the new data, which is computationally costly, does not align with the human learning process, and imposes storage constraints and privacy concerns. Alternatively, continual learning has emerged as a crucial approach for developing unified and sustainable deep models to deal with new classes, tasks, and the drifting nature of data in non-stationary environments for various application areas. Continual learning techniques enable models to adapt and accumulate knowledge over time, which is essential for maintaining performance on evolving datasets and novel tasks. Owing to its popularity and promising performance, it is an active and emerging research topic in the medical field and hence demands a survey and taxonomy to clarify the current research landscape of continual learning in medical image analysis. 
This systematic review paper provides a comprehensive overview of the state-of-the-art in continual learning techniques applied to medical image analysis. We present an extensive survey of existing research, covering topics including catastrophic forgetting, data drifts, stability, and plasticity requirements. Further, an in-depth discussion of key components of a continual learning framework, such as continual learning scenarios, techniques, evaluation schemes, and metrics, is provided. 
Continual learning techniques encompass various categories, including rehearsal, regularization, architectural, and hybrid strategies. We assess the popularity and applicability of continual learning categories in various medical sub-fields like radiology and histopathology. Our exploration considers unique challenges in the medical domain, including costly data annotation, temporal drift, and the crucial need for benchmarking datasets to ensure consistent model evaluation. 
The paper also addresses current challenges and looks ahead to potential future research directions.

%%%%% A thorough analysis of various sources of drifts in medical data and hence various continual learning scenarios are discussed for various research topics in medical image processing. Continual learning techniques can be broadly categorized into rehearsal, regularization, architectural, and hybrid strategies-based methods. The popularity and suitability of each category of approach for different medical sub-fields including radiology, histopathology, etc., are discussed. Furthermore, we discuss the unique challenges posed by the medical domain, such as costly data annotation, temporal drift, and the critical need for benchmarking datasets for homogeneous model evaluation. The paper also examines promising directions for future research, including the adaptation of smooth data drift scenarios, active learning, self-supervised learning, unsupervised learning, and privacy-preserving continual learning. Finally, we conclude this survey by discussing existing challenges and potential future research directions.

%%%%%%%%%%%%%%%%%By synthesizing and evaluating the existing works, this systematic review aims to provide a clear understanding of the current landscape of continual learning in medical imaging analysis. It serves as a valuable resource for researchers, clinicians, and healthcare practitioners interested in leveraging continual learning techniques to improve diagnostic accuracy, patient care, and the overall performance of medical imaging systems.

%% file: extra/keywords/arxiv.tex
Continual Learning \and
Medical Data Drift \and
Domain Shift \and
Concept Drift \and
Medical Image Analysis \and
Histopathology \and
Radiology

%% file: content/sections/introduction/index.tex
\section{Introduction}\label{sec:introduction}

In the evolving field of medical image analysis, the dynamic nature of healthcare data poses a critical challenge for the generalizability of the machine learning/deep learning models to new data/domains \cite{miotto, 9363915}. The data-driven approaches have challenges due to the limited availability and accessibility of sufficiently large and diverse medical data for training \cite{miotto, Hao, 9363915, JOC1, JOC2}. 
% \st{Additionally, the source variability, scanner manufacturer, imaging protocol, and diverse population groups make the medical data heterogeneous, this adds a bias and discrepancies between training and testing datasets, thus leading to performance degradation.} 
{Additionally, the source variability due to different scanner manufacturers, staining and imaging protocols, slice thickness, different patient cohorts, etc., makes the medical data heterogeneous. This introduces bias and discrepancies between the training and test datasets if they originate from different data sources, thus leading to performance degradation} \cite{Luca, 9363915}. 
In order to handle the model generalizability issues, domain adaptation methods become popular and aim to transfer knowledge from one domain to other unseen data sources or domains \cite{jeff01, Jeff02, Hao, CW, becker2014domain, FENG2023102664}. However, domain adaptation poses unique challenges due to the sensitive and complex nature of healthcare data \cite{azad2022medical}. The most common associated issues are the limited availability of labeled medical data for training, heterogeneity in data sources contributing to significant domain shifts, clinical disparities and population variances, and inter-rater variabilities. Also, biases present in the source domain data can propagate to the target domain, so ensuring fair and unbiased predictions across diverse patient populations becomes a critical concern. Moreover, medical data encompasses various modalities, including imaging, electronic health records, and genomic data, making the task of adapting models to handle multimodal data and ensuring interoperability exceptionally complex. 
{Further, the accessibility of source data may be limited to a short period of time or may be prohibited altogether due to strict privacy regulations in the medical domain \mbox{~\cite{thandiackal2024multi}}. Thus, domain adaptation approaches that require simultaneous availability of source and target data may not be feasible.}

{Another related learning paradigm, transfer learning}, has been widely adopted in the medical domain to address challenges related to limited data availability \cite{YU, Gha}. It transfers knowledge gained from the source task to the target task to improve its learning or performance. {Unlike domain adaptation, where only the data distribution changes, transfer learning covers changes in the feature space, label space, as well as in the data distribution of the source and target domain \mbox{~\cite{kouw2018introduction,zhou2022domain,Hao}}. In a small-scale medical disease classification dataset, it can be beneficial to include knowledge gained from a model trained on a large-scale labeled natural image dataset (ImageNet). The model performance on the medical disease dataset may be better as compared to training the same model solely on the medical disease dataset from scratch. However, at the same time the performance on the ImageNet dataset cannot be guaranteed by this model (which is also not intended in transfer learning).} 
{In transfer learning, the focus is on leveraging prior knowledge rather than retaining it, hence performance on the source data may be compromised. Generally, learning a new dataset with shifted distribution results in a sharp decrease in performance on the source dataset, also known as ``catastrophic forgetting'' of deep neural networks \mbox{~\cite{mccloskey1989catastrophic, goodfellow2013empirical}}. A detailed description of catastrophic forgetting is provided in Section {\ref{sec:catastrophic-forgetting}.}}

{In the real world, sequential adaptation to more than one target domain, without necessitating the availability of source data can be desired. In this direction,} Continual learning (CL) - continuous adaptation to new information, has emerged as an important dimension in enhancing the performance and reliability of medical image analysis systems. Unlike transfer learning, CL focuses on both the source domain and the target domain.
A CL approach aims to retain knowledge from previously seen tasks while adapting to new tasks and avoiding catastrophic forgetting issues. CL models are employed for predictive analytics, especially in situations where clinical outcomes can be automatically obtained and incorporated into the algorithm {\mbox{\cite{lee2020clinical}}}. This capability enhances the model's predictive power by learning from real-time patient data. CL methods can also be employed to utilize the multi-modality dataset for better interpretability and analysis. In recent years, an increasing number of CL methods have been explored and proposed in various subareas of computer vision tasks. \refx{fig:pieChartPaperOverYear} presents the paper distributions based on different aspects of CL and shows the growth in exploration in {medical} domain through the rising number of publications over the years. In this paper, we discuss the various aspects of CL, particularly considering its application and implications in the medical domain. Here, we aim to contribute to the ongoing discourse on adapting and improving machine learning models for sustained effectiveness in the dynamic healthcare landscape. By emphasizing CL, we recommend models that not only demonstrate robust performance initially but also possess the ability to evolve and improve over time. CL in medical image analysis represents a powerful approach to developing intelligent systems that can evolve, learn, and adapt to the complexities of healthcare, ultimately contributing to improved patient outcomes and enhanced clinical decision-making. We discuss the limitations and challenges of the existing methods and explore the methods/techniques that can be utilized for developing a robust algorithm. {A thorough search of the existing body of literature highlights the uniqueness of our work as the first comprehensive survey of CL techniques applied to medical image analysis.} 
% \st{A thorough search of the existing body of literature reveals the groundbreaking nature of our work as the first comprehensive survey of CL techniques applied to medical image analysis.} 
This scholarly endeavor aims not only to contribute novel insights but also to establish a foundational reference for researchers, offering a roadmap that can guide future exploration and incite scholarly interest in the academic community. The primary contributions of our academic pursuit are elucidated below:

$\bullet$~Pioneering in its scope, this survey paper provides the first comprehensive exploration of CL applications in the field of medical image analysis. Our focus extends to delivering a thorough overview encompassing all pertinent papers and elucidating details regarding well-known methods in medical image analysis.

$\bullet$~We introduce a rigorous categorization of CL models within the academic community, presenting a systematic taxonomy that categorizes research based on different CL strategies. Our classification discerns between various CL techniques, such as rehearsal, regularization, architectural, and hybrid methods. Additionally, we contextualize these techniques within various medical sub-fields, offering a nuanced academic perspective.

$\bullet$~Beyond application-centric discussions, our exploration delves into the scholarly challenges and open issues surrounding CL in medical image analysis. By addressing academic intricacies, including data annotation costs, temporal drift, and the necessity for benchmarking datasets, we contribute to the scholarly discourse. Additionally, we identify emerging academic trends that give rise to open questions, shaping the trajectory of future academic research in CL applied to medical image analysis.

\subsection*{Motivation and Uniqueness of this Survey}
\noindent
Over the past few decades, CL approaches in computer vision tasks have seen substantial progress, leading to numerous survey papers exploring deep CL models for computer vision tasks \cite{qu2021recent,wang2023comprehensive,de2021continual,mai2022online}. While some of these surveys focus on specific applications, such as classifications~\cite{de2021continual,mai2022online} others take a more general approach to {evaluation policies~\hbox{\cite{mundt2021cleva}} or} concepts and practical perspectives \cite{wang2023comprehensive}. Notably, none of these surveys specifically addresses the applications of CL techniques in medical image analysis, leaving a significant gap in the literature. We believe that insights from successful CL models in vision can be beneficial for the medical community, guiding the retrospective analysis of past and future research directions in CL \cite{verma2023privacy,lee2020clinical}.
CL has proven its potential in developing unified and sustainable deep models capable of handling new classes, tasks, and the evolving nature of data in non-stationary environments across various application areas. Our survey aims to bridge the gap by providing valuable insights that can assist medical researchers, including radiologists, in adopting up-to-date methodologies in their fields. In our survey, we analyze various sources of drifts in medical data, defining CL scenarios in medical images. We present a multi-perspective view of CL by categorizing techniques into rehearsal, regularization, architectural, and hybrid strategies-based methods. The discussion extends beyond applications, encompassing underlying working principles, challenges, and the imaging modality of the proposed methods. We emphasize how this additional information can aid researchers in consolidating literature across the spectrum. A concise overview of our paper is illustrated in \refx{fig:CLtaxonomy}.

%%%%Introduce the field of bioinformatics.
%%%Explain the concept of data drift and its significance.
%%%Present the need for continual learning-based drift adaptation in bioinformatics.
%%%State the objectives and scope of the systematic review.
%%%check these papers:
%https://pubs.rsna.org/doi/epdf/10.1148/radiol.2020200038
%https://www.thelancet.com/journals/landig/article/PIIS2589-7500(20)30102-3/fulltext
%%https://link.springer.com/chapter/10.1007/978-3-030-87234-2_16

%%%%%%%%%%%%%%%%%%%%%%%%%%%%%%%%%%%%%%%%%%%%%%%%%%%%%%%%%%%%%%
% Search strategy
\input{content/sections/introduction/search-strategy}

%%%%%%%%%%%%%%%%%%%%%%%%%%%%%%%%%%%%%%%%%%%%%%%%%%%%%%%%%%%%%%
% Paper organization
\input{content/sections/introduction/paper-organization}

%%%%%%%%%%%%%%%%%%%%%%%%%%%%%%%%%%%%%%%%%%%%%%%%%%%%%%%%%%%%%%

%===================================
% \input{content/tables/abbreviations}
\ifthenelse{\boolean{useSingleColumn}}{

\input{content/tables/abbreviations-onecolumn}

}{
    % Two column image
    \input{content/tables/abbreviations}

}
%===================================

%% file: content/sections/introduction/search-strategy.tex
\subsection*{Search strategy}
\noindent 
\label{sec:search-strategy}
% To conduct a thorough literature search, we followed the same strategy presented in \tabrefcite{azad2023advances,azad2023foundational} and utilized DBLP, Google Scholar, and Arxiv Sanity Preserver, employing custom search queries to retrieve scholarly publications related to our topic-CL. 
% Our search query was\texttt{
%     (continual learning | medical | sequence of tasks) 
%     (segmentation | classification | medical | lifelong learning)}.
% These platforms allowed us to filter results into categories such as peer-reviewed journal papers, conference or workshop proceedings, non-peer-reviewed papers, and preprints. We carefully filtered our search results through a two-step process: initial screening based on the title and abstract, followed by a subsequent screening of the full text. This approach was instrumental in eliminating false positives, ensuring that only relevant content was included in our study.

To conduct a thorough literature search, we followed the same strategy presented in \tabrefcite{azad2023advances,azad2023foundational} and utilized DBLP, Google Scholar, and Arxiv Sanity Preserver, employing custom search queries to retrieve scholarly publications related to our topic-CL. 
Our search query was\texttt{
    (continual learning | medical | sequence of tasks) 
    (segmentation | classification | medical | lifelong learning)}.
These platforms allowed us to filter results into categories such as peer-reviewed journal papers, conference or workshop proceedings, non-peer-reviewed papers, and preprints. 
%%%%%%\st{We carefully filtered our search results through a two-step process: initial screening based on the title and abstract, followed by a subsequent screening of the full text. This approach was instrumental in eliminating false positives, ensuring that only relevant content was included in our study.}
{We filtered our search results through a two-step process: first by screening titles and abstracts, and then by reviewing the full text based on specific criteria as follows:
1. Relevance to Continual Learning: Focus on significant contributions to continual learning or related areas such as lifelong learning, incremental learning, and online learning.
2. Publication Venue: Preference for papers published in reputable journals and conferences known for high-quality research.
3. Novelty and Contribution: Inclusion of papers presenting novel research, methodologies, or applications.
4. Experimental Rigor: Papers must include comprehensive experiments and results validating their claims.
5. Theoretical Foundation: Preference for papers with a strong theoretical foundation, including formal definitions and analysis.}
% This systematic screening ensured our survey includes only high-quality, relevant papers that provide valuable insights into continual learning.

%% file: content/sections/introduction/paper-organization.tex
\subsection*{Paper organization}\label{sec:paper-organization}
\noindent
All the abbreviations used in this manuscript along with their expansion are tabulated in \refx{tab:fullform}.
A background about various sources of drift in medical data, the catastrophic forgetting issue, CL, a pipeline for CL framework, and CL applications is provided in \refx{sec:background}. Then a thorough description of various kinds of continual learning scenarios {explored} in the medical domain is elaborated in \refx{sec:CLscenario}. Further, we cover various categories of continual learning techniques and their applicability in the medical domain via \refx{sec:CLtechnique}. \refx{sec:CLsupervision} provides details about the level of supervision required in different kinds of proposed frameworks. \refx{sec:evaluation} provides comprehensive practical information such as the experimental setups, training process, and evaluation metrics for measuring the plasticity and stability of continual learning frameworks. \refx{sec:futureDirection} discusses the current challenges in the continual learning literature and future directions. Eventually, the last section provides the conclusion of the survey.

%% file: content/tables/abbreviations-onecolumn.tex
\begin{table}[!ht]
	\caption{
         Abbreviations and their expansion.
         Green rows signify abbreviations related to CL (scenarios and techniques), cyan corresponds to imaging modalities and techniques, while orange is associated with evaluation metrics.
    }
	\centering
	\begin{tabular}{ r | l }
        \rowcolor{gray!30}
        \bottomrule
        \textbf{Acronym} & \textbf{Full name}  \\ 
        \hline
        
        \rowcolor{green!10}
        CL & Continual Learning \\
        
        \rowcolor{green!5}
		CIS & Class Incremental Scenario \\
		\rowcolor{green!10}
        DIS & Domain Incremental Scenario \\
		\rowcolor{green!5}
        TIS & Task Incremental Scenario \\
	    \rowcolor{green!10}
        IIS & Instance Incremental Scenario \\	
        
        \rowcolor{green!5}
        EWC & Elastic Weight Consolidation \\
		\rowcolor{green!10}
        DWC & Distributed Weight Consolidation \\
        \rowcolor{green!5}
        MAS & Memory Aware Synapses \\
        \rowcolor{green!10}
        SI & Synaptic Intelligence \\

        \rowcolor{cyan!10}
        sMRI & structural Magnetic Resonance Imaging \\
		\rowcolor{cyan!5}
        CMR & Cardiac Magnetic Resonance \\
        % CMR & Cardiovascular Magnetic Resonance \\
        \rowcolor{cyan!10}
        CT & Computed Tomography \\
        \rowcolor{cyan!5}
        VQA & Visual Question-Answering \\
        \rowcolor{cyan!10}
        H\&E & Hematoxylin and Eosin \\
        % WBC & White Blood Cells \\
        \rowcolor{cyan!5}
            sEMG& Surface electromyography \\
        \rowcolor{cyan!10}
            WSI& Whole Slide Images \\

        \rowcolor{orange!5}
            wCC&weighted Correlation  \\
        \rowcolor{orange!10}
            PCC&Pearson Correlation Coefficient  \\
        \rowcolor{orange!5}
            rMSE&root Mean Square Error \\
        \rowcolor{orange!10}
            FLOPS&Floating Point Operations Per Second\\
        \rowcolor{orange!5}
            SRC&Spearman's rank correlation coefficient  \\
        \rowcolor{orange!10}
            KRC&Kendall rank correlation coefficient \\
        \rowcolor{orange!5}
            AUPRC& Area Under the Precision-Recall Curve \\
        \rowcolor{orange!10}
            SSD & Symmetric Surface Distance \\
	
        \rowcolor{orange!5}
        IoU & Intersection over Union \\
		\rowcolor{orange!10}
        AUC & Area Under ROC Curve \\
		\rowcolor{orange!5}
        MAE & Mean Absolute Error \\
		\rowcolor{orange!10}
        MCR & Mean Class Recall \\
		\rowcolor{orange!5}
        AP & Average Precision \\
        \rowcolor{orange!10}
        DSC & Dice Similarity Coefficient \\
		\rowcolor{orange!5}
        HD95 & 95\% Hausdorff Distance \\
		\rowcolor{orange!5}
        FROC & Free-response Receiver-Operating Characteristic  \\
		\rowcolor{orange!10}
        ROC & Receiver-Operating Characteristic \\
		\rowcolor{orange!5}
        MSD & Mean Surface Distance \\
		\rowcolor{orange!10}
        ASSD & Average Symmetric Surface Distance \\
        \rowcolor{orange!5}
        BWT & Backward transfer \\
        \rowcolor{orange!10}
        FWT & Forward transfer \\
	\rowcolor{orange!5}
        TPR &True Positive Rate \\
        \rowcolor{orange!10} 
        SNR& Signal-to-Noise Ratio \\ 

	\hline
	\end{tabular}
	\label{tab:fullform}
\end{table}

%% file: content/tables/abbreviations.tex
\begin{table}[!ht]
	\caption{
         Abbreviations and their expansion.
         Green rows signify abbreviations related to CL (scenarios and techniques), cyan corresponds to imaging modalities and techniques, while orange is associated with evaluation metrics.
    }
    \vspace{-1em}
	\centering
	\begin{tabular}{ r | l }
        \rowcolor{gray!30}
		\bottomrule
		\textbf{Acronym} & \textbf{Full name}  \\ 
		\hline
        
        \rowcolor{green!10}
        CL & Continual Learning \\
        
        \rowcolor{green!5}
		CIS & Class Incremental Scenario \\
		\rowcolor{green!10}
        DIS & Domain Incremental Scenario \\
		\rowcolor{green!5}
        TIS & Task Incremental Scenario \\
	    \rowcolor{green!10}
        IIS & Instance Incremental Scenario \\	
        
        \rowcolor{green!5}
        EWC & Elastic Weight Consolidation \\
		\rowcolor{green!10}
        DWC & Distributed Weight Consolidation \\
        \rowcolor{green!5}
        MAS & Memory Aware Synapses \\
        \rowcolor{green!10}
        SI & Synaptic Intelligence \\

        \rowcolor{cyan!10}
        sMRI & structural Magnetic Resonance Imaging \\
		\rowcolor{cyan!5}
        CMR & Cardiac Magnetic Resonance \\
        % CMR & Cardiovascular Magnetic Resonance \\
        \rowcolor{cyan!10}
        CT & Computed Tomography \\
        \rowcolor{cyan!5}
        VQA & Visual Question-Answering \\
        \rowcolor{cyan!10}
        H\&E & Hematoxylin and Eosin \\
        % WBC & White Blood Cells \\
        \rowcolor{cyan!5}
            sEMG& Surface electromyography \\
        \rowcolor{cyan!10}
            WSI& Whole Slide Images \\

        \rowcolor{orange!5}
            wCC&weighted Correlation  \\
        \rowcolor{orange!10}
            PCC&Pearson Correlation Coefficient  \\
        \rowcolor{orange!5}
            rMSE&root Mean Square Error \\
        \rowcolor{orange!10}
            FLOPS&Floating Point Operations Per Second\\
        \rowcolor{orange!5}
            SRC&Spearman's rank correlation coefficient  \\
        \rowcolor{orange!10}
            KRC&Kendall rank correlation coefficient \\
        \rowcolor{orange!5}
            AUPRC& Area Under the Precision-Recall Curve \\
        \rowcolor{orange!10}
            SSD & Symmetric Surface Distance \\
	
        \rowcolor{orange!5}
        IoU & Intersection over Union \\
		\rowcolor{orange!10}
        AUC & Area Under ROC Curve \\
		\rowcolor{orange!5}
        MAE & Mean Absolute Error \\
		\rowcolor{orange!10}
        MCR & Mean Class Recall \\
		\rowcolor{orange!5}
        AP & Average Precision \\
        \rowcolor{orange!10}
        DSC & Dice Similarity Coefficient \\
		\rowcolor{orange!5}
        HD95 & 95\% Hausdorff Distance \\
		\rowcolor{orange!5}
        FROC & Free-response Receiver-Operating Characteristic  \\
		\rowcolor{orange!10}
        ROC & Receiver-Operating Characteristic \\
		\rowcolor{orange!5}
        MSD & Mean Surface Distance \\
		\rowcolor{orange!10}
        ASSD & Average Symmetric Surface Distance \\
        \rowcolor{orange!5}
        BWT & Backward transfer \\
        \rowcolor{orange!10}
        FWT & Forward transfer \\
	\rowcolor{orange!5}
        TPR &True Positive Rate \\
        \rowcolor{orange!10} 
        SNR& Signal-to-Noise Ratio \\ 

	\hline
	\end{tabular}
	\label{tab:fullform}
\end{table}

%% file: content/sections/background/index.tex
\section{Background}
\label{sec:background}

% Medical data drifts in non-stationary environment
\input{content/sections/background/medical-data-drifts-in-non_stationary-environment}

% Catastrophic forgetting
\input{content/sections/background/catastrophic-forgetting}

% The Essence of continual learning
\input{content/sections/background/essence-of-cl}

% Continual learning pipeline
%===========================================
\begin{figure}[!ht]
    \centering
    \ifthenelse{\boolean{useSingleColumn}}{
        % Single column image
        \includegraphics[width=.65\textwidth]{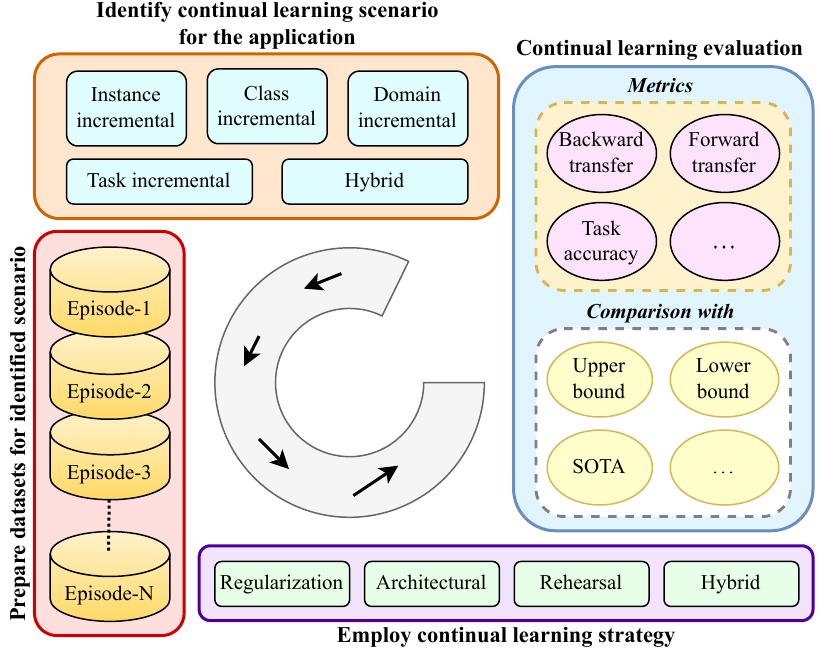}
    }{
        % Two column image
        \includegraphics[width=.48\textwidth]{content/figures/drawn/cl-pipeline.pdf}
    }
    \vspace{-1em}
    \caption{A coarse level flowchart for designing a CL pipeline}
    \label{fig:pipeline}
\end{figure}
%===========================================
\input{content/sections/background/cl-pipeline}

% Application of continual learning
%===========================================
\begin{figure}[!ht]
\centering
    \ifthenelse{\boolean{useSingleColumn}}{
        % Single column image
        \includegraphics[width=.55\textwidth]{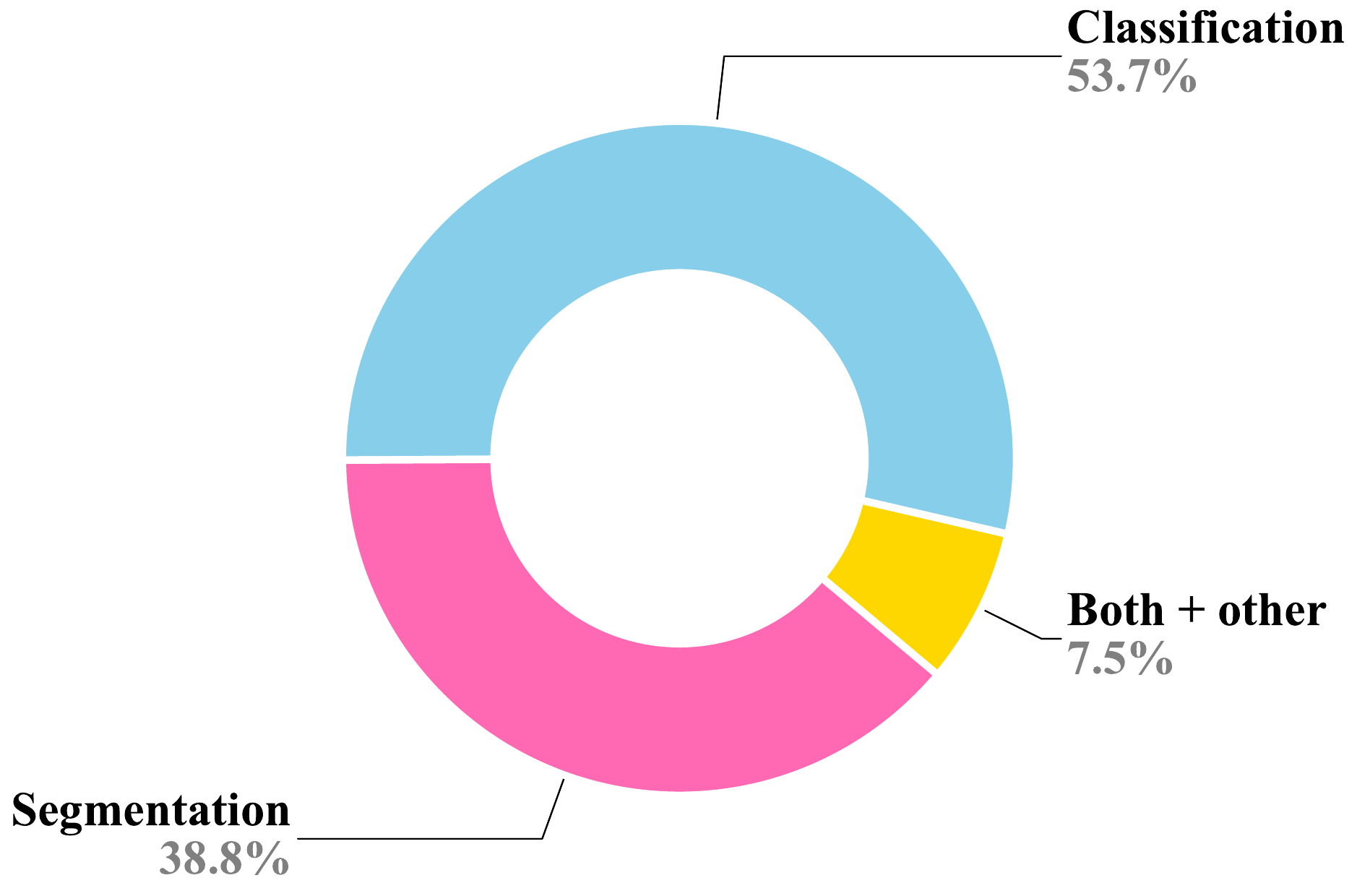}
    }{
        % Two column image
        \includegraphics[width=.48\textwidth]{content/figures/charts/chart_segmentation.pdf}
    }
    \vspace{-1em}
    \caption{Ratio of CL-based research for downstream applications}
    \label{fig:pieChartSegmentation}
\end{figure}
%===========================================
%===========================================
\begin{figure}[!ht]
    \centering
    \includegraphics[width=.48\textwidth]{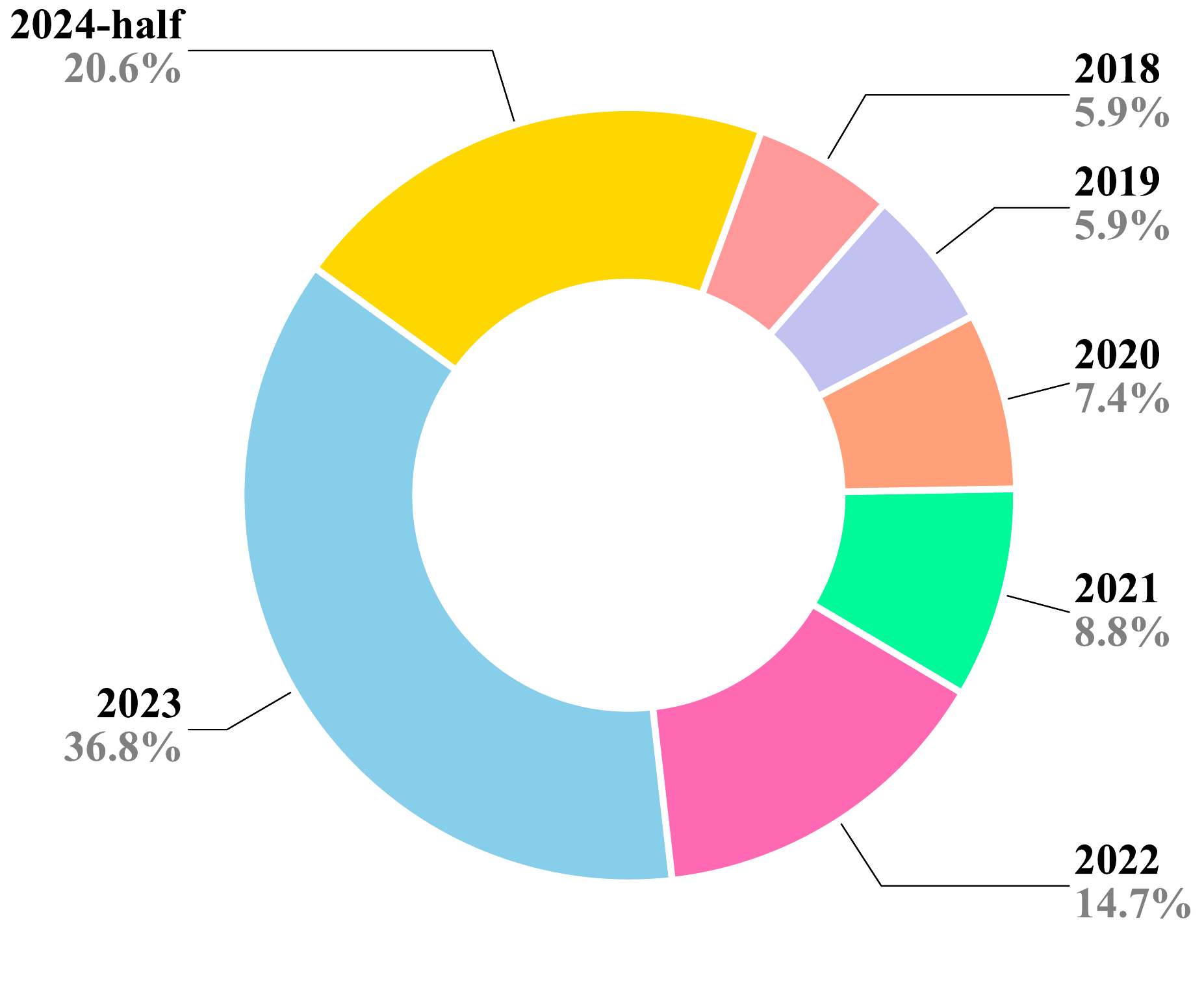}
    % \caption{CL-based research contributions over the year}
    \vspace{-1em}
    \caption{CL-based research contributions over the years. Percentages represent the number of CL papers in the medical domain each year, showing the increasing trend and growing importance of CL research.}
    \label{fig:pieChartPaperOverYear}
\end{figure}
%===========================================
\input{content/sections/background/application-of-cl}

%% file: content/sections/background/medical-data-drifts-in-non_stationary-environment.tex
\subsection{Medical data drifts}
\label{sec:medical-data-drifts}
In clinical practices, the data distribution evolves over time, reflecting the dynamic nature of the healthcare environment \cite{sahiner2023data, LACS, MORE}. Inconsistencies in data collection procedures across different healthcare settings or institutions contribute to data drift. Also, the introduction of new medical imaging devices, diagnostic tools, and data acquisition techniques leads to a shift in the technological landscape \cite{derakhshani2022lifelonger, LACS}. Moreover, the advancement in medical research and the discovery of new diseases/treatment methods raise the understanding of healthcare. This new knowledge can influence the characteristics of medical data that cause shifts in the underlying distribution \cite{LACS, MORE}. Also, data sources can have their dynamics and, therefore, are inherently non-constant. For instance, cardiac CT images are captured under time-varying factors such as breathing and heart rates. Non-homogeneous data is another challenge, as individual health differences among patients can vary over time due to factors like genetics, age, occupation, and lifestyle \cite{LACS, BER}.
Additionally, variations in sample preparation or pre-processing methods contribute to distinctions among imaging datasets. In digital histopathology, differences in staining policies across labs introduce undesired stain variances in whole-slide images \cite{AIPath, AM}. The imaging solution may also influence the final digital visualization throughout the entire learning process over time. Nonlinear augmentation of computed radiography occurs at various degrees due to the differing physical and chemical properties of contrast mediums from different brands. Variables like sensor signal-to-noise ratio (SNR), customized parameter settings in imaging software, and storage-friendly distortions can all impact the quality of the resultant image. For instance, in digital histopathology, billion-pixel whole slide images (WSI) at a fixed magnification have seen variations in storage size, ranging from megabytes to gigabytes per image over the years, consequently enhancing the dataset quality in terms of micron-per-pixel (MPP) for continual learning tasks \cite{AIPath}.

Data drifts can be broadly categorized as the covariate, label, and concept shift. We provide their explanation along with examples in Table~{\ref{tab:driftExample}}. The medical data drifts can have significant implications for the performance and reliability of machine learning and deep learning models \cite{BER}. Traditional machine learning often relies heavily on static data and feature engineering, where human experts manually select the relevant features. In the case of data drift, these handcrafted features may become less informative, and the model may struggle to adapt to new patterns \cite{BAYRAM2022108632}. More specifically, for both machine learning and deep learning, this issue is particularly prevalent in dynamic and non-stationary environments, where the statistical properties of the data evolve. Understanding the sources of drift, such as inconsistencies in data collection and technological advancements, is crucial for developing robust models \cite{BER}. Proactive strategies, including regular model updates and continuous monitoring, are essential to ensure that machine learning models remain effective and reliable in navigating the evolving healthcare landscape by addressing medical data drift and developing machine learning models that can adapt to the ever-changing nature of clinical data, ultimately enhancing patient care and outcomes.

%%%%\textcolor{red}{Discuss the virtual drift, real drift. what drift may occur in this field of study Explain the concept of data drift and its causes in bioinformatics applications. Discuss the challenges and limitations of traditional machine learning approaches in handling data drift.}

%%%%%%%%%%%%%%%%%%%%%%%%%%%%%%%%%%%%%%%%%%%%%%%%%%%%%%%%%%%%%%%
\begin{table*}[!ht]
    \caption{Data drift categorization}
    \vspace{-1em}
    \centering
    \resizebox{.99\textwidth}{!}{
    \begin{tabular}{ p{1.5cm} | p{17cm} } 
        % header			
        \bottomrule
        \rowcolor{gray!20}
        \textbf{Drift type} & \textbf{Detail with example}  \\ \hline
    
        Covariate drift & It refers to situations when input distribution $p(x)$ differs and conditional distribution $p(y|x)$ remains same between source and target data.
        \textit{Example:} Change in WSI staining type induces covariate drift. The model trained on H\&E stained breast cancer data may not perform well on the same tissue with CD8 staining. \\\hline

        \rowcolor{gray!05}
        Label drift & It refers to situations when output distribution $p(y)$ changes, but the conditional distribution $p(x|y)$ remains same.
        \textit{Example:} Inter-annotators difference may induce this kind of drift as some experts may be biased toward annotating a particular disease class. Thus, some classes may be undersampled or oversampled in target compared to the source data. \\\hline
        
        Concept drift & It refers to situations when input and output distribution remains same but the conditional relation $p(y|X)$ differs between source and target data.
        \textit{Example:} association of chest X-ray to COVID +ve class changed over time with new findings.
         %%%%\\(b) changing the diabetes diagnostic criteria for fasting plasma glucose may change the meaning of \\automatically derived labels
        \\\hline
    \end{tabular}
    }
    
    \label{tab:driftExample}
\end{table*}
%%%%%%%%%%%%%%%%%%%%%%%%%%%%%%%%%%%%%%%%%%%%%%%%%%%%%%%%%%%%%%%
%%%%https://pubs.rsna.org/doi/epdf/10.1148/rg.220107

%% file: content/sections/background/catastrophic-forgetting.tex
\subsection{Catastrophic forgetting}
\label{sec:catastrophic-forgetting}
Throughout a lifetime, a human brain continuously acquires knowledge, and learning new concepts or tasks has no detrimental effect on previously learned ones. Instead, learning several closely related concepts even boosts the learning of all associated ones. In contrast, artificial neural networks, although inspired by the human brain, often suffer from 'catastrophic forgetting', a tendency to overwrite or forget the knowledge acquired in the past upon learning new concepts~\cite{mccloskey1989catastrophic,ratcliff1990connectionist}. This can be attributed to the fact that the model entirely optimizes for the given dataset. In other words, a model with optimized weights for a task \textbf{T1}, when trained on a new task \textbf{T2}, will freely optimize the existing weights to meet the objectives in task \textbf{T2}, which may now no longer be optimal for the previous task \textbf{T1}. This can be a significant challenge, especially in scenarios where an AI system is expected to learn and adapt to a stream of tasks or datasets over time. Catastrophic forgetting in neural networks is an interesting phenomenon that has attracted lots of attention in recent research~\cite{goodfellow2013empirical,KUMARI2024117100}. Medical data often come from different sources with varying imaging protocols, equipment, and patient populations. For example, MRI scans from different hospitals may have distinct characteristics, leading to domain shifts that can cause a model to forget previously learned features when introduced to new data. Also, medical datasets are often limited in size and can be highly imbalanced, with some conditions being much more common than others. This imbalance can exacerbate catastrophic forgetting, as the model may overly specialize in newly introduced, more frequent classes at the expense of older, less frequent ones. Addressing catastrophic forgetting is crucial for the development of reliable and effective medical AI systems. By implementing strategies such as regularization, rehearsal, generative replay, dynamic architectures, and domain adaptation, researchers can enhance the robustness of models against forgetting, ensuring consistent and accurate performance in medical applications. This is essential for maintaining diagnostic consistency, adapting to new medical knowledge, and providing trustworthy clinical decision support.

%% file: content/sections/background/essence-of-cl.tex
\subsection{Continual learning overview}
\label{sec:essence-of-cl}
A naive solution to deal with catastrophic forgetting can be retraining the model collectively on old and new data from scratch each time new data with drifted distribution or classes are encountered \cite{lee2020clinical}. This process mostly gives the desired classification or segmentation performance; however, it causes an intense burden on computing and storage requirements and, hence, is impractical for deployment. Additionally, the retraining process requires storing the past data and thus causes privacy violations, which can be a major bottleneck of such a strategy in healthcare applications. 

Alternately, CL, also termed as `continuously learning', `incremental learning', `sequential learning' or `lifelong learning', has emerged as a promising solution in various fields to deal with the catastrophic forgetting issue \cite{de2021continual,mai2022online}. 
%%%%%In dynamic real-world environments, where data distributions may change or new tasks emerge, a continuous adaptation of the environment is crucial for AI systems to be robust, reliable, and efficient. 
It helps in efficiently leveraging existing knowledge and incorporating new information without the need for extensive retraining. The primary goal of CL is to develop techniques and strategies that allow a neural network to learn new tasks while retaining knowledge of previous tasks. In other words, it aims to enable the network to continually adapt to new information without completely erasing or degrading its performance on earlier tasks. Overall, CL helps to address the issue of catastrophic forgetting and minimizes the need for additional resources to store historical data. 
CL offers a range of strategies and methods, such as regularization (constraining weight update to avoid forgetting learned concepts), rehearsal (partially using some form of old data to replay with current data), and architectural modifications (reserving or partitioning network for different tasks), to help neural networks remember and consolidate knowledge from past tasks. These strategies help prevent or reduce catastrophic forgetting and improve the generalization ability of the model.

%% file: content/sections/background/cl-pipeline.tex
\subsection{Continual learning pipeline}
\label{sec:cl-pipeline}
The design of a CL pipeline is illustrated in ~\refx{fig:pipeline} which provides an overview of the key components and stages involved in the construction of the CL pipeline. Given a problem statement, first, we need to identify which CL scenarios it falls under, i.e., whether there is a possibility of domain shifts in future data, inclusion of new classes, or the end application, i.e., the task itself may change. For example, if we want to develop a breast cancer classification model to be able to work on the H\&E dataset from different centers then the datasets across centers may have drift and hence fall into the domain incremental scenario of CL. Detailed information about CL scenarios is provided in \refx{sec:CLscenario}. Once we have identified the CL scenario, training, and testing, datasets need to be prepared to mimic a continual stream of datasets arriving one after another. The sequence of datasets is frequently referred to as tasks, experiences, or episodes in literature. We also use these terms interchangeably in this manuscript. 
%%%%Since the term `task' is also used for end application and thus may be challenging to get the true interpretation of the term, hence, we use the term `episode' to refer to the sequence of datasets in a CL pipeline. 
For each episode, separate training and testing data is prepared; thus, for a given sequence of four datasets, the pipeline requires four train-test pairs to develop and evaluate a CL model. Once the datasets are ready, a CL strategy suitable to the application at hand is identified and deployed in any off-the-shelf deep classification or segmentation model. There are various CL strategies, some offering privacy-preserved learning, while some offer better performance but at the cost of more resources, storage, and privacy violations as they store some past data. Typically, the model is trained on the first training episode and evaluated on testing data from all the episodes. After this, the training shifts to the next episode, where the inclusion of partial training data from the previous episode is possible. Here, updating the model with new training data, the evaluation is again done on all the testing data, and the process repeats until the last episode. Application-specific performance metrics (e.g., accuracy, dice similarity coefficient, etc.) computed on testing data of each episode are observed and analyzed over the sequence. Then, we can compute various metrics on top of it to quantify forgetting and forward transfer. Lastly, the CL framework is evaluated against state-of-the-art works and non-CL methods offering upper and lower bounds of performance. Joint or cumulative training gives the highest average performance, whereas naively finetuning on the current episode gives the lowest performance~\cite{kaustaban2022characterizing,lenga2020continual}.

%% file: content/sections/background/application-of-cl.tex
\subsection{Application of continual learning}\label{sec:application-of-cl}

CL has numerous applications across various domains due to its ability to mitigate catastrophic forgetting. The efficiency and robustness of CL in real-time scenarios derive from its ability to adapt, reduce computational overhead, and address the challenges of dynamic data. This makes it a valuable approach for applications where time constraints, adaptability, and efficiency are of great importance. In the medical domain, CL has been widely explored for various segmentation and classification applications (\refx{fig:pieChartSegmentation}) and continuously exhibited its merits over static models as reflected by the increasing number of research contributions over the year (\refx{fig:pieChartPaperOverYear}).
CL can improve the diagnosis and decision-making ability in a resource-constrained environment. Further, it is beneficial in real-time monitoring of patients, telemedicine, and maintaining a dynamic knowledge base. This section presents the most prominent applications of CL in medical settings in terms of the number of publications.

%%%\textbf{Applications of Domain incremental settings: }
\textbf{Radiology and imaging}:
Radiological imaging techniques are continually advancing, requiring models to adapt to new technologies and methodologies. For CT, research has investigated the impact of scanners and reconstruction parameters on both machine learning predictions and human annotations. The findings indicate that the variability introduced by different scanners has a detrimental effect on radiomics \cite{radio1,radio2}  and other imaging features. Therefore, it is imperative to take into account this scanner variability when developing machine learning models. CL facilitates the dynamic adjustment of models to evolving imaging techniques, ensuring accurate interpretations ~\cite{radio3} ~\cite{radio4}. \refx{tab:classificationCL} and \refx{tab:segmentationCL} present the selected works for CL-based classification and segmentation in the medical domain.

\textbf{Disease progression modeling}: CL is crucial in tracking the progression of diseases such as Alzheimer's Disease (AD). Models must adapt to new patient data over time, incorporating the latest diagnostic criteria and treatment strategies~\cite{li2020continual}. Recent work was proposed for modeling AD progression in a CL manner that respects the longitudinal data sets coming in sequence and ensures equal prediction accuracy for future visits~\cite{Alz}.

%%%%\textbf{Applications of class incremental settings:}
\textbf{Drug discovery}: CL can accelerate drug discovery by continuously integrating newly discovered chemical compounds, pharmacological data, and clinical trial results to predict drug interactions and efficacy. A recent work presented a CL-based model called Multi-Scale Temporal Convolutional Networks-based AntiBacterial Peptide Prediction (MSTCN-ABPpred). This model is designed for the classification and discovery of antibacterial peptides (ABPs). While MSTCN-ABPpred can dynamically adapt and retrain based on predicted ABPs and non-ABPs within protein sequences, it does have limitations in terms of providing information about the identified ABPs' targets, haemotoxicity, cytotoxicity, and minimum inhibitory concentration (MIC) against various bacteria~\cite{DD}.

%% file: content/sections/cl-scenarios/index.tex
\begin{figure}[!ht]
    \centering
    \ifthenelse{\boolean{useSingleColumn}}{
        % Single column image
        \includegraphics[width=.56\textwidth]{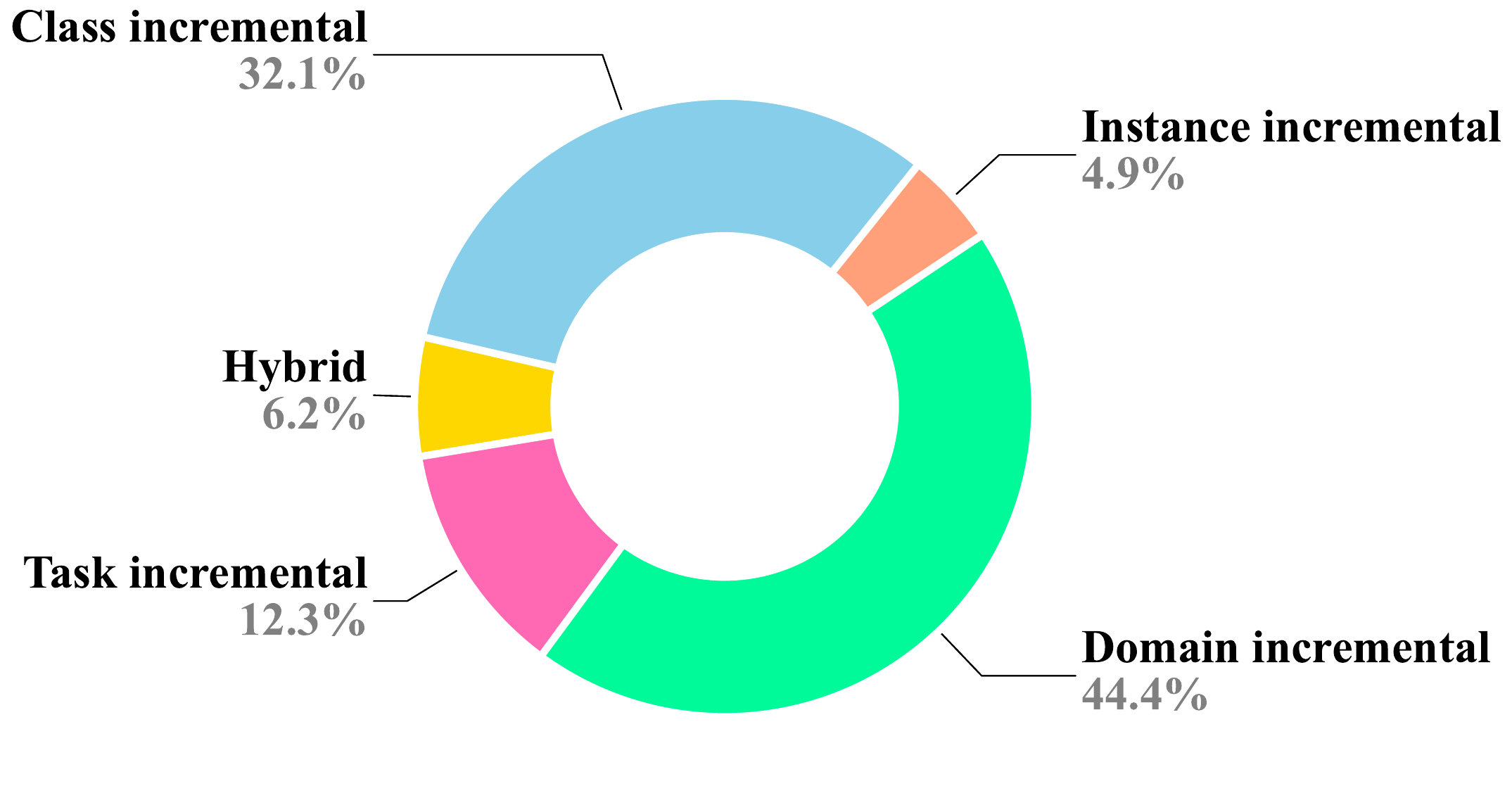}
    }{
        % Two column image
        \includegraphics[width=.48\textwidth]{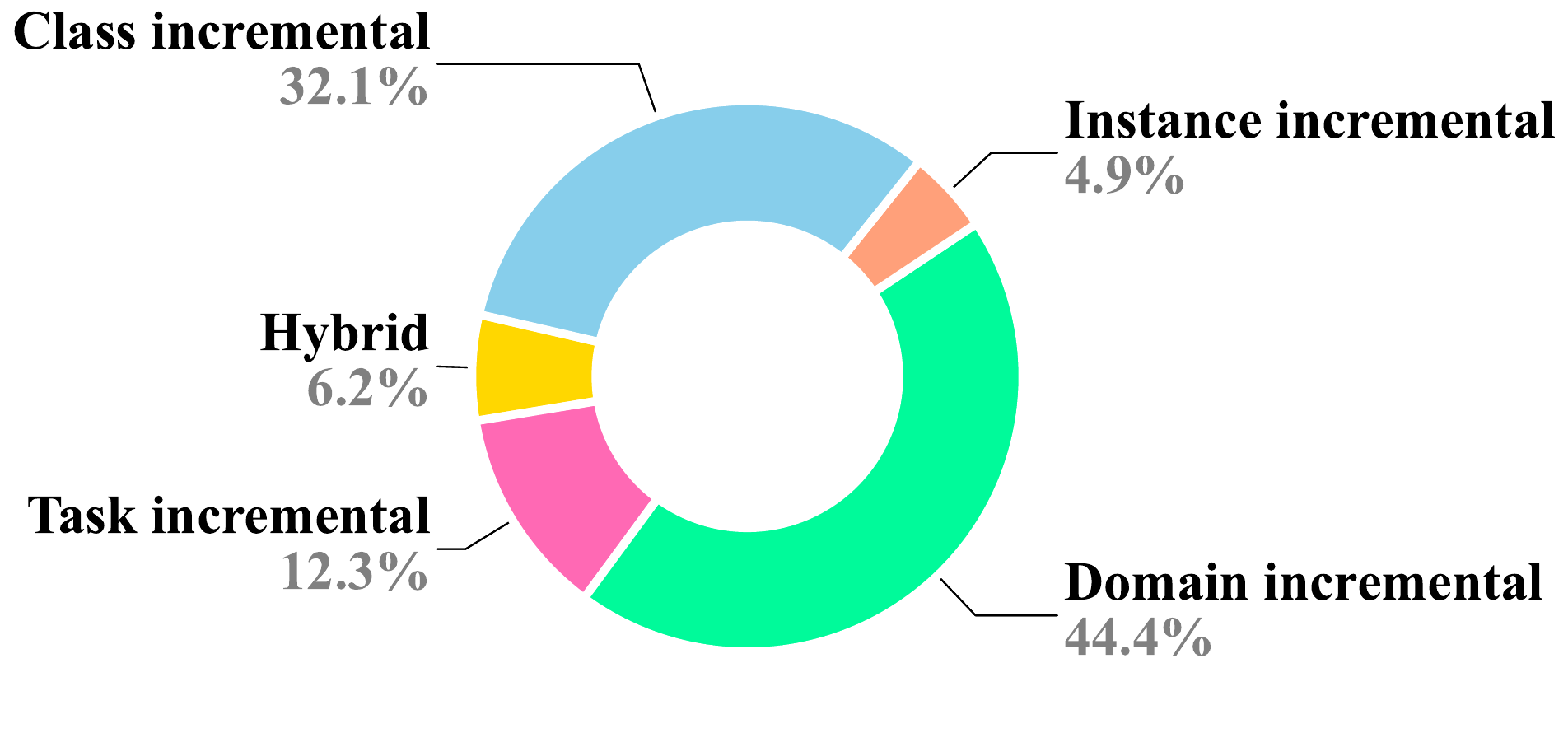}
    }
    \vspace{-1em}
    \caption{Ratio of CL-based works for different incremental scenarios}
    \label{fig:pieChartCLscenarios}
\end{figure}
%===========================================

%%%%%%%%%%%%%%%%%%%%%%%%%%%%%%%%%%%%%%%%%%%%%%%%%%%%%%%%%%%%%%%
\section{Continual learning scenarios}
\label{sec:CLscenario}
Depending upon what kind of change in the sequence of data is expected, the existing CL scenarios can be broadly categorized into five categories, viz., (a) {instance}-incremental, (b) class-incremental, (c) task-incremental, (d) domain-incremental, and (e) hybrid. \refx{fig:pieChartCLscenarios} shows the percentages of the above-mentioned incremental CL scenarios exploited in recent years. The statistics reflect that domain-incremental is highly explored with a 44.4\% ratio out of five major settings, followed by the class-incremental settings having 32.1\% works and {instance}-incremental being the least challenging and rarely explored for different medical image analysis applications.

\input{content/tables/data-incremental-detail}
\input{content/sections/cl-scenarios/data-incremental}

\input{content/tables/class-incremental-detail}

\input{content/sections/cl-scenarios/class-incremental}

\input{content/tables/task-incremental-detail}
\input{content/sections/cl-scenarios/task-incremental}

\input{content/tables/domain-shift-detail}
\input{content/sections/cl-scenarios/domain-incremental}
\input{content/tables/hybrid-detail}
\input{content/sections/cl-scenarios/simulated-or-hybrid-settings}

%% file: content/tables/data-incremental-detail.tex
% %------------dataIncrementalDetailTable-------------------
\begin{table*}[!ht]
\caption{List of various \textbf{instance} incremental scenarios in literature}
\vspace{-1em}
\centering
\resizebox{.99\textwidth}{!}{
\begin{tabular}{ p{3.5cm} | p{5cm} | p{14cm} } 
% header			
\bottomrule
\rowcolor{gray!20}
\textbf{Reference (year)} 
& \textbf{Application (\# Episodes)} 
& \textbf{Description}
\\ \hline

% body
\rowcolor{datainccolor!\tabcellci}
\tabrefcite{ravishankar2019feature}
& Pneumothorax identification from X-ray images (2 Ep.) 
& 4 batches each of 2K were created from a subset of ChestXRay \tabdescite{wang2017chestx} 
\\ \hline

\rowcolor{datainccolor!\tabcellcis}
\tabrefcite{kaustaban2022characterizing}
& Histopathology data-based tumor classification (4 Ep.) 
& Tasks were created in instance incremental fashion from colorectal cancer dataset CRC \tabdescite{kather2019predicting} having a total of 9 classes, (i.e., all 9 classes are present in each episode)
\\ \hline

\rowcolor{datainccolor!\tabcellci}
\tabrefcite{wei2023representative} & 
Brain tumor segmentation (8 Ep.)
& Starting from 160 samples, authors incrementally add 40 more samples after each training session from the LGG Segmentation dataset \tabdescite{buda2019association} 
\\ \hline

\rowcolor{datainccolor!\tabcellcis}
\tabrefcite{bringas2024cladsi} 
&  Alzheimer's disease stage identification using motion-sensor (2 Ep.,3 Ep.,4 Ep.)& 3 experiments were designed with 2, 3, and 4 episodes having per-episode 374, 249, and 187 samples, respectively.
\\ \hline

\end{tabular}
}
\label{tab:dataIncrementalDetailTable}
\end{table*}
%----------------------------------------

%% file: content/sections/cl-scenarios/data-incremental.tex
\subsection{Instance incremental scenarios}
\label{sec:data-incremental-scenarios}

Instance Incremental Scenario (IIS), also termed as data incremental scenario, involves the continuous learning process where the model keeps adapting to incoming data streams that come from the same data distribution. This scenario doesn't involve dealing with entirely new categories or significant changes in data patterns, making it generally the least challenging among all other CL scenarios. In the medical domain, especially digital pathology, a common practice is for expert pathologists to annotate a dataset in multiple stages or batches~\cite{kaustaban2022characterizing}. Each of these annotation batches can be thought of as representing samples from the same underlying data distribution, ensuring some consistency in the data source across different batches. Thus, IIS scenarios deal with the condition where the samples of a dataset are sequentially made available. Owning to its simplicity, this category is not very popular. Only a few classification applications are explored in this scenario, as mentioned in \refx{tab:dataIncrementalDetailTable}. For example, in the breast cancer classification application, \citet{chen2022breast} set up IIS on the BreakHis~\cite{spanhol2015dataset} dataset. More and more samples from each of the malignant and benign classes are added over the subsequent episodes. 
\citet{ravishankar2019feature} try to exhibit a scenario for pneumothorax detection in a hospital where data arrives in an incremental fashion. They sequentially fed 2000 samples in 4 subsequent episodes from an X-ray dataset for Pneumothorax identification consisting of a total of 8000 samples. A label drift is also possible in real-world IIS settings. However, the majority of previous works deliberately curate a similar amount of samples for a particular class over episodes, and hence, label drift phenomena are not considered. One potential solution to produce the desired number of samples for classes affected by label-shift from past episodes would be to use a generator model such as GAN  \mbox{\cite{byunconditional}}.

%%%%Ravishankar et al.~\cite{ravishankar2019feature} considered Pneumothorax identification from X-ray images in data incremental setting.They emulated incremental learning by dividing the 8,000 training images into incremental groups of 2,000 and assessed the model's performance using a separate test set of approximately 2,000 images. This simulation closely resembles a real-world scenario where a model is deployed for pneumothorax detection in a hospital, and data arrives in an incremental fashion.
%%%%%%Chen et al.~\cite{chen2022breast} setup data incremental scenario for breast cancer classification application on BreakHis dataset. Authors include more samples from each of malignant and benign classes over the episodes.

%% file: content/tables/class-incremental-detail.tex
% %-------------classIncrementalDetailTable------------------
\begin{table*}[!ht]
\caption{List of various class incremental scenarios in literature}
\vspace{-1em}
\centering
\resizebox{.99\textwidth}{!}{
\begin{tabular}{ p{3.2cm} | p{5.3cm} | p{14cm} } 
% header			
\bottomrule
\rowcolor{gray!20}
\textbf{Reference (year)} 
& \textbf{Application (\# Episodes)} 
& \textbf{Description}
\\ \hline 

% body
\rowcolor{classinccolor!\tabcellci}
\tabrefcite{ozdemir2018learn}
& MRI based humerus and scapula segmentation (2 Ep.) 
& Incrementally adding anatomical structure, private data
\\ \hline

\rowcolor{classinccolor!\tabcellcis}
\tabrefcite{ozdemir2019extending}
& MRI based knee segmentation (2 Ep.)  
& SKI10 MICCAI Grand Challenge \tabdescite{heimann2010segmentation} 
\\ \hline

\rowcolor{classinccolor!\tabcellci}
\tabrefcite{li2020continual}
& Dermoscopic images based skin disease classification (4-20 Ep.) 
& (a) CIS for Skin8 dataset having total 8 classes on which 2 classes per episode used, (b) CIS for Skin40 dataset from ISIC2019 \tabdescite{tschandl2018ham10000} consists of 40 classes of skin disease images collected from the internet, here 2/5/10 classes per episode used in 3 experiments, (c) non-medical dataset CIFAR100 with 100 classes,  here 5/10/20 classes per episode used in 3 experiments. 
\\ \hline

\rowcolor{classinccolor!\tabcellcis}
\tabrefcite{liu2022learning}
& Class incremental segmentation of abdomen organs (4 Ep.) 
& 4 episodes were created for incremental segmentation of liver, spleen, pancreas, right kidney, and left kidney organs from CT datasets \tabdescite{MultiAtl34online,simpson2019large}, KiTS \tabdescite{heller2019kits19} + private
\\ \hline

\rowcolor{classinccolor!\tabcellci}
\tabrefcite{derakhshani2022lifelonger}
&Disease classification (4 Ep.) 
&  TissueMNIST, OrganaMNIST, PathMNIST, and BloodMNIST datasets from MedMNIST repository \tabdescite{yang2023medmnist} were used in 4 experiments, each with 4 episodes. 
\\ \hline

\rowcolor{classinccolor!\tabcellcis}
\tabrefcite{akundi2022incremental}
& Chest X-ray classification in CIS (5 Ep.) 
& 5 classes from CheXpert dataset  
\\ \hline

\rowcolor{classinccolor!\tabcellci}
\tabrefcite{kaustaban2022characterizing}
& Histopathology data based tumor classification (4 Ep.) 
& 4 episodes were created in CIS from Colorectal cancer dataset CRC \tabdescite{kather2019predicting} having a total 9 classes
\\ \hline

\rowcolor{classinccolor!\tabcellcis}
\tabrefcite{chen2022breast}
& Histopathology data-based breast cancer classification (4 Ep.) 
& 4 episodes created in CIS from BreakHis dataset
\\ \hline

\rowcolor{classinccolor!\tabcellci}
\tabrefcite{zhang2023continual}
& (a) CT based abdomen, gastrointestinal, and other organ segmentation (7, 3, 4 classes in 3 Ep.), (b) abdomen segmentation to liver tumor segmentation task (13 classes, 1 class in 2 Ep.) 
& (a) Abdomen segmentation from JHH \tabdescite{xia2022felix} (private) 
(b) 13 class abdomen segmentation dataset as BTCV \tabdescite{landman2015miccai} and liver tumor segmentation data as LiTS \tabdescite{bilic2023liver}
\\ \hline

\rowcolor{classinccolor!\tabcellcis}
\tabrefcite{ji2023continual}
& 3D CT scan based whole-body organs segmentation (4 Ep. in 2 permutations of episodes) 
& TotalSegmentator \tabdescite{wasserthal2023totalsegmentator} (103 classes: various organs), 3 private datasets: ChestOrgan (31 classes: chest scans), HNOrgan (13 classes: head and neck scans), EsoOrgan (1 class: esophageal cancer) 
\\ \hline

\rowcolor{classinccolor!\tabcellci}
\tabrefcite{chee2023leveraging}
&(a) Cancer classification (3,5 Ep.), (b) diabetic retinopathy classification (2 Ep.), (c) skin lesions classification (3,4 Ep.) 
& (a) Histopathology data based human colorectal cancer classification: CCH5000 \tabdescite{kather2016multi} (total 8 classes, Exp=(4,1,1,1,1) classes, Exp=(4,2,2) classes), (b) diabetic retinopathy (DR) classification using retinal images: EyePACS \tabdescite{kaggleDiabeticRetinopathy} (total 5 classes: no DR, mild DR, moderate DR, severe DR and proliferative DR, Exp=(3,2) classes), (c) pigmented skin lesions classification: HAM10000 \tabdescite{tschandl2018ham10000} (total 7 classes: Exp=(4,1,1,1) classes, Exp=(4,2,1) classes)
\\ \hline

\rowcolor{classinccolor!\tabcellcis}
\tabrefcite{zhang2023adapter}
& Disease classification (4-10 Ep.) 
& (a) skin lesions classification with Skin8 dataset (International Skin Imaging Collaboration (ISIC) \tabdescite{tschandl2018ham10000}) having 8 classes distributed over 4 episodes, (b) Path16, a pathology image collection from various public sources: total 16 classes distributed over 7 episodes, (c) CIFAR100: non-medical images, 100 classes distributed over 5-10 episodes  
\\ \hline

\rowcolor{classinccolor!\tabcellci}
\tabdescite{bai2023revisiting} 
& Surgical visual-question localized-answering 
& EndoVis18, EndoVis17, M2CAI  
\\ \hline

\rowcolor{classinccolor!\tabcellcis}
\tabrefcite{sadafi2023continual}
& Microscopic images based WBC classification (4 Ep.) 
& With 3 different datasets, 3 separate CIS experiments each having 4 episodes: (a) (3,3,3,4) classes from Matek-19 \tabdescite{matek2019single}, (b) (3,3,3,4) classes from INT-20, and (c) (3,2,2,3) classes from Acevedo-20 \tabdescite{acevedo2020dataset}
  \\ \hline

\rowcolor{classinccolor!\tabcellci}
\tabrefcite{wang2023rethinking} 
& (a) segmentation for endoscopy (2-3 Ep.), (b) surgical instrument segmentation (2 Ep.) 
& Incrementally adding segmentation structure from EDD2020 \tabdescite{ali2021deep,ali2020endoscopy} dataset which includes 5 classes as Barrett's esophagus, cancer, high-grade dysplasia, polyp, and suspicious. Initially, 3-4 classes are in 1st episode, and then 2 or 1 classes are incrementally added in subsequent episodes. (b) EndoVis18 \tabdescite{allan20202018} and EndoVis17 \tabdescite{allan20192017} datasets as two episodes having some overlapping classes 
\\ \hline 

% \rowcolor{yellow}
\rowcolor{classinccolor!\tabcellcis}
\tabrefcite{hua2023incremental} 
&sEMG-based gesture classification (4 Ep.) 
&2 experiments were created, each with 4 episodes using Ninapro DB2 sEMG dataset for gesture recognition. The first experiment has (8, 11, 14, 17) gesture classes and the second experiment has (10, 20, 30, 40) gesture classless.
\\ \hline 

% \rowcolor{yellow}
\rowcolor{classinccolor!\tabcellci}
\tabrefcite{huang2023conslide} 
& Tumor subtype classification at WSI-level (4 Ep.)&Incrementally learning tumor classes using 4 datasets (NSCLC, BRCA, RCC, ESCA) from TCGA\footnote{https://www.cancer.gov/ccg/research/genome-sequencing/tcga} project. Each dataset has 2 distinct tumor subtypes leading to a total 8 subtypes.
\\ \hline 

% \rowcolor{yellow}
\rowcolor{classinccolor!\tabcellcis}
\tabrefcite{xiao2023fs3dciot} 
&Skin disease classification (18 Ep.)&
Incrementally learning 5 new skin disease classes in each episode  
\\ \hline 

% \rowcolor{yellow}
\rowcolor{classinccolor!\tabcellci}
\tabrefcite{li2023doctor} 
&Disease classification on physiological signals (2 Ep.)& 3 separate experiments were designed, each with 2 episodes. (a) using CovidDeep\tabdescite{hassantabar2021coviddeep}, 1st episode has healthy and symptomatic patients and 2nd episode only asymptomatic patients,
(b) using DiabDeep\tabdescite{yin2019diabdeep} dataset: 1st episode has healthy and Type-I diabetic patients and 2nd episode only Type-II diabetic patients, and (c) using MHDeep\tabdescite{hassantabar2022mhdeep} dataset: 1st episode has participants with healthy condition and major depressive disorder and 2nd episode has bipolar depressive disorder patients.
\\ \hline

\end{tabular}
}
\label{tab:classIncrementalDetailTable}
\end{table*}
%----------------------------------------
%%%%%%%%%%%%%%%%%%%%%%%%%%%%%%%%%%%%%%%%%%%%%%%%%%%%%%%%%%%%%%%%%

% %-------------classIncrementalDetailTable------------------
\begin{table*}[!ht]
\ContinuedFloat
\caption{List of various class incremental scenarios in literature (\textit{Continued})}
\vspace{-1em}
\centering
\resizebox{.99\textwidth}{!}{
\begin{tabular}{ p{3.2cm} | p{5.3cm} | p{14cm} } 
% header			
\bottomrule
\rowcolor{gray!20}
\textbf{Reference (year)} 
& \textbf{Application (\# Episodes)} 
& \textbf{Description}
\\ \hline 

% \rowcolor{yellow}
\rowcolor{classinccolor!\tabcellci}
\tabrefcite{sun2023few} 
& Multi-class classification on time-series signals (5 Ep., 10 Ep., 5 Ep., 4 Ep.)& 4 separate experiments were designed as follows: 
(a) using Mit-BIH dataset \tabdescite{goldberger2000physiobank} 4 episodes were created where 1st episode has 4 classes and the 4 new classes are added one by one in the next 4 episodes. 
(b) using FaceAll dataset \tabdescite{dau2019ucr}, 10 episodes were created where 1st episode has 5 classes and the next 9 episodes introduce 9 new classes, one by one. 
(c) using Wave dataset \tabdescite{dau2019ucr} a total of 5 episodes were created where 1st episode has 12 lasses and the next 4 classes introduce new 12 classes, 3 classes each. 
(d) using Mit-BIH Long-Term ECG dataset \tabdescite{goldberger2000physiobank} 4 episodes were created where 3 classes were in 1st episode and the next 3 episodes contained 3 new classes, 1 class in each.
\\ \hline 

% \rowcolor{yellow}
\rowcolor{classinccolor!\tabcellcis}
\tabrefcite{yang2023continual} 
&Skin disease classification on dermoscopic and clinical images: (4-20 Ep.)&2 datasets were used separately as follows: (a) Skin7~\tabdescite{codella2018skin} containing a total of 7 classes is used in CIS where 1 and 2 classes are incrementally added leading to 6 and 4 episodes, respectively. (b) Skin40 dataset~\tabdescite{sun2016benchmark} containing a total of 40 classes is used in CIS where 2,5, and 10 classes are incrementally added leading to 20, 8, and 4 episodes, respectively.
\\ \hline

% \rowcolor{yellow}
\rowcolor{classinccolor!\tabcellci}
\tabrefcite{verma2023privacy} 
& Disease classification on (a) fundus (3 Ep.), (b) pathology images: (3 Ep.)& (a) OCT dataset~\tabdescite{kermany2018large} with total 4 classes used to create 3 episodes: episode-1 has two classes (Normal and Choroidal Neovascularization), episode-2 has 1 class (Diabetic Macular Edema) and episode-3 contains 1 class (Drusen). 
(b) PathMNIST dataset~\tabdescite{yang2023medmnist} from the MedMNIST repository offers 9 classes for colon pathology. It is used to create 3 episodes with 3 classes each.
\\ \hline

% \rowcolor{yellow}
\rowcolor{classinccolor!\tabcellcis}
\tabrefcite{ceccon2024fairness} 
&Chest X-ray based disease classification (5 Ep.) &2 datasets (ChestX-ray14 \tabdescite{wang2017chestx} with 14 classes and CheXpert \tabdescite{irvin2019chexpert}  with 12 classes) used separately to create two experiments, each with 5 episodes. 
 \\ \hline

% \rowcolor{yellow}
\rowcolor{classinccolor!\tabcellci}
\tabrefcite{bayasi2024gc} 
& (a) Skin lesion classification (2-3 Ep.), (b) blood cell classification (4 Ep.), (c) colon tissue classification (4 Ep.)   &  (a) HAM10000~\tabdescite{tschandl2018ham10000},
Dermofit~\tabdescite{ballerini2013color}, and Derm7pt~\tabdescite{kawahara2018seven} datasets with total 7, 7, and 6 classes used to create 3, 3, and 2 episodes each with 2-3 non-overlapping skin disease classes, respectively, 
(b) PBS-HCB~\tabdescite{acevedo2020dataset} dataset with a total of 8 classes used to create 4 episodes each with 2 non-overlapping blood cell classes, 
(c) NCT-CRC-HE~\tabdescite{kather2019predicting} dataset with a total of 9 classes used to create 4 episodes each with 2-3 non-overlapping colon tissue classes.
\\ \hline

% \rowcolor{yellow}
\rowcolor{classinccolor!\tabcellcis}
\tabrefcite{qazi2024dynammo} 
& (a) Disease classification in histopathology images: (7 Ep.), (b) Skin lesion classification: (4 Ep.) & (a) Using Path16 dataset 7 episodes were created as different sources of dataset with different classes from histopathology, (b) using skin8 dataset, a total of 4 episodes were created by adding 2 new classes per episode.
\\ \hline

% \rowcolor{yellow}
\rowcolor{classinccolor!\tabcellci}
\tabrefcite{zhu2024lifelong} &Histopathology WSI retrieval: (4 Ep.)
&Authors curate a sequence of 4 WSI datasets (NSCLC, RCC, BRCA and GAST) from TCGA. Each episode introduces a new dataset with 2 new classes of cancer, showing a CIS setting. 
\\ \hline 

\end{tabular}
}
\label{tab:classIncrementalDetailTable2}
\end{table*}
%----------------------------------------

%% file: content/sections/cl-scenarios/class-incremental.tex
\subsection{Class incremental scenarios}
\label{sec:cl-scenarios}

%%% from a paper: It refers to continuously extend an existing model to unseen classes with new streams of data. Since the knowledge to learn from each batch of unseen classes is largely non-overlapping, this scenario is expected to be the hardest among CL scenarios. For example, adding new tissue types in a tumor classification model or adding new cell types into a cell detection model for AI-assisted diagnosis.

Class Incremental Scenario (CIS) refers to a situation where the aim is to adapt the model to accommodate novel classes from new data streams. Given that each batch of these unseen classes offers knowledge that largely differs from the previous batches, this scenario is anticipated to be the most challenging among continual learning scenarios. An instance of this challenge is incorporating new types of tissues into a tumor classification model or introducing novel cell types into a model designed for AI-assisted diagnosis in cell detection. In this scenario, there is only one incremental task that gets additions of new classes in subsequent episodes. Some works assume mutually exclusive classes, whereas some classes are overlapping.

Multi-class datasets with more than 5-6 classes are mainly considered to create a class incremental scenario. There can be an equal or variable number of classes across the episodes. \citet{chen2022breast} propose a breast cancer classification model where they iteratively include malignant and benign classes with one sub-class from each category. This scenario is very popular for segmentation applications where the model may be demanded to segment more and more classes over time. 
An exhaustive list of class incremental scenarios explored in literature is provided via \refx{tab:classIncrementalDetailTable}. References, along with detailed segmentation or classification applications, are tabulated.

%%% References are provided along with detailed segmentation or classification applications mention the class incremental scenarios are tabulated.  

%%%Chen et al.~\cite{chen2022breast} curates a class incremental scenario brom BreakHis breast cancer classification dataset. The dataset has 4 sub-classes for malignant and benign cases; the authors iteratively include these classes in subsequent episodes.
%%%%%%Ozdemir et al.~\cite{ozdemir2018learn} without an assumption on mutual exclusivity of classes. Our experimental dataset consists of 9 Dixon sequences of left shoulder collected with 1.5 T at resolution of 0.91 mm × 0.91 mm × 3 mm, corresponding to 192 × 192 × 64 voxel resolution. Humerus and scapula bones were annotated by an expert. Our goal is to combine knowledge from different data for a network that can segment both anatomical structures. We evaluated our proposed method for the 3 scenarios shown in the table in Fig. 2. One volume each was fixed randomly for validation and testing each of all scenarios. The first scenario (Case 1) tests a typical setting where different anatomies were of interest and thus annotated in separate studies. The second scenario (Case 2) aims to observe advantages of incremental training with minimal effort, i.e., incrementally annotated data, giving insight on an extreme case where a single volume annotation is provided. The last scenario (Case 3) studies the feasibility of combining learned segmentation information from different anatomy and images of different contrast.

%% file: content/tables/task-incremental-detail.tex
% %------------taskIncrementalDetailTable---------------------
\begin{table*}[!ht]
\caption{List of various task incremental scenarios in literature}
\vspace{-1em}
\centering
\resizebox{.99\textwidth}{!}{
\begin{tabular}{ p{3.5cm} | p{5cm} | p{14cm} } 
% header			
\bottomrule
\rowcolor{gray!20}
\textbf{Reference (year)} 
& \textbf{Application (\# Episodes)} 
& \textbf{Description}
\\ \hline 

% body
\rowcolor{taskinccolor!\tabcellci}
\tabrefcite{baweja2018towards}
& MRI-based normal brain structures segmentation (2 Ep.)
& UK Biobank \tabdescite{miller2016multimodal} dataset used to create two tasks: 1$^{\text{st}}$) segmentation of cerebrospinal fluid, grey matter, white matter and 2$^{\text{nd}}$) segmentation of white matter lesions.
\\ \hline

\rowcolor{taskinccolor!\tabcellcis}
\tabrefcite{ravishankar2019feature}
& Chest X-ray view classification (2 Ep.) 
& Shift from task (4ch vs. PLAX) to task (2ch vs. PSAX) 
\\ \hline 

\rowcolor{taskinccolor!\tabcellci}
\tabrefcite{zhang2019continually}
&Longitudinal MRI-based Alzheimer’s
disease progression modeling (7 Ep.) 
& Tasks are MR images belonging to different time points
\\ \hline 

\rowcolor{taskinccolor!\tabcellcis}
\tabrefcite{kaustaban2022characterizing}
& Histopathology data-based tumor classification (4 Ep.) 
& 4 episodes were created in TIS from Colorectal cancer dataset CRC \tabdescite{kather2019predicting} having a total of 9 classes 
\\ \hline

\rowcolor{taskinccolor!\tabcellci}
\tabrefcite{derakhshani2022lifelonger}
& Disease classification (4 Ep.) 
& 4 episodes were created from each of the datasets in TissueMNIST, OrganaMNIST, PathMNIST, BloodMNIST (MedMNIST repository \tabdescite{yang2023medmnist}) 
\\ \hline 

\rowcolor{taskinccolor!\tabcellcis}
\tabrefcite{bera2023memory}
&MRI-based binary segmentation (3 Ep.) 
&3 episodes were curated as binary segmentation application using MRI data on 3 organs (prostate, spleen, hippocampus): Promise12~\tabdescite{litjens2014evaluation} (prostate) $\rightarrow$ MSD~\tabdescite{NCIISBI291online} (spleen) $\rightarrow$ Drayd~\tabdescite{denovellis2021hippocampal} (hippocampus) 
\\ \hline

% \rowcolor{yellow}
\rowcolor{taskinccolor!\tabcellci}
\tabrefcite{wu2024modal} & Image super-resolution (4 Ep.)
& First 3 episodes are curated from IXI~\tabdescite{brainixidataset} dataset referring to (PD, T1, and T2) weighted brain MRI, and the last episode refers to chest X-ray from Chest X-ray~\tabdescite{wang2017chestx} dataset.
\\ \hline 

% \rowcolor{yellow}
\rowcolor{taskinccolor!\tabcellcis}
\tabrefcite{li2023doctor} 
&Disease classification on physiological signals (2-3 Ep.)& A task incremental scenario with
2-3 different disease classification tasks were created using 2-3 datasets (CovidDeep\tabdescite{hassantabar2021coviddeep}, DiabDeep\tabdescite{yin2019diabdeep}, and MHDeep\tabdescite{hassantabar2022mhdeep})
\\ \hline

% \rowcolor{yellow}
\rowcolor{taskinccolor!\tabcellci}
\tabrefcite{verma2024confidence} 
& Disease classification on (a) Fundus (2 Ep.), (b) pathology images (3 Ep.)& (a) OCT~\tabdescite{kermany2018large} with total 4 classes used to create 2 episodes, each having 2 classes
(b) PathMNIST~\tabdescite{yang2023medmnist} from the MedMNIST repository offers 9 classes for colon pathology. It is used to create 3 episodes, each with 3 classes.
\\ \hline

% \rowcolor{yellow}
\rowcolor{taskinccolor!\tabcellcis}
\tabrefcite{ye2024continual} &Multi-modality representation learning (5 Ep.) & 5 different medical data modalities including medical report, MRI, X-ray, CT, and histopathology data were learned in 5 episodes for SSL-based representation learning
\\ \hline

\end{tabular}   
}
\label{tab:taskIncrementalDetailTable}
\end{table*}
%----------------------------------------

%% file: content/sections/cl-scenarios/task-incremental.tex
\subsection{Task incremental scenarios}
\label{sec:task-incremental-scenarios}
Task Incremental Scenario (TIS) comes into the picture when we have a multi-task problem and a single adaptive model is desired. Each task is considered an episode. Thus each episode has disjoint label space. %%%Task ids can be provided or not during training and testing.
However, there is ambiguity in the literature with task incremental scenarios as the other scenarios (class and domain) are frequently referred to as task incremental scenarios by different research communities. For example, \citet{ravishankar2019feature} create the first episode having 2 chest x-ray views (2ch vs. PLAX) and then the next episode with 2 other views (4ch vs. PSAX) as a task incremental scenario, which can also be a class incremental scenario. Further, another example in this line can be found in the work by \citet{baweja2018towards}. Here, the authors create a task incremental scenario with 2 episodes for brain MRI segmentation application where the $1^{st}$ episode was multi-class segmentation of cerebrospinal fluid, grey matter, and white matter, and the $2nd$ episode was the segmentation of white matter lesions. If we have multiple tasks that are very different, then it would certainly be the task incremental; however, if the tasks are more close then it is merely a design choice whether to treat them as class incremental or task incremental. Another work where \citet{kaustaban2022characterizing} explore domain incremental scenario as organ shift for tumor classification problem where the first episode is a colon cancer dataset with nine classes (CRC dataset), and the next episode is a breast cancer dataset with 2 classes (PatchCam dataset). Similar is the case with \citet{sadafi2023continual} where White Blood Cells (WBC) classification from 3 datasets, each having a different number of classes, is regarded as domain incremental, which can also be regarded as task-incremental. \citet{derakhshani2022lifelonger} curate class incremental and task incremental scenarios from the same dataset taken from MedMNIST \cite{yang2023medmnist} where a dataset is divided into non-overlapping classes leading to 4 episodes. If the model is evaluated only on the learned class in the specific episode, then they regard it as task incremental; otherwise, if evaluated on cumulative classes from all seen episodes, then class incremental scenario.
\citet{kaustaban2022characterizing} say that any of the scenarios among instance, class, and domain incremental can be treated as task incremental, provided that each incoming data stream (episode) is treated as a distinct task. For a comprehensive overview, we have tabulated all the works along with their claimed task incremental scenario in \refx{tab:taskIncrementalDetailTable}.

Further, it is essential to consistently provide prior information about which specific task (referred to as task identity, i.e., task ID) the test data pertains to, and predictions are made accordingly. However, if task IDs are not provided, then this category is termed as a task-free scenario which is a more challenging one.
%%%%%%\cite{ravishankar2019feature} formulated new task scenarios by having 2 different classes in the next episode. Specifically shifting from 2ch vs. PLAX to 4ch vs. PSAX was considered task incremental. 
%%% %%%%%%from a paper: Any of the scenarios in data, class, and domain incremental can be considered as task incremental if each data stream is defined as a new task and during inference prior knowledge is always given as to which task (task identity, i.e. task ID) the test data should be predicted with.

%% file: content/tables/domain-shift-detail.tex
% %-----------domainShiftDetailTable------------------
\begin{table*}[!ht]
\caption{List of various domain shift scenarios in literature}
\vspace{-1em}
\centering
\resizebox{.99\textwidth}{!}{
\begin{tabular}{ p{3.2cm} | p{5cm} | p{15cm} } 
% header			
\bottomrule
\rowcolor{gray!20}
\textbf{Reference (year)} 
& \textbf{{Shift source} (\# Episodes)} 
& \textbf{Description}
\\ \hline 

% body
\rowcolor{domaininccolor!\tabcellci}
\tabrefcite{mcclure2018distributed} 
& Cross-site (4 Ep.) 
& sMRI based Axial and sagittal brain segmentation: datasets from 4 sites: HCP \tabdescite{van2013wu}, NKI \tabdescite{nooner2012nki}, Buckner \tabdescite{biswal2010toward}, WU120 \tabdescite{power2017temporal} in different order 
\\ \hline 

\rowcolor{domaininccolor!\tabcellcis}
\tabrefcite{karani2018lifelong} 
& Cross-scanners, hospitals, or acquisition protocols (4 Ep.) 
& MR brain segmentation on various combinations using 5 datasets from different sources: Human Connectome Project \tabdescite{van2013wu}, Alzheimer's Disease Neuroimaging Initiative~\tabdescite{uscADNIAlzheimerapossDisease}, Autism Brain Imaging Data Exchange \tabdescite{di2014autism} and Information eXtraction from Images~\tabdescite{brainixidataset} 
\\ \hline

\rowcolor{domaininccolor!\tabcellci}
\tabrefcite{van2019towards} 
& Low vs. high-grade (2 Ep.) 
& MR data based glioma segmentation: 2018 BraTS Challenge (low and high-grade), an in-house dataset with non-enhancing low-grade 
\\ \hline

\rowcolor{domaininccolor!\tabcellcis}
\tabrefcite{venkataramani2019towards} 
& Varying in disease type, intensity patterns (2) \& contrast 
& X-ray lung segmentation: Montgomery \tabdescite{jaeger2014two} as source, JSRT \tabdescite{shiraishi2000development} and Pneumoconiosis (private) as two target domains 
\\ \hline

\rowcolor{domaininccolor!\tabcellci}
\tabrefcite{ravishankar2019feature} 
& Inter-subject variability (2 Ep.) 
& (a) Chest X-ray view classification (4ch vs. PLAX) from adult to pediatric as two domains (b) chest X-ray view classification (2ch vs. PSAX) from adult to pediatric as two domains 
\\ \hline 

\rowcolor{domaininccolor!\tabcellcis}
\tabrefcite{lenga2020continual} 
& Cross-sites (2 Ep.) 
& Chest X-ray classification on datasets from 2 sources: ChestX-ray14 dataset from NIH Clinical Center and MIMIC-CXR dataset from the Beth Israel Deaconess Medical Center 
\\ \hline 

\rowcolor{domaininccolor!\tabcellci}
\tabrefcite{ozgun2020importance} 
& Different age ranges, MRI field strengths \& presence of pathologies or motion artifacts (4 Ep.) 
& Brain MRI segmentation datasets: 4 Ep. were created from 3 datasets: CANDI \tabdescite{kennedy2012candishare} (1 Ep.), ADNI \tabdescite{jack2008alzheimer} (1 Ep.), and MALC \tabdescite{asman2013non} (2 Ep. based on age) 
\\ \hline

\rowcolor{domaininccolor!\tabcellcis}
\tabrefcite{hofmanninger2020dynamic} 
&  Different scanner parameters and target-shift (3 Ep.)
& Chest CT synthetic classification task: the second episode have change in scanner parameter as compared to first episode \tabdescite{hofmanninger2020automatic}, then target shift is introduced intentionally in the third episode by imprinting a synthetic target structure in the form of a cat on random locations, rotations and varying scale in 50\% of the samples. 
\\ \hline

\rowcolor{domaininccolor!\tabcellci}
\tabrefcite{morgado2021incremental} 
& Appearance and body-parts (2 Ep.) 
& Dermatological imaging modality classification (full-body, anatomic, macroscopic, and dermoscopic classes): For each of the classes, author considered disjoint images in the two episodes, for example, in the second episode, legs and arms images were considered for full-body class, images containing hands or feet only considered for anatomic, images containing face region for macroscopic class, and pink colored images for dermoscopic class. 
\\ \hline

\rowcolor{domaininccolor!\tabcellcis}
\tabrefcite{srivastava2021continual} 
& Cross-sites (3 Ep.) 
& Chest X-ray classification: NIH Chest-X-rays14, PadChest, and CheXpert 
\\ \hline

\rowcolor{domaininccolor!\tabcellci}
\tabrefcite{memmel2021adversarial} 
& Multi-scanner (2 Ep.) 
& MRI-based Hippocampal Segmentation: 3 datasets as 2018 Medical Segmentation Decathlon challenge \tabdescite{simpson1902large}, Scientific Data \tabdescite{kulaga2015multi}, and Alzheimer's Disease Neuroimaging Initiative \tabdescite{boccardi2015training}: 2 datasets together in 1st episode and then 3rd dataset in 2nd episode, different such combinations were experimented  
\\ \hline

\rowcolor{domaininccolor!\tabcellcis}
\tabrefcite{perkonigg2021continual} 
&  Multi-scanners (3 Ep.)  
& Brain age estimation: IXI\tabdescite{brainixidataset} (Philips Gyroscan Intera 1.5T, Philips Intera 3.0T scanner), OASIS3 \tabdescite{lamontagne2019oasis} (Siemens TrioTim 3.0T scanner) 
\\ \hline

\rowcolor{domaininccolor!\tabcellci}
\tabrefcite{bayasi2021culprit} 
& Multi-site (6 Ep.) 
& Skin lesion image classification: HAM10000 \tabdescite{tschandl2018ham10000}, Dermofit \tabdescite{ballerini2013color}, Derm7pt \tabdescite{kawahara2018seven}, MSK \tabdescite{codella2018skin}, PH2 \tabdescite{mendoncca2013ph},  UDA \tabdescite{codella2018skin}  
\\ \hline

\rowcolor{domaininccolor!\tabcellcis}
\tabrefcite{zhang2021comprehensive} 
& Cross-sites (6 Ep.), (4 Ep.) 
& (a) Prostate segmentation: MRI datasets across 6 sites: RUNMC \tabdescite{bloch2015nci}, BMC \tabdescite{bloch2015nci}, HCRUDB \tabdescite{lemaitre2015computer}, UCL \tabdescite{litjens2014evaluation}, BIDMC \tabdescite{litjens2014evaluation}, HK \tabdescite{litjens2014evaluation}, (b) optic cup and disc segmentation: public multi-site fundus image datasets from 4 sources \tabdescite{sivaswamy2015comprehensive,fumero2011rim,orlando2020refuge} 
\\ \hline

\rowcolor{domaininccolor!\tabcellci}
\tabrefcite{perkonigg2022continual} 
& Multi-scanner (4 Ep.) 
& (a) CMR based Cardiac segmentation (4 Ep.): Cardiac \tabdescite{campello2021multi}, (b) CT based Lung nodule detection (4 Ep.): LIDC \tabdescite{setio2017validation}, + LNDb challenge \tabdescite{pedrosa2019lndb}, (c) MRI based Brain Age Estimation segmentation (4 Ep.): IXI\tabdescite{brainixidataset} + OASIS-3 \tabdescite{lamontagne2019oasis} 
\\ \hline

\rowcolor{domaininccolor!\tabcellcis}
\tabrefcite{derakhshani2022lifelonger} 
& Across organs or modality (4 Ep.) &Disease classification: 4 episodes using 4 datasets ordered as: BloodMNIST, OrganaMNIST, PathMNIST, TissueMNIST (MedMNIST repository) 
\\ \hline

\rowcolor{domaininccolor!\tabcellci}
\tabrefcite{karthik2022segmentation} 
& Cross-center (8 Ep.) 
& Sclerosis lesions segmentation from brain MRI datasets described in Kerbrat et al. \tabdescite{kerbrat2020multiple} 
\\ \hline

\rowcolor{domaininccolor!\tabcellcis}
\tabrefcite{gonzalez2022task} 
& (a) Cross-domain (5 Ep.), (b) manual image contrast change (5 Ep.) 
& MRI hippocampus segmentation: (a) HarP \tabdescite{boccardi2015training}, Dryad \tabdescite{kulaga2015multi}, Decathlon \tabdescite{simpson2019large}, (b) data transformation applied using TorchIO library on Decathlon \tabdescite{simpson2019large} as intensity rescaling, affine transformations, rotation, translation 
\\ \hline

\rowcolor{domaininccolor!\tabcellci}
\tabrefcite{ranem2022continual} 
& Cross-site (3 Ep.) 
& MRI based binary hippocampus segmentation: Decathlon \tabdescite{antonelli2022medical}, Drayd \tabdescite{denovellis2021hippocampal}, HarP \tabdescite{boccardi2015training}
\\ \hline

\rowcolor{domaininccolor!\tabcellcis}
\tabrefcite{kaustaban2022characterizing}
& (a) Multi-organ (2 Ep.), (b) scanning protocol (5 Ep.) 
& (a) Tumor classification: CRC (a colon cancer dataset with 9 classes), PatchCam (a breast cancer dataset with 2 classes), (b) Tumor classification: 5 domain shift scenarios were simulated by changing H\&E composition in CRC dataset 
\\ \hline

\rowcolor{domaininccolor!\tabcellci}
\tabrefcite{shu2022replay}
& (a) Imaging protocol (2 Ep.), (b) multi-source (3 Ep.) 
& (a) Low and high-quality fundus images (retinal images) from EyeQ dataset as domain shift condition \tabdescite{fu2019evaluation}, (b) fundus disease classification (AMD, DR, glaucoma, myopia, and normal classes) across 3 datasets: ODIR~\tabdescite{grandchallengeODIR2019Grand}, R\&R (RIADD~\tabdescite{grandchallengeRIADDISBI2021} + REFUGE~\tabdescite{grandchallengeREFUGEGrand}), and iSee \tabdescite{fang2020attention} 
\\ \hline

\rowcolor{domaininccolor!\tabcellcis}
\tabrefcite{li2022domain} 
& Multi-scanner vendors (4 Ep.) 
& CMR based Cardiac segmentation of left ventricle, right ventricle, and left ventricle myocardium: 4 episodes as 4 scanner vendors (Siemens, Philips, General Electric, Cannon) from M\&Ms \tabdescite{campello2021multi} dataset 
\\ \hline

\rowcolor{domaininccolor!\tabcellci}
\tabrefcite{sadafi2023continual} 
& Cross-sites (3 Ep.) 
& WBC classification across 3 datasets: Matek-19 \tabdescite{matek2019single}, INT-20, and Acevedo-20 \tabdescite{acevedo2020dataset} 
\\ \hline

\rowcolor{domaininccolor!\tabcellcis}
\tabrefcite{bera2023memory} 
& (a) Cross-center (4 Ep.), (b) cross-center (2 Ep.) 
& (a) Binary prostate segmentation: Prostate158 \tabdescite{adams2022prostate158} $\rightarrow$ NCI-ISBI \tabdescite{NCIISBI291online} $\rightarrow$ Promise12 \tabdescite{litjens2014evaluation} $\rightarrow$ Decathlon \tabdescite{antonelli2022medical}, (b) binary hippocampus segmentation: Drayd \tabdescite{denovellis2021hippocampal} $\rightarrow$ HarP \tabdescite{boccardi2015training}) 
\\ \hline 

\rowcolor{domaininccolor!\tabcellci}
\tabrefcite{zhu2023uncertainty} 
& (a) Cross-site (6 Ep.), (b) cross-site \& cross-modality (2 Ep.), (c) same-site \& cross-modality (2 Ep.) 
& (a) Binary prostate segmentation from T2-weighted MRI scans collected from 6 sites (12–30 scans/site) \tabdescite{liu2020shape,bloch2015nci,lemaitre2015computer}, (b) multi-class (liver, left and right kidneys, and spleen) abdominal segmentation between 30 CT and 20 MRI T2-SPIR scans, (c) muscle segmentation of 13 lower-leg muscles and bones between 30 MRI T1 and 30 mDixon scans 
\\ \hline

\rowcolor{domaininccolor!\tabcellcis}
\tabrefcite{bandi2023continual} 
& Cross-organ (3 Ep.) 
& Histopathology data based cancer detection from breast (CAMELYON16, CAMELYON17), colon (private), and head-neck (private) datasets 
\\ \hline
\end{tabular}
}
\label{tab:domainShiftDetailTable}
\end{table*}
%-----------------------------------------------
%%%%%%%%%%%%%%%%%%%%%%%%%%%%%%%%%%%%%%%%%%%%%%%%%%%%%%%%%%%%%%%%%%%%%%%%%
% %-----------domainShiftDetailTable------------------
\begin{table*}[!ht]
\ContinuedFloat
\caption{List of various domain shift scenarios in literature (\textit{Continued})}
\vspace{-1em}
\centering
\resizebox{.99\textwidth}{!}{
\begin{tabular}{ p{3.2cm} | p{5cm} | p{15cm} } 
% header			
\bottomrule
\rowcolor{gray!20}
\textbf{Reference (year)} 
& \textbf{{Shift source} (\# Episodes)} 
& \textbf{Description}
\\ \hline 

% \rowcolor{yellow}
\rowcolor{domaininccolor!\tabcellci}
\tabrefcite{byunconditional} 
& Various demographics, collection periods, camera types, and image quality (2-3 Ep.) 
& (a) Diabetic retinopathy severity classification across 2 datasets (Messidor-2, APTOS), (b) dermoscopy skin lesion detection across 3 datasets (BCN2000, PAD-UEFS-20, HAM10000)
\\ \hline

% \rowcolor{yellow}
\rowcolor{domaininccolor!\tabcellcis}
\tabrefcite{sun2023adaptive} 
&Data distribution shifts in time series vital signals (10 time-steps.) & Mortality prediction (COVID-19 \tabdescite{yan2020interpretable} datasets), (b) Sepsis situation identification (SEPSIS \tabdescite{seymour2017time} dataset)
\\ \hline 
    
% \rowcolor{yellow}
\rowcolor{domaininccolor!\tabcellci}
\tabrefcite{li2023doctor} 
& Distributional-drift (2 Ep.)& Disease classification on physiological signals: for each of the 3 datasets (CovidDeep\tabdescite{hassantabar2021coviddeep}, DiabDeep\tabdescite{yin2019diabdeep}, and MHDeep\tabdescite{hassantabar2022mhdeep}), 2 episodes were curated by splitting patient into 2 groups.
\\ \hline

% \rowcolor{yellow}
\rowcolor{domaininccolor!\tabcellcis}
\tabrefcite{chen2023generative} & Multi-site, multi-vendor (3 Ep., 3 Ep., 6 Ep.) & 3 separate applications including (a) Optic disc, (b) cardiac, (c) prostate segmentation were considered in domain shift condition \\ \hline

% \rowcolor{yellow}
\rowcolor{domaininccolor!\tabcellci}
\tabrefcite{li2024dual} 
& Multi-vendor (5 Ep.) & MR Cardiac segmentation (LV, RV, MYO) with 2 datasets (ACDC and M\&M). Episode-1 is from ACDC dataset which is collected from Siemens scanners and the next 4 episodes are from M\&M dataset which is collected from 4 vendors (Siemens, Philips, General Electric, and Cannon).
\\ \hline

% \rowcolor{yellow}
\rowcolor{domaininccolor!\tabcellcis}
\tabrefcite{kim2024continual} 
&Multi-site (4 Ep.)& Arrhythmia detection on ECG datasets: 4 datasets (\tabdescite{zheng202012}, \tabdescite{wagner2020ptb}, \tabdescite{alday2020classification}, \tabdescite{liu2018open}) used as 4 episodes
\\ \hline

% \rowcolor{yellow}
\rowcolor{domaininccolor!\tabcellci}
\tabrefcite{aslam2024cel} 
&Data distribution shifts in time series vital signals (10 contexts with significant change in mean and standard deviation)
& Disease outbreak detection via time series signal:  
3 datasets (Mpox~\tabdescite{mathieu2022mpox}, Influenza~\tabdescite{cdcFluViewInteractive},   Measles~\tabdescite{europaSurveillanceAtlas})) were used separately to curate 10 episodes.
\\ \hline

% \rowcolor{yellow}
\rowcolor{domaininccolor!\tabcellcis}
\tabrefcite{bayasi2024gc} 
&Multi-source (4 Ep.)&Skin lesion classification: 4 publicly available skin disease classification datasets (HAM10000~\tabdescite{tschandl2018ham10000},
Dermofit~\tabdescite{ballerini2013color}, Derm7pt~\tabdescite{kawahara2018seven}, MSK~\tabdescite{codella2018skin})  were used as 4 episodes, all with same 4 classes of skin disease\\ \hline

% \rowcolor{yellow}
\rowcolor{domaininccolor!\tabcellci}
\tabrefcite{zhu2024boosting}& Multi-acquisition, multi-equipment & (a) Prostate segmentation (6 Ep.), (b) Cardiac
segmentation (3 Ep.) 
\\ \hline

% \rowcolor{yellow}
\rowcolor{domaininccolor!\tabcellcis}
\tabrefcite{thandiackal2024multi} 
&Multi-source (3 Ep.)   & Histopathology tissue classification: 3 different datasets (K-19~\tabdescite{kather2016multi}, K-16~\tabdescite{kather2019predicting}, CRC-TP~\tabdescite{javed2020cellular}) are considered as 3 episodes which contain 7 medically relevant patch classes from H\&E stained WSIs of colorectal biopsies 
\\ \hline

\end{tabular}
}
\label{tab:domainShiftDetailTable2}
\end{table*}
%-----------------------------------------------

%% file: content/sections/cl-scenarios/domain-incremental.tex
\subsection{Domain incremental scenarios}
\label{sec:domain-incremental-scenarios}
Domain Incremental Scenario (DIS) is the most popular and frequently observed category of CL scenarios for medical applications. 
Similar to IIS, here also, the task remains the same over time. However, in contrast to IIS, where data arrives from a single domain, here, the episodic data originates from a different domain or context. This scenario aligns with the idea of learning in a changing environment where the datasets from different domains (e.g., research sites, hospitals, imaging modality, image acquisition protocol, etc.) are incrementally encountered over time, which thus involve covariance shift-induced discrepancies in data. 
%%%%Various environmental factors can cause change in the data distribution over time. 

In clinical applications, this might encompass changes in factors like the methods used for tissue processing and staining, the demographics of the patient population, or the types of scanning instruments employed, among other variables. Thus, continuous modifications in diagnostic techniques result in alterations in the appearance of medical images. Factors such as the brand of the scanning machine, the method used to create the images, radiation dosage, and specific settings in the scanning process, including the use of contrast agents, can all influence how the images look. These changes in image characteristics, which occur independently of the actual biological content being scanned lead to domain shifts.

Even if we have a single task and a fixed number of classes, these shifts can pose a challenge for the static deep model used in clinical settings because these shifts can quickly make existing models outdated and less effective. Here, a CL model aims to continuously update itself by incorporating new data streams that come from distributions that have shifted over time. 

There have been plenty of attempts to develop CL models and evaluate different kinds of domain incremental scenarios. All the domain incremental scenarios are not equi-hard; some works consider a simpler level of domain shift, whereas some have a severe domain shift. To provide a comparative view, we tabulate all the domain incremental scenarios considered in literature through \refx{tab:domainShiftDetailTable}.

%% file: content/tables/hybrid-detail.tex
\begin{table*}[!ht]
\caption{List of various hybrid CL scenarios in literature}
\vspace{-1em}
\centering
\resizebox{.99\textwidth}{!}{
\begin{tabular}{ p{3cm} | p{4cm} | p{14cm} } 
% header			
\bottomrule
\rowcolor{gray!20}
\textbf{Reference (year)} 
& \textbf{Application (\# Episodes)} 
& \textbf{Description}
\\ \hline

\rowcolor{hybridinccolor!\tabcellci}
\tabrefcite{yang2023few} 
& MedMNIST \tabdescite{yang2023medmnist} based disease classification (4 Ep.) 
& CIS+DIS: 3 datasets (PathMNIST, DermaMNIST, and OrganAMNIST) as source domain and then 3 more datasets (RetinaMNIST, BreastMNIST, BloodMNIST) as domain shift but classes are added in the class incremental fashion with 1 class at a time 
\\ \hline 

% body
\rowcolor{hybridinccolor!\tabcellcis}
\tabrefcite{liu2023incremental} 
& Brain tumor segmentation (3 Ep.) 
& CIS with domain shift conditions: incrementally learn tumor core (BraTS2013 \tabdescite{menze2014multimodal} dataset), the enhancing tumor (TCIA \tabdescite{clark2013cancer} dataset), and edema (CBICA \tabdescite{bakas2018identifying} dataset) structures via three datasets, each following different data distributions 
\\ \hline

\rowcolor{hybridinccolor!\tabcellci}
\tabrefcite{sadafi2023continual} 
& WBC classification (3 Ep.) 
& DIS (multi-site) + CIS using 3 WBC classification datasets having different number of classes: CIS on Matek-19 \tabdescite{matek2019single} then shift to INT-20 dataset and use it in CIS manner then shift to Acevedo-20 \tabdescite{acevedo2020dataset} dataset and use in CIS manner 
\\ \hline

%%it is repeated from first row                                                                
% \rowcolor{hybridinccolor!\tabcellci}
%          \tabrefcite{liu2023incremental} 
% & MRI-based brain tumor segmentation (4 Ep.) 
% & CIS+DIS Incrementally adding more structures to segmentation task from dataset acquired from different vendors/centers 
% \\ \hline 

% \rowcolor{yellow}
\rowcolor{hybridinccolor!\tabcellcis}
\tabrefcite{ceccon2024multi} 
&Chest X-ray based disease classification (7 Ep.) & Authors curate a novel scenario termed NIC by interleaving new classes and new instances (from a new dataset as a new domain) in a sequence. The first episode contains a fixed set of classes from ChestX-ray14 \tabdescite{wang2017chestx} dataset and then the same classes from CheXpert \tabdescite{irvin2019chexpert} dataset were considered as second episode. Then some new disjoint classes from ChestX-ray14 were considered in the third and fourth episodes. Thus, authors interleave between new classes and new domains to curate a hybrid scenario containing a total of 7 episodes. 
\\ \hline

% \rowcolor{yellow}
\rowcolor{hybridinccolor!\tabcellci}
\tabrefcite{bayasi2024gc} 
&Skin lesion classification (5 Ep.) &  CIS+DIS: 5 skin lesion classification datasets ( HAM10000~\tabdescite{tschandl2018ham10000},
Dermofit~\tabdescite{ballerini2013color}, Derm7pt~\tabdescite{kawahara2018seven}, MSK~\tabdescite{codella2018skin},
UDA~\tabdescite{codella2018skin}, BCN~\tabdescite{combalia2019bcn20000}, PH2~\tabdescite{mendoncca2013ph}) with different number of classes (overlapping) were sequentially arranged as 5 episodes which exhibit incremental class and domain simultaneously. 
\\ \hline

\end{tabular}
}
\label{tab:hybridDetailTable}
\end{table*}
%----------------------------------------

%% file: content/sections/cl-scenarios/simulated-or-hybrid-settings.tex
\subsection{Simulated or Hybrid Settings}
\label{sec:simulated-or-hybrid-settings}
It may not always be the case that the model observes a particular kind of incremental scenario, such as a domain incremental condition. The CL model developed for handling domain incremental conditions may not perform well for a class incremental condition. Thus, there are attempts to evaluate the same model on different CL scenarios; however, a separate evaluation is followed for each kind of incremental CL scenario. In contrast, having a set of episodes with a mix of CL scenarios is close to real-life conditions. Thus it becomes important to also test the performance in hybrid settings. {\hbox{\refx{tab:hybridDetailTable}} provides a comprehensive overview of works that designed hybrid incremental scenarios.}

An attempt in direction is made by \citet{sadafi2023continual} where other than pure domain incremental and class incremental, a hybrid incremental scenario is also considered. The authors create a sequence of episodes with both class and domain-incremental cases on multi-site WBC classification datasets. They create a long sequence of episodes with a total of 12 episodes. First, they set up a class incremental scenario within the Matek-19~\cite{matek2019single} dataset by adding 3-4 classes in each subsequent episode up to a total of 4 episodes. After this a new domain, i.e., the INT-20 dataset was considered in the same fashion, contributing 4 more episodes. Finally, a third domain i.e., a new dataset Acevedo-20~\cite{acevedo2020dataset} was introduced with 4 episodes in class incremental fashion. 

{Compared to the natural image domain, novel class appearance along with domain shifts is more frequent in medical applications owing to inherent heterogeneity in staining agents, protocols, imaging techniques, vendors, etc. Therefore, there has been recent interest and development of a novel CL scenario that consider new classes as well as new domains/instances, also termed as `New Instances and New Classes' (NIC) scenario \mbox{\cite{ceccon2024multi}}. The NIC scenario is particularly relevant for medical image analysis as it addresses the simultaneous occurrence of new types of medical conditions (new classes) and new patient data (new instances), which is common in clinical practice. This scenario emphasizes the need for CL methods that effectively handle class and domain incremental learning. 
}
%%%%%A novel direction for having a mix of scenarios is considered by \citet{liu2023incremental}. 
In contrast to traditional class-incremental settings, which often do not account for data shift, \citet{liu2023incremental} propose a method for brain tumor segmentation that incrementally adds structures under varying domain shift conditions, such as distinct sites, scanners, or MRI modalities. This approach addresses class-incremental learning by incorporating domain-specific variations.
{\mbox{\citet{bayasi2024gc}} tackle a hybrid continual learning (CL) scenario using five skin lesion datasets from different sources and domains, each containing overlapping and varying numbers of classes. By treating these five datasets as five sequential episodes, their approach exemplifies the integration of both class-incremental and domain-incremental challenges. Similarly, \mbox{\citet{ceccon2024multi}} addresses a hybrid CL scenario in the context of chest X-ray-based disease classification. They interleave disease classes and domains by utilizing datasets from two hospitals, each containing 18 classes, to curate a total of seven episodes. Each episode introduces either new disease classes or a new domain, thus combining class-incremental and domain-incremental learning within their methodology. Medical applications frequently encounter novel classes and domain shifts due to the diversity in staining agents, imaging protocols, and vendor-specific techniques. This has led to the development of new CL scenarios or hybrid learning scenarios.}

Further, there are some attempts to curate shifts in datasets and various incremental scenarios instead of collecting real datasets reflecting the situation. For example, \citet{kaustaban2022characterizing} simulated domain shift in H\&E stain exhibiting distinct appearance caused by different staining protocols. This dataset is designed to simulate real-world scenarios, where data shifts occur due to differences in scanners, stainers, reagents, and other factors. They considered 9 classes from CRC \cite{kather2019predicting} which is a H\&E stain-based colorectal cancer classification dataset. Each class is divided into 5 disjoint sets for 5 episodes. The first set, regarded as Domain1 is without any alteration and the next 4 sets for each class undergo various changes to create domain shift. 
%%%%A visualization with 3 classes is shown in Fig~\ref{fig:kaustaban_domain_inc}. 
A short description of 4 domain shifts is as follows: \textbf{Domain 2} (increased stain intensity, simulating, for example, concentration increase of the eosin and/or hematoxylin solutions, each with a different extent of change). \textbf{Domain 3} (decreased eosin stain intensity, simulating, for example, slides prepared from many years ago with fading stain). \textbf{Domain 4} (change of hue, simulating, for example, change of reagent manufacturer, scanner or stainer) \textbf{Domain 5} (change of hue and saturation, simulating, for example, change of reagent manufacturer, scanner or stainer). Finally, authors used this augmented dataset for setting class-incremental, domain-incremental, data-incremental, as well as task-incremental scenarios.

Another line of research popularly uses MNIST or similar well-defined datasets instead of collecting real datasets for the evaluation of the CL model. In the medical field, there is a recently released (year 2021) MNIST-like collection of biomedical images offering dozens of datasets, termed MedMNIST \cite{yang2023medmnist}. \citet{derakhshani2022lifelonger} consider 4 MedMNIST datasets, including BloodMNIST, OrganaMNIST, PathMNIST, and TissueMNIST as 4 domains and curate 3 CL scenarios with it.  
%%%Authors considered 4 datasets from MedMNIST repository, viz., BloodMNIST, OrganaMNIST, PathMNIST, and TissueMNIST which are also treated as 4 domains. 
These datasets are multi-class (8-11 classes in each) disease classification datasets from different imaging modalities and organs. Each dataset was split to have disjoint classes leading to a total of 4 episodes in each dataset. Task and class incremental scenarios were set up for each dataset separately. In the task incremental scenario, the aim was to evaluate only the classes from the current episode whereas in class incremental, the aim was to evaluate all seen classes. Thus the class incremental scenario is difficult than task incremental. They term the datasets generated from different sites/imaging protocols as \textbf{cross-domain incremental} scenario, instead of domain-incremental. Here the evaluation is done across the 4 datasets (BloodMNIST, OrganaMNIST, PathMNIST, and TissueMNIST), i.e., each dataset is treated as an episode. Further authors presented domain-agnostic and domain-aware settings in this. Domain-agnostic case is more difficult than domain-aware as the domain ID is not explicitly provided in domain-agnostic. 
%%%%%%%A pictorial representation by the authors is presented in Fig.~\ref{fig:Derakhshani_cl_scenarios}.

%% file: content/sections/cl-technique/index.tex
%===========================================
\begin{figure}[!ht]
    \centering
    \ifthenelse{\boolean{useSingleColumn}}{
        % Single column image
        \includegraphics[width=.55\textwidth]{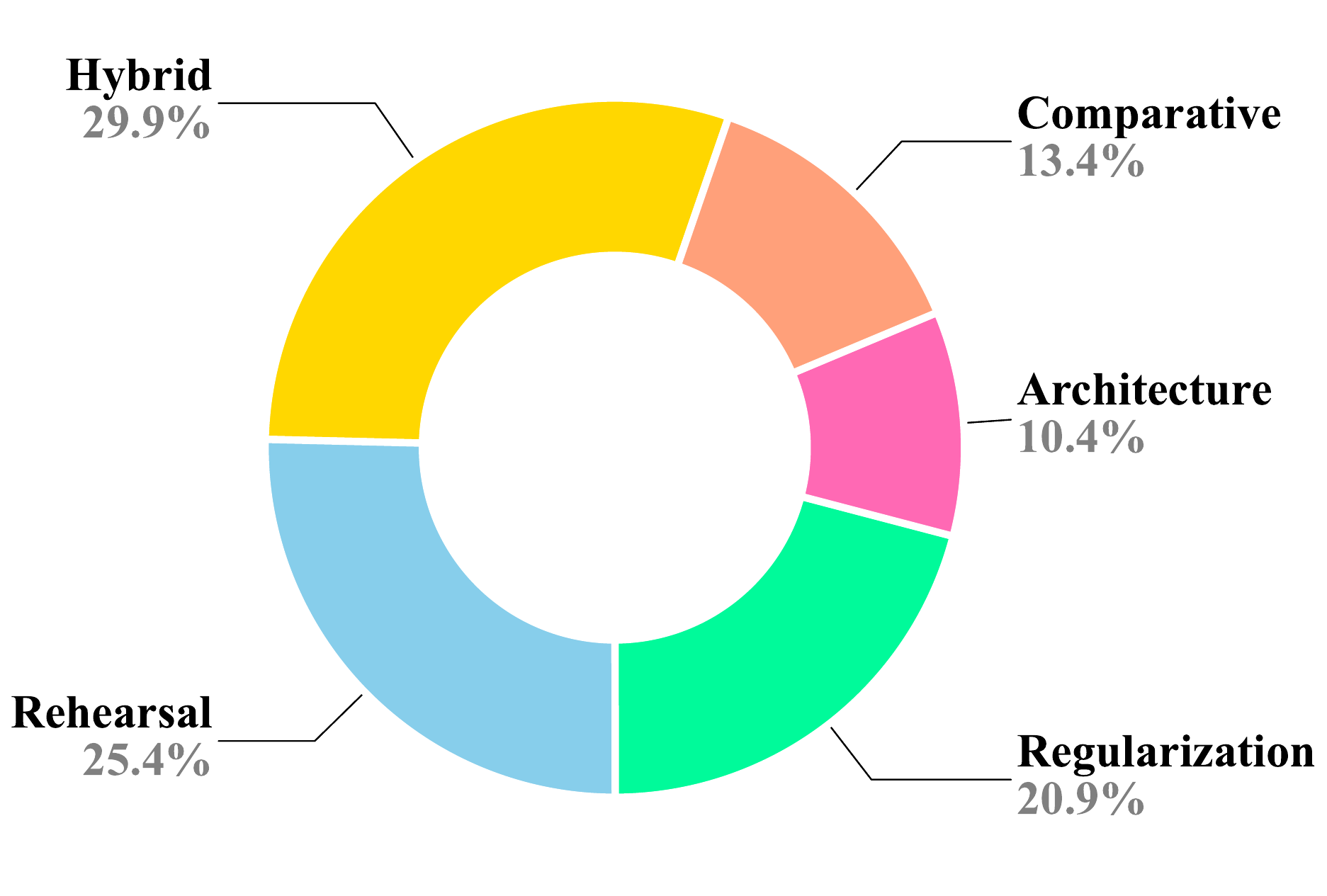}
    }{
        % Two column image
        \includegraphics[width=.48\textwidth]{content/figures/charts/chart_cl_method.pdf}
    }
    \vspace{-1em}
    \caption{Popularity of different CL strategies for medical image analysis}
    \label{fig:pieChartCLmethod}
\end{figure}
%===========================================

\begin{figure*}[!ht]
\centering
\includegraphics[width=\textwidth]{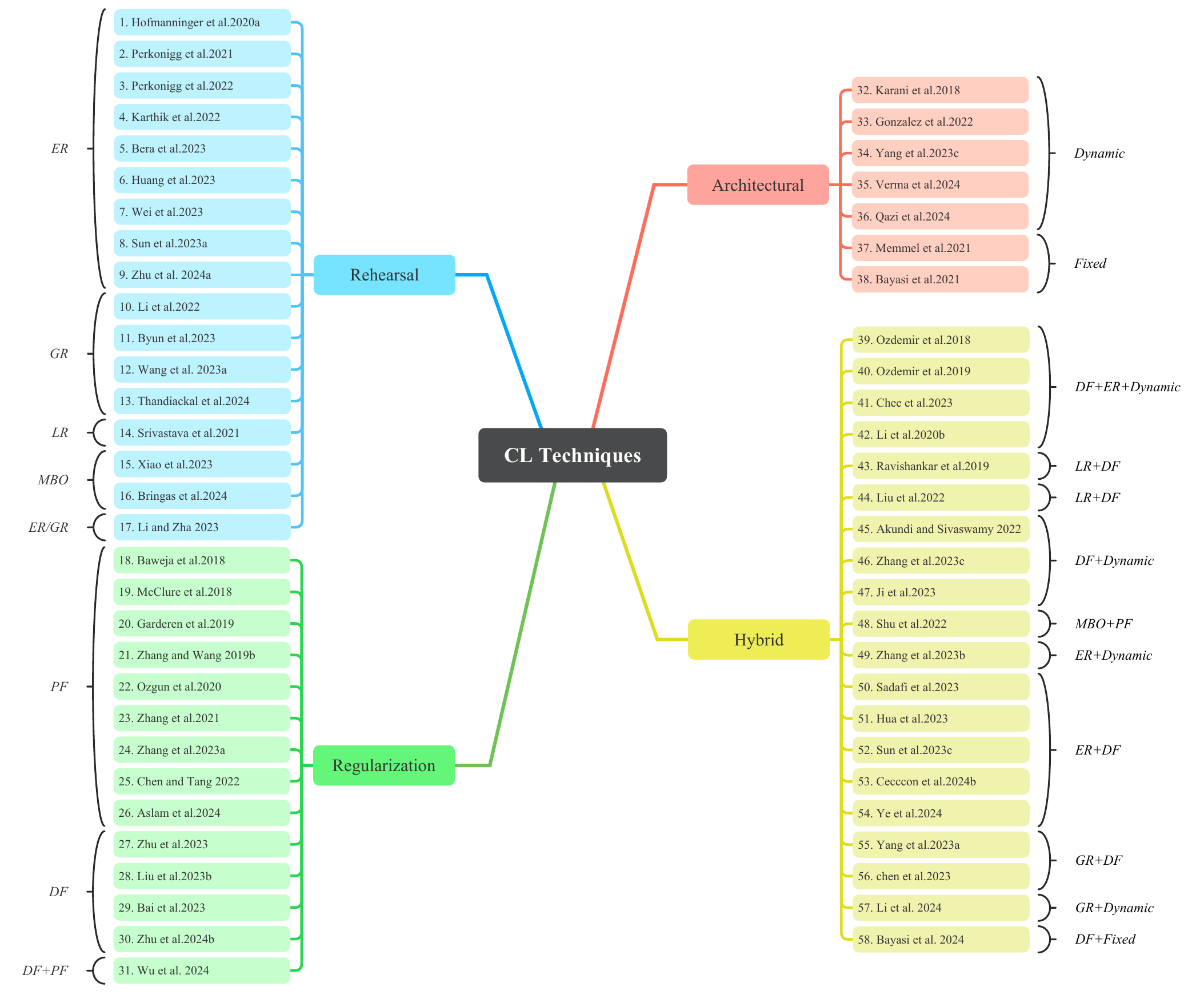}
\caption{The proposed CL taxonomy categorizes various CL models according to their underlying design principles. The abbreviations used are as follows: \textit{ER} for Experience-{replay}, \textit{GR} for Generative-{replay}, \textit{LR} for Latent-{replay}, \textit{PF} for Prior-focused, \textit{DF} for Data-focused, and \textit{MBO} for Memory Buffer Optimization. 
1. \cite{hofmanninger2020dynamic},
2. \cite{perkonigg2021continual},
3. \cite{perkonigg2022continual},
4. \cite{karthik2022segmentation},
5. \cite{bera2023memory},
6. \cite{huang2023conslide},
7. \cite{wei2023representative},
8. \cite{sun2023adaptive},
9. \cite{zhu2024lifelong},
10. \cite{li2022domain},
11. \cite{byunconditional}
12. \cite{wang2023rethinking}
13. \cite{thandiackal2024multi}
14. \cite{srivastava2021continual}
15. \cite{xiao2023fs3dciot}
16. \cite{bringas2024cladsi}
17. \cite{li2023doctor}
%%%
18. \cite{baweja2018towards},
19. \cite{mcclure2018distributed},
20. \cite{van2019towards},
21. \cite{zhang2019continually}
22. \cite{ozgun2020importance},
23. \cite{zhang2021comprehensive},
24. \cite{zhang2023s},
25. \cite{chen2022breast},
26. \cite{aslam2024cel}
27. \cite{zhu2023uncertainty},
28. \cite{liu2023incremental},
29. \cite{bai2023revisiting}
30. \cite{zhu2024boosting}
31. \cite{wu2024modal}
%%%%%%%%%%%%5
32. \cite{karani2018lifelong},
33. \cite{gonzalez2022task},
34. \cite{yang2023continual}
35. \cite{verma2024confidence}
36. \cite{qazi2024dynammo}
37. \cite{memmel2021adversarial},
38. \cite{bayasi2021culprit},
%%%%%%%%%%%%%%%%%%%%5
39. \cite{ozdemir2018learn},
40. \cite{ozdemir2019extending},
41. \cite{chee2023leveraging},
42. \cite{li2020continual},
43. \cite{ravishankar2019feature},
44. \cite{liu2022learning},
45. \cite{akundi2022incremental},
46. \cite{zhang2023continual}
47. \cite{ji2023continual}
48. \cite{shu2022replay}
49. \cite{zhang2023adapter},
50. \cite{sadafi2023continual},
51. \cite{hua2023incremental}
52. \cite{sun2023few}
53. \cite{ceccon2024multi}
54. \cite{ye2024continual}
55. \cite{yang2023few},
56. \cite{chee2023leveraging},
57. \cite{li2024dual},
58. \cite{bayasi2024gc}
}
\label{fig:CLtaxonomy}
\end{figure*}
%===========================================

%%%%%%%%%%%%%%%%%%%%%%%%%%%%%%%%%%%%%%%%%%%%%%%%%%%%%%
\section{Continual learning strategy}
\label{sec:CLtechnique}
Owing to frequently encountered domain shifts in medical applications, CL strategies aim to tackle performance drop due to domain shifts and thus might become a necessity in clinical applications~\cite{pianykh2020continuous, lee2020clinical}. All the available strategies for preventing catastrophic forgetting can be broadly categorized into three main categories: (a) rehearsal-based approaches where a small memory is used to store previous episode data in some form, (b) regularization-based methods where the aim is to control weight update to minimize forgetting the previous learning, (c) architectural-based methods which mainly aim to keep some network parameters isolated for each episode, and (d) hybrid category which offers various combination of any of these three categories. \refx{fig:CLtaxonomy} presents a taxonomy of CL techniques and their adaptation in the medical field by various works. Further, the sub-technique category for each work is also mentioned in the figure.
{More detail of CL techniques in each work can be found in \mbox{~\refx{tab:rehearsalLiterature}}, 
\mbox{~\refx{tab:regularizationLiterature}},
\mbox{~\refx{tab:architectureLiterature}},
\mbox{~\refx{tab:hybridLiterature}}, and
\mbox{~\refx{tab:comparativeLiterature}} for rehearsal, regularization, architectural, hybrid, and comparative studies, respectively.}
Additionally, a pie chart showing the ratio of works falling in the above-mentioned categories is shown via \refx{fig:pieChartCLmethod}. It indicates the popularity of using one or more strategies together in hybrid settings owing to better classification or segmentation performances in sequential learning. Now, we discuss each category in detail in the following sections.

%%%%%%%%%%%%%%%%%%%%%%%%%%%%%%%%%%%%%%%%%%%%%%%%%%%%%%%%%%%%%%%%%%%%%%%%

% Rehearsal based
\input{content/tables/rehearsal-literature}

\input{content/sections/cl-technique/rehearsal-based}

% Regularization based
\input{content/tables/regularization-literature}

\input{content/sections/cl-technique/regularization-based}

% Architecture based
\input{content/tables/architecture-literature}

\input{content/sections/cl-technique/architecture-based}

% Hybrid techniques
\input{content/tables/hybrid-literature}
\input{content/sections/cl-technique/hybrid-techniques}

\input{content/sections/cl-technique/comparative-studies}

%% file: content/tables/rehearsal-literature.tex
%-----------------rehearsalLiterature--------------------
\begin{table*}[!ht]
\caption{List of various rehearsal based works in literature}
\vspace{-1em}
\centering
\resizebox{.99\textwidth}{!}{
\begin{tabular}{ p{3.2cm} | p{8.5cm} | p{8.7cm} } 
    % header			
\bottomrule
\rowcolor{gray!20}
\textbf{Reference (year)} 
& \textbf{Application \& CL scenario } 
& \textbf{CL technique}
\\ \hline 

% body
\rowcolor{rehearsalcolor!\tabcellci}
\tabrefcite{hofmanninger2020dynamic} 
& Chest CT classification: DIS (3 Ep.)
& Experience replay with a dynamic memory size where samples are ranked by computing Gram matrix on the last activation maps 
\\ \hline 

%%not a cl paper        
% \tabrefcite{venkataramani2019towards} & X-ray lung segmentation: DIS (2 Ep.) & \\ \hline 
\rowcolor{rehearsalcolor!\tabcellcis}
\tabrefcite{srivastava2021continual} 
& Chest X-ray classification: DIS(3 Ep.) 
& Latent replay 
\\ \hline 

\rowcolor{rehearsalcolor!\tabcellci}
\tabrefcite{perkonigg2021continual} 
& Brain age estimation: DIS (3 Ep.)
& Experience replay 
\\ \hline 

\rowcolor{rehearsalcolor!\tabcellcis}
\tabrefcite{perkonigg2022continual} 
&  3 separate experiments for 3 applications: (a) cardiac segmentation: DIS (4 Ep.), (b) lung nodule detection: DIS (4 Ep.), and (c) brain age estimation: DIS (4 Ep.) 
& Experience replay 
\\ \hline 

\rowcolor{rehearsalcolor!\tabcellci}
\tabrefcite{karthik2022segmentation} 
& MRI data for brain sclerosis lesions segmentation: DIS (8 Ep.) 
& Random samples from each episode kept for experience replay 
\\ \hline 

\rowcolor{rehearsalcolor!\tabcellcis}
\tabrefcite{li2022domain} 
& CMR based Cardiac segmentation across 4 scanners: DIS (4 Ep.) 
& Generative replay: style-oriented replay 
\\ \hline 

\rowcolor{rehearsalcolor!\tabcellci}
\tabrefcite{bera2023memory} 
& (a) Binary segmentation of prostate, spleen, and hippocampus: TIS, (b) binary hippocampus segmentation across 2 datasets: DIS, (c) binary prostate segmentation across 4 datasets: DIS 
& Experience replay based rehearsal strategy: store max K exemplars (half of the exemplars based on fraction of positive class content inside the image and other half based on gradient variation) 
\\ \hline 

\rowcolor{rehearsalcolor!\tabcellcis}
\tabrefcite{wang2023rethinking} 
& (a) Segmentation for endoscopy: CIS (2-3 Ep.) (b) surgical instrument segmentation (2 Ep.) 
& Generative replay 
\\ \hline 

% \rowcolor{yellow}
\rowcolor{rehearsalcolor!\tabcellci}
\tabrefcite{byunconditional} 
& (a) Diabetic retinopathy severity classification: DIS (2 Ep.), (b) dermoscopy skin lesion detection: DIS (3 Ep.) 
& Generative replay: The generator model is a conditional text-to-image diffusion model that is updated with current episode data and then used to generate labeled data for the current episode and stored in the buffer. Current episode data along with labeled data for each episode kept in the buffer, is utilized for fine-tuning the classifier.
\\ \hline 

% \rowcolor{yellow}
\rowcolor{rehearsalcolor!\tabcellcis}
\tabrefcite{huang2023conslide} 
&Tumor subtype classification at WSI-level: CIS (4 Ep.) & Experience replay
\\ \hline 

% \rowcolor{yellow}
\rowcolor{rehearsalcolor!\tabcellci}
\tabrefcite{sun2023adaptive} 
& Disease classification in time-series signals: DIS (10 time-steps) & Experience replay (adaptive importance based replay) 
\\ \hline 

% \rowcolor{yellow}
\rowcolor{rehearsalcolor!\tabcellcis}
\tabrefcite{wei2023representative} & 
Brain tumor segmentation: IIS (8 Ep.)
& Experience replay
\\ \hline

% \rowcolor{yellow}
\rowcolor{rehearsalcolor!\tabcellci}
\tabrefcite{xiao2023fs3dciot} 
&Skin disease classification with dermoscopic and clinical images: CIS (18 Ep.)& Rehearsal (Memory buffer optimization: extension of GEM approach) 
\\ \hline  

% \rowcolor{yellow}
\rowcolor{rehearsalcolor!\tabcellcis}
\tabrefcite{li2023doctor} 
&Disease classification on physiological signals: DIS (2 Ep.), CIS (2 Ep.), TIS (2-3 Ep.)& (a) Experience replay: training loss based sample selection for the buffer, (b)  generative replay: parametric (GMM) or non-parametric (KDE) models as a generator for past episodes.      
\\ \hline

% \rowcolor{yellow}
\rowcolor{rehearsalcolor!\tabcellci}
\tabrefcite{thandiackal2024multi} 
& Histopathology colorectal tissue classification: DIS (3 Ep.)   
& GAN-based generative replay
\\ \hline

% \rowcolor{yellow}
\rowcolor{rehearsalcolor!\tabcellcis}
\tabrefcite{bringas2024cladsi} 
& Alzheimer's disease stage identification using motion-sensor: IIS (2 Ep.,3 Ep.,4 Ep.)&Rehearsal (A-GEM)
\\ \hline

% \rowcolor{yellow}
\rowcolor{rehearsalcolor!\tabcellci}
\tabrefcite{zhu2024lifelong} &Histopathology WSI retrieval: CIS (4 Ep.)
& Rehearsal (experience replay: memory bank with reservoir sampling) \\ \hline 

\end{tabular}
}
\label{tab:rehearsalLiterature}
\end{table*}
%-----------------------------------------------

%% file: content/sections/cl-technique/rehearsal-based.tex
\subsection{Rehearsal based}
\label{sec:rehearsal-based}
This category of methods aims to approximate and recover old data distributions to augment with the new task data. Typically, there is a memory buffer to store data samples from past tasks, which are then used for replaying during the learning of a new task in order to retain previously learned knowledge and hence mitigate catastrophic forgetting. 
Here, the samples can be original images~\cite{rebuffi2017icarl,karthik2022segmentation} or deep features~\cite{van2019three} or generated pseudo samples~\cite{shin2017continual} and can be selected via various heuristics and stored in memory buffer.

Based on the content of the memory buffer, the methods in this category can be broadly divided into the following sub-categories viz., (i) experience replay-based, (ii) generative replay-based, (iii) latent replay-based, and (iv) memory buffer optimization.

\subsubsection{Experience replay-based} 
%%%%\textcolor{red}{ iCaRL: The Incremental Classifier and Representation Learning method maintains a set of exemplar examples for each observed class. For each class, an exemplar set is a subset of all examples of the class, aiming to carry the most representative information of the class. The method uses a CNN as a feature extractor and classifies by comparing the feature vector of the input to the feature vectors of the examples ~\cite{rebuffi2017icarl}. }
%%%\textcolor{red}{With iCaRL~\cite{rebuffi2017icarl}, authors extend on LwF by proposing a strategy for selecting an exemplar dataset, which keeps a “representative” subset of the earlier training data for the existing classes, and put an upper bound on required memory requirements. }
%%%%\textcolor{red}{iCaRL:This method stores a subset of most representative examples per class in memory, which are chosen according to approximate class means in the learned feature space.}

In experience replay, a few past exemplars (images) are stored in a small memory buffer. The main challenge here is to design a strategy to select important exemplars that will be stored in the limited storage to fully exploit the memory buffer.

For natural image classification, \citet{rebuffi2017icarl} proposed a popular experience-based method called incremental Classifier and Representation Learning (iCaRL) which is also highly explored for various medical applications, e.g., histopathology tumor classification \cite{kaustaban2022characterizing}, disease classification \cite{derakhshani2022lifelonger}, etc.
The iCaRL strategy has a fixed memory buffer condition where a subset of the most representative examples is maintained for each class, aiming to carry the most representative information of the class in the learned feature space. The distance between data instances in the latent feature space is used to update the memory buffer. During representation learning, both the stored samples and current task samples are utilized for training. During inference, a nearest-mean-of-exemplars classification strategy is used to assign the label to the class with the most similar prototype. The original iCaRL method requires all data from the new task to be trained together. To address this limitation and enable the new instances from a single task to come at different time steps, \citet{chaudhry2019tiny} proposed Experience Replay (ER), which uses reservoir sampling to randomly sample a certain number of data instances from a data stream of unknown length, and store them in the memory buffer.

Further exploration has been conducted to select the most representative samples for replay in medical applications. \citet{bera2023memory} propose a simple sample selection technique for binary segmentation with a memory bank of $K$ samples, equi-distributed over the classes. Half of the exemplars are selected based on the occupancy of the positive class inside the image, i.e., the higher the content of the positive class, the higher the chance of being selected; the other half are selected based on their contribution to the learning process, i.e., if the gradient variation is more for a sample, then it is hard to learn and hence important. \citet{hofmanninger2020dynamic} propose to store samples based on their degree of uniqueness inferred using the Gram matrix computed on activations from the last convolution layer of the deep model. They evaluated the framework on chest CT data with synthetically generated domain shifts for classification application. Further, \citet{perkonigg2021continual,perkonigg2022continual} propose to combine rehearsal-based CL and active learning, where the model indicates important samples that need to be annotated rather than annotating all the samples. They detect domain shifts using a memory buffer for outliers, prompting the labeling of informative samples for model adaptation.
%%%%%%%%%%%%%%%%%%%%%%%%%%%%%%%%%%%%%%%%%%%%%%%%%%%%%%%%%%%%%%%%%%%%%%%%%
\subsubsection{Latent replay-based} 
While experience replay requires storing the past samples in raw form, it can cause serious privacy violations in critical medical applications {\mbox{\cite{thandiackal2024multi,zhu2024boosting,bayasi2024gc}}}. Therefore, a less concerning direction, i.e., storing features instead of raw images, is also explored. For continual chest X-ray classification application in domain shift conditions, \citet{srivastava2021continual} explore leveraging vector-quantization to store and replay hidden representations under memory constraints. 
{Although not directly accessible, sharing of latent representations involves possible privacy threats through manipulation and reconstruction of actual sensitive medical data~\mbox{\cite{pennisi2023privacy}}. 
Given access to a model and its latent space, a malicious entity could reverse-engineer a patient's medical record, violating privacy. Attackers might create adversarial instances to alter latent representations, potentially reconstructing crucial information from seemingly benign data.
Reconstruction attacks~\mbox{\cite{newaz2020adversarial}}, membership inference attacks~\mbox{\cite{shokri2017membership}}, model inversion attacks~\mbox{\cite{fredrikson2015model}}, etc., can be potential challenges associated with sharing of features. Models like GAN, auto-encoder, variational auto-encoder, etc, inherently possess the capability to regenerate raw data from latent spaces.
Appropriate privacy measures like differential privacy~\mbox{\cite{abadi2016deep}}, privacy-preserved neural networks~\mbox{\cite{jovanovic2022private}}, adversarial training~\mbox{\cite{yi2019generative}}, encryption and secure multi-party computation~\mbox{\cite{spini2024privacy}}, etc. need to be considered to protect latent features during processing and sharing.}

%%%" storing the past samples in raw form, it can cause serious privacy violations in critical medical applications." -> will be good to add more references to this section.... Also, talk about the risks of raw data regeneration from latent representations here.
%%%%%%%%%%%%%%%%%%%%%%%%%%%%%%%%%%%%%%%%%%%%%%%%%%%%%%%%%%%%%%%%%%%
\subsubsection{Generative replay-based}
This category of approaches emerged as an alternative to experience replay, which highly violates privacy concerns with storing past samples and high memory buffer demands. Here, instead of a memory buffer of actual samples, there is a generative model that can generate samples, latent representation, both, or other information related to past tasks. 
This category of methods is also related to incremental learning of generative models where their incremental update is required. Hence, an additional requirement here is the continuous updating of the generative adversarial network. Since actual samples are not stored for replay, this category of approaches is also termed pseudo-rehearsal-based approaches. 

To generate the past data, \citet{li2022domain} employ a style-oriented replay module, which includes a base generative model trained on the first arrived domain and a style bank to record style adjustments for successive domains. Then, they incorporate the replayed past data to jointly optimize the model with current data to alleviate catastrophic forgetting. For continual semantic segmentation of MRI data emerging from different institutions, \citet{memmel2021adversarial} train a model by using all the simultaneously available datasets. Their method disentangles content from domain information through adversarial training, resulting in domain-invariant content representations. Other methods follow the sequential arrival of a dataset, i.e., one dataset at a time, whereas this work requires at least two datasets to start the feature disentanglement learning.

\subsubsection{Memory buffer optimization} 

Experience replay-based approaches might overfit the stored sub-samples and seem to be bounded by joint training. Alternatively, memory buffer optimization-based approaches following constrained optimization provide solutions that have more scope for both plasticity and stability. {The core idea is to guide the training of the current episode by utilizing the stored buffer samples from the past. Typically, the gradients of the current task are projected such that they do not negatively impact the gradients computed on the buffer samples. Thus, it is ensured that the model performs well on all previous episodes while learning a new episode.} For example, the Gradient Episodic Memory (GEM) \cite{lopez2017gradient} approach corrects the gradient computed on a mini-batch during stochastic gradient descent by utilizing an exemplar {sample set from past episodes}. This prevents changes that could degrade the performance of the network on the exemplar set. Further, the A-GEM approach \cite{chaudhry2018efficient} relaxes the projection in one direction computed by randomly selected samples from the memory buffer where both the buffer and sampling size are tunable hyper-parameters. 
There are some early attempts in this area in the medical realm. For example, for the fundus disease diagnosis problem, \citet{shu2022replay} presents a gradient regularisation approach for preventing forgetting and a replay-oriented consistency calculation method combined with a subspace weighting strategy to promote model adaptability. The authors employ gradient regularisation in conjunction with a replay-oriented technique. The replay-oriented technique adaptively improves the update for incremental domains without carefully choosing exemplars for replay, whereas gradient regularisation preserves information from previous ones.

%%%%%\textcolor{red}{ GEM: The Gradient Episodic Memory (GEM) approach uses an exemplar set to correct the gradient calculated on a mini-batch during stochastic gradient descent preventing changes that would harm the performance of the network on this exemplar set~\cite{lopez2017gradient} . }
%%%%%\textcolor{red}{A-GEM: In this online constrained replay-based method~\cite{chaudhry2018efficient}, model updates are constrained to prevent forgetting by projecting the estimated gradient on the direction determined by randomly selected samples from a replay buffer. Buffer size and sampling size from the buffer are both tunable hyper-parameters. In the literature, A-GEM was tested only in CIS and TIS as reported in ~\cite{chaudhry2018efficient}.}

%%%%%%\textit{\cite{shu2022replay} propose a gradient regularization approach to suppress forgetting, and a replay-oriented consistency calculation method combined with a subspace weighting strategy to improve the model plasticity. The authors use gradient regularization with a replay-oriented strategy. The replay-oriented strategy adaptively boosts the update for incremental domains without delicately selecting exemplars for replay, while the gradient regularization preserves the knowledge from preceding ones.}

%% file: content/tables/regularization-literature.tex
%----------------regularizationLiterature-------------------
\begin{table*}[!ht]
\caption{List of various regularization based works in literature}
\vspace{-1em}
\centering
\resizebox{.99\textwidth}{!}{
\begin{tabular}{ p{3.5cm} | p{8.5cm} | p{8cm} } 
% header			
\bottomrule
\rowcolor{gray!20}
\textbf{Reference (year)} 
& \textbf{Application \& CL scenario } 
& \textbf{CL technique}
\\ \hline 

% body
\rowcolor{regularizationcolor!\tabcellci}
\tabrefcite{baweja2018towards} 
& MRI-based normal brain structures segmentation: TIS (2 Ep.) 
& EWC 
\\ \hline 

\rowcolor{regularizationcolor!\tabcellcis}
\tabrefcite{mcclure2018distributed} 
& Axial and sagittal brain segmentation: DIS (4 Ep.) 
& Distributed Weight Consolidation (DWC) 
\\ \hline 

\rowcolor{regularizationcolor!\tabcellci}
\tabrefcite{zhang2019continually}
& Longitudinal MRI-based Alzheimer’s
disease progression modeling: TIS (7 Ep.) 
& EWC 
\\ \hline 

\rowcolor{regularizationcolor!\tabcellcis}
\tabrefcite{van2019towards} 
& MR data-based glioma segmentation: DIS (2 Ep.) 
& EWC 
\\ \hline 

\rowcolor{regularizationcolor!\tabcellci}
\tabrefcite{ozgun2020importance} 
& Brain MRI segmentation: DIS (4 Ep.)
& Learning rate regularization built over memory aware synapses technique 
\\ \hline 

\rowcolor{regularizationcolor!\tabcellcis}
\tabrefcite{chen2022breast} 
& Histopathology-based breast cancer classification: CIS (4 Ep.) 
& Regularization: EWC 
\\ \hline 

\rowcolor{regularizationcolor!\tabcellci}
\tabrefcite{zhang2023s,zhang2021comprehensive} & 
(a) MRI-based prostate segmentation: DIS, (b) retinal image-based optic cup and disc segmentation: DIS 
& Shape and semantics-based selective regularization to penalize changes of parameters with high joint shape and semantics-based importance 
\\ \hline 

\rowcolor{regularizationcolor!\tabcellcis}
\tabrefcite{zhu2023uncertainty} 
& (a) Binary prostate segmentation from T2-weighted MRI scans collected from 6 sites: DIS, (b) cross-site and cross-modality 4-class abdominal segmentation between CT and MRI scans: DIS, (c) same-site cross-modality muscle segmentation of muscles and bones between MRI and mDixon scans: DIS
& Knowledge distillation 
\\ \hline 

\rowcolor{regularizationcolor!\tabcellci}
\tabrefcite{liu2023incremental} 
& Brain tumor segmentation: CIS with domain shift (3 Ep.) 
& Pseudo label-based knowledge distillation 
\\ \hline 

\rowcolor{regularizationcolor!\tabcellcis}
\tabrefcite{bai2023revisiting} 
& Surgical question answering: CIS 
& Regularization: knowledge distillation 
\\ \hline

% \rowcolor{yellow}
\rowcolor{regularizationcolor!\tabcellci}
\tabrefcite{wu2024modal} & Image super-resolution: TIS (4 Ep.)& Regularization (knowledge distillation and parameter importance based gradient update)
\\ \hline 

% \rowcolor{yellow}
\rowcolor{regularizationcolor!\tabcellcis}
\tabrefcite{aslam2024cel} 
& Disease outbreak detection in time-series signal: DIS (10 Ep.) &Regularization (EWC)  
\\ \hline

% \rowcolor{yellow}
\rowcolor{regularizationcolor!\tabcellci}
\tabrefcite{zhu2024boosting} & (a) MRI Prostate segmentation: DIS (b) MRI Cardiac
segmentation: DIS&  Regularization (knowledge distillation) 
\\ \hline

\end{tabular}
}
\label{tab:regularizationLiterature}
\end{table*}
%-----------------------------------------------

%% file: content/sections/cl-technique/regularization-based.tex
\subsection{Regularization-based}
\label{sec:regularization-based}
Rehearsal methods are quite popular due to their comparative better performance than other categories; however, the assumption of the availability of past data makes them less suitable for medical applications. In contrast, regularization-based approaches~\cite{kirkpatrick2017overcoming,schwarz2018progress,zenke2017continual} avoid storing examples and mainly add a regularization term in the loss function or regularize the learning rate to penalize model updates that could lead to large deviation from an existing model, thus avoiding forgetting of learned knowledge.

Regularization-based methods can be categorized into data-focused and prior-focused approaches. Data-focused approaches~\cite{silver2002task} distill the knowledge of old tasks to enhance the CL capabilities of the present model, whereas prior-based approaches such as~\cite{zenke2017continual,kirkpatrick2017overcoming,aljundi2018memory} define importance weights for the network's parameters. Based on these weights, a regularization loss is introduced that penalizes the shift of important parameters.

%%%%\textcolor{red}{Regularization-based approach preserves old knowledge by using additional regularization term to penalize changes in previously learned features. Previous works \cite{kirkpatrick2017overcoming,zenke2017continual,aljundi2018memory} estimate the importance of each weight in a previously learned model and apply penalty if there are updates to the important weights. Other works use distillation loss to ensure that the features learned of the old classes are preserved \cite{li2017learning,douillard2020podnet,kim2021split} . Another interesting direction introduced in recent work \cite{tao2020topology} is the use of topology-preserving loss to maintain the feature space topology. However, one difficulty faced by approaches in this category is balancing the regularization term such that learning of new classes would not be hindered.}
%=================================================================
\subsubsection{Prior-focused regularization}
Prior-focused methods estimate the importance of all neural network parameters, used as prior when learning from new data. During the training of subsequent tasks, larger changes to important parameters are penalized. Elastic Weight Consolidation (EWC) \cite{kirkpatrick2017overcoming}, initially designed for reinforcement learning in Atari games emerged as the first to establish the technique. EWC aims to constrain parameters of the model that are critical for performing previous tasks during the learning of the new tasks. It uses the Fisher information matrix to calculate parameter importance for a given domain. These parameters are then regularized to prevent catastrophic forgetting.

In the medical domain, \citet{baweja2018towards} are the first to explore EWC  to address the issue of catastrophic forgetting in neural networks when sequentially learning two distinct segmentation tasks (normal brain structure and white matter lesion). Specifically, the first task consists of multi-class segmentation of cerebrospinal fluid, grey matter, and white matter, and the second task consists of segmentation of white matter lesions. The study demonstrates that EWC effectively reduces catastrophic forgetting in this challenging medical imaging context. Further, \citet{van2019towards} adopt EWC for glioma segmentation on different datasets in domain-shift arising from low and high-grade glioma in different datasets. In the context of histopathology breast cancer classification task, \citet{chen2022breast} showcase EWC capabilities for class incremental setting. \citet{mcclure2018distributed} introduce Distributed Weight Consolidation (DWC) as a CL method to consolidate weights of separate neural networks trained on independent datasets. Further, \citet{zhang2023s} propose to compute the importance matrix jointly based on shape and semantics information in context for continual medical segmentation. The shape-based importance measures how sensitive a parameter is to shape properties in the images and semantics-based importance measures a parameter's sensitivity to reliable semantic predictions, ensuring that noisy or uncertain semantics do not influence the learning process. Then, the selective regularization scheme is applied to penalize updates to model parameters with high joint importance weights. This ensures that critical shape and semantic knowledge from previous sites is not overwritten or forgotten.

Another prior-focused strategy is Synaptic Intelligence (SI), which alleviates catastrophic forgetting by allowing individual synapses (i.e., neurons) to estimate their importance for solving a learned task. Similarly to EWC, the approach penalizes changes to the most relevant synapses so that new tasks can be learned with minimal forgetting \cite{zenke2017continual}. A disadvantage is the need to distribute some extra parameters per weight in addition to their value, but in terms of data size, this is far less than providing the exemplars.

Apart from EWC and SI, various other methods have been contributed. Popularly, the Memory Aware Synapses (MAS) approach \cite{aljundi2017expert} calculates the importance of weights with a model of Hebbian learning in the biological system, which relies on the sensitivity of the output function and can hence be utilized in an unsupervised manner. For medical segmentation, a MAS-inspired learning rate regularization approach was contributed by \citet{ozgun2020importance} for sequential training on different domains. It involves reducing the learning rate for important parameters to prevent forgetting, offering a more direct approach compared to the surrogate loss used in MAS.
{Yet another novel emerging direction is Orthogonal Weight Modification (OWM) \mbox{\citet{zeng2019continual}}, where the gradient of the current task is projected into the orthogonal direction to the subspace spanned by gradients of all previous tasks. Then parameter updates for the current task are permitted only in the orthogonal direction of the past episodes, thus protecting against interference with already learned knowledge.}

\subsubsection{Data-focused regularization}
The data-focused regularization method \cite{li2017learning} aims to distill knowledge from a model trained on the previous tasks to the model trained on the new task in order to consolidate previously learned knowledge. Typically, the previous model acts as a teacher and the current model as a student while adopting knowledge-distillation (KD) \cite{gou2021knowledge} to avoid catastrophic forgetting. 
Usually, all the old data should be available in knowledge-distillation, which is not the case with CL; therefore, as an alternative, few old data, current data, or the generated old data are explored for the same. The loss function has an additional distillation loss for replayed data. Each input is replayed with a soft target obtained using the stored model.

%%%%\textcolor{red}{Later on, learning without forgetting (LwF)~\cite{li2017learning} has been proposed, which utilizes distillation loss [1] such that when new classes are being added to a network, final activation response of the previous classes are also used for back-propagation. LwF leverages distillation to regularize the current loss with soft targets taken from a previous version of the model.}

\citet{silver2002task} first proposed to use previous task model outputs given new task input images, mainly for improving new task performance. Later, in natural image classification application, \citet{li2017learning} re-introduced the concept as the LwF technique. A copy of the previous model parameters is stored before learning the new task, and then it is used to get the soft labels for the new task as the target {from} the classifiers of the previous tasks. The available ground truth is used as the target for the new task classifier. LwF has been adopted in various medical applications like incremental brain MRI segmentation \cite{ozdemir2018learn}, chest X-ray classification \cite{lenga2020continual}, disease classification \cite{derakhshani2022lifelonger}, histopathology tumor classification \cite{kaustaban2022characterizing}, etc.

Other works in the natural image processing domain \cite{jung2016less,zhang2020class} have been introduced with LwF-related ideas; however, it has been shown that this strategy is vulnerable to domain shift between tasks \cite{aljundi2017expert}. It was pointed out that LwF would not result in good performance if the data distributions between different tasks are quite diverse \cite{rannen2017encoder}. To overcome this, \citet{rannen2017encoder} facilitate incremental integration of shallow auto-encoders to constrain task features in their corresponding learned low dimensional space. Hence, they further trained an autoencoder for each task to learn the most important features corresponding to the task and used it to preserve knowledge. 
{Such multiple expert-based knowledge-distillation have proven effective in domain generalization application too \mbox{\cite{niu2023knowledge}}. Student expert modules possess domain-specific information. Each expert also learns from all other students by knowledge-distillation technique to facilitate domain-invariant features. Also, when more emphasis is attributed toward improving unseen classes, i.e., forward transfer, CL based sequential learning of domains helps in domain generalization \mbox{\cite{li2020sequential}}. }

%%%%%%%%%%%%%%%%%%%\textcolor{red}{REVIEWER ASKS ...Will be good to highlight how it helps in domain generalization. }

%%%%%%%%%%%%%%%%%%%%%%%%%%%%%%%%%%%%%%%%%%%%%%%%%%%%%%%%%%%%%%%%%%%%%%
%%%%%\textcolor{red}{ Besides, dynamically expandable networks augment architecture with new modules (e.g., gating autoencoders~\cite{aljundi2017expert} and batch normalization layers ~\cite{karani2018lifelong}) to accommodate new knowledge, contributing to zero forgetting yet causing quadratic parameter increase and requiring task label for each sample at test time. In a task-agnostic manner with fixed network architecture, selective regularization methods ~\cite{aljundi2018memory,kirkpatrick2017overcoming} explore model parameters that are important for preserving old knowledge, and then minimize their alterations when learning new knowledge. }
%%%The work by \cite{yan2021dynamically} introduces the notion of expandable representation learning by adding new branches to the feature extractor of existing network to learn novel concepts for the incoming data while fixing the weights of old branches to preserve previously learned features. However, there is minimal exploitation of old knowledge since each feature extractor branch is independent of each other.
%%%%%%%%%%%%%%%%%%%%%%%%%%%%%%%%%%%%%%%%%%%%%%%%%%%%%%%%%%%%%%%%%%%%%%

%% file: content/tables/architecture-literature.tex
%---------------architectureLiterature--------------------
\begin{table*}[!ht]
\caption{List of various architecture based works in literature}
\vspace{-1em}
\centering
\resizebox{.99\textwidth}{!}{
\begin{tabular}{ p{3.5cm} | p{7.5cm} | p{8cm} } 
% header			
\bottomrule
\rowcolor{gray!20}
\textbf{Reference (year)} 
& \textbf{Application \& CL scenario } 
& \textbf{CL technique}
\\ \hline 

% body
\rowcolor{architecturecolor!\tabcellci}
\tabrefcite{karani2018lifelong} 
& MR brain segmentation: DIS (4 Ep.) 
& Domain specific batch normalization layer 
\\ \hline 

\rowcolor{architecturecolor!\tabcellcis}
\tabrefcite{bayasi2021culprit} 
& Skin lesion image classification: DIS (6 Ep.) 
& Architecture (fixed but partitioned network: culprit unit pruning mechanism) 
\\ \hline 

\rowcolor{architecturecolor!\tabcellci}
\tabrefcite{memmel2021adversarial} 
& Brain MRI hippocampal segmentation: DIS (2 Ep.) 
& Architecture (feature disentanglement through adversarial training) 
\\ \hline 

\rowcolor{architecturecolor!\tabcellcis}
\tabrefcite{gonzalez2022task} 
& MRI-based Hippocampus Segmentation:  DIS (2 Ep.)
& Architecture based approach based on out-of-distribution detection concept: maintain multivariate Gaussians for all past learning created on batch normalization layer 
\\ \hline

% \rowcolor{yellow}
\rowcolor{architecturecolor!\tabcellci}
\tabrefcite{yang2023continual} 
&Skin disease classification on dermoscopic, clinical images: CIS (4-20 Ep.)& Pretrained feature extractor + class specific GMMs built on each deep feature

\\ \hline

% \rowcolor{yellow}
\rowcolor{architecturecolor!\tabcellcis}
\tabrefcite{verma2024confidence} 
& Disease classification on (a) Fundus: TIS (2 Ep.), (b) pathology images: TIS (3 Ep.)
& Architectural (Task specific model is trained, task id is inferred during inference)
\\ \hline

% \rowcolor{yellow}
\rowcolor{architecturecolor!\tabcellci}
\tabrefcite{qazi2024dynammo} 
& (a) Disease classification in histopathology images: CIS (7 Ep.), (b) Skin lesion classification: CIS (4 Ep.) & Task specific adapter with merging facility to increase computational efficiency \\ \hline

\end{tabular}
}
\label{tab:architectureLiterature}
\end{table*}
%-----------------------------------------------

%% file: content/sections/cl-technique/architecture-based.tex
\subsection{Architecture-based}
\label{sec:architecture-based}
Architecture-based methods, also termed parameter isolation-based methods typically assign different parameters in a network to each task. This can be achieved by either fixing the architecture or dynamically extending the network \cite{mallya2018packnet,hung2019compacting,mallya2018packnet,fernando2017pathnet,yoon2017lifelong}. Fixed architectures are limited by the network's capacity, whereas dynamic architectures need more memory with every new task. Most fixed architectures-based methods assign different parts of the network for each task, which requires task identity during inference, but this identity information is usually unavailable. 

In the fixed architecture category, \citet{bayasi2021culprit} proposes a cl strategy where a subset of the network units is assigned to learn each domain separately. Their approach introduces a novel pruning criterion that allows a fixed network to learn new data domains sequentially over time. They identify culprit units associated with wrong classifications in each domain and use them to learn the new domain while freezing the non-culprit nodes. Similarly, \citet{mallya2018packnet} propose to prune the weights with low magnitude and reuse them for the next task, while the remaining weights that are responsible for the previous tasks are kept unchanged. In architecture-based approaches, typically the old knowledge is not exploited to learn the new, thus preventing knowledge transfer from a related task which is an important factor in CL.

Some works aim to dynamically extend the feature extractor module by learning task-specific branches in it \cite{yan2021dynamically}. \citet{rusu2016progressive} propose to use a dynamic architecture that blocks any changes to the network trained on previous knowledge and expands the architecture by allocating sub-networks with a fixed capacity to be trained with the new data, and thus it keeps a pool of pre-trained models, one for each learned task. In contrast, \citet{aljundi2017expert} propose a network of experts where each expert is a model trained given a specific task and a set of gating autoencoders that learn a representation for the task at hand, and, at inference time, automatically forward the test sample to the relevant expert. Another dynamic expandable model direction given by \citet{yoon2017lifelong} expands the network using group sparse regularization to decide how many neurons to add at each layer and perform selective retraining. Similarly, \citet{karani2018lifelong} propose to use domain/task-specific batch norm layers to adapt to new MRI protocols while learning the segmentation of various brain regions. Entire batch-normalization layers were dedicated to modeling domain differences. However, this causes quadratic increase in parameters with new tasks. Moreover, as this approach dedicates specific batch-normalization parameters to each dataset/domain/task, task labels are necessary to determine to which dataset each sample belongs.

In contrast, \citet{gonzalez2022task} investigate the performance of CL methods in a task-agnostic setting, which better simulates dynamic clinical environments characterized by gradual population shifts. They propose an out-of-distribution detection-based solution that signals when to expand the model and select the best parameters during inference. Specifically, they learn a multivariate Gaussian on the last batch normalization layer of the deep model and store it in the memory. When there is out-of-distribution detection or domain shift is alerted based on a threshold on Mahalanobis distance from existing Gaussian in the memory and the new data, a new Gaussian is added to the memory. The closest Gaussian is used for inference and thus eliminates the need for domain ID during inference.

%% file: content/tables/hybrid-literature.tex
%------------------hybridLiterature-------------------
\begin{table*}[!ht]
\caption{List of various hybrid CL technique based works in literature}
\vspace{-1em}
\centering
\resizebox{.99\textwidth}{!}{
\begin{tabular}{ p{4cm} | p{7cm} | p{9cm} } 
% header			
\bottomrule
\rowcolor{gray!20}
\textbf{Reference (year)} 
& \textbf{Application \& CL scenario} 
& \textbf{CL technique}
\\ \hline 

% body
\rowcolor{hybridcolor!\tabcellci}
\tabrefcite{ozdemir2018learn} 
& MRI-based Humerus and scapula segmentation: CIS (2 Ep.)
& Knowledge distillation + experience replay + multi-head 
\\ \hline

\rowcolor{hybridcolor!\tabcellcis}
\tabrefcite{ozdemir2019extending} 
& MRI based knee segmentation: CIS (2 Ep.) 
& Knowledge distillation + experience replay + multi-head 
\\ \hline

\rowcolor{hybridcolor!\tabcellci}
\tabrefcite{ravishankar2019feature} 
& Chest X-ray view classification: DIS (2 Ep.), TIS (2 Ep.), Pneumothorax identification from X-ray: IIS (4 batches, each of 2K) 
& Latent replay where samples are selected either randomly or importance based (farther from cluster centroids) + dynamic network (task specific feature transformer layers) 
\\ \hline

\rowcolor{hybridcolor!\tabcellci}
\tabrefcite{li2020continual} 
& Dermoscopic images based skin disease classification: CIS (4-20 Ep.) 
& Dual knowledge distillation (from old classifier (previous classes) and the fine-tuned classifier (a new FC layer fine-tuned on new classes)) + experience replay (fixed subset of data for each class is stored), during inference nearest-mean-of-exemplars method used for class prediction \tabdescite{rebuffi2017icarl} (where the mean is often calculated by averaging the feature vectors of the stored or selected subset of data in the feature space for each class) 
\\ \hline

\rowcolor{hybridcolor!\tabcellci}
\tabrefcite{liu2022learning} 
&CT data based segmentation of abdomen organs: CIS (4 Ep.)  
& Rehearsal (latent replay) + regularization (knowledge distillation) 
\\ \hline

\rowcolor{hybridcolor!\tabcellcis}
\tabrefcite{akundi2022incremental} 
& Chest X-ray classification: CIS (5 Ep.) 
& Regularization (knowledge distillation) + architecture (task specific head) 
\\ \hline

\rowcolor{hybridcolor!\tabcellci}
\tabrefcite{shu2022replay} 
& Fundus disease classification: DIS (2-4 Ep.) 
& Replay (memory buffer optimization) + regularization (gradient regularization) 
\\ \hline

\rowcolor{hybridcolor!\tabcellci}
\tabrefcite{chee2023leveraging} 
& 3 applications as (a) Cancer classification: CIS, (b) diabetic retinopathy classification: CIS, (c) skin lesions classification: CIS 
& Architecture (separate high-level descriptor for each task) + regularization (distillation loss) + replay (data based rehearsal, fixed memory per task) 
\\ \hline 

\rowcolor{hybridcolor!\tabcellcis}
\tabrefcite{zhang2023continual} 
& (a) Multi-organ segmentation: CIS, (b) abdomen segmentation to liver tumor segmentation: CIS 
& Architecture: organ specific segmentation head + pseudo labelling based knowledge distillation 
\\ \hline

\rowcolor{hybridcolor!\tabcellci}
\tabrefcite{ji2023continual} 
& Multi-organ segmentation: CIS
& Architecture: organ specific decoder + pseudo labelling based knowledge distillation 
\\ \hline

\rowcolor{hybridcolor!\tabcellcis}
\tabrefcite{zhang2023adapter}  
& Continual disease classification in different image modality settings: CIS
& Experience replay + Architecture based (keeps task specific adapters and classifier heads) 
\\ \hline 

\rowcolor{hybridcolor!\tabcellcis}
\tabrefcite{sadafi2023continual} 
& Dermoscopic WBC classification: DIS, CIS, DIS+CIS 
& Experience replay based rehearsal (representativeness of examples for each class and model uncertainty) + regularization (binary-cross entropy based distillation loss) 
\\ \hline

\rowcolor{hybridcolor!\tabcellci}
\tabrefcite{yang2023few} 
& MedMNIST based disease classification: CIS+DIS (4 Ep.) 
& Generative replay+ data-focused regularization 
\\ \hline

% \rowcolor{yellow}
\rowcolor{hybridcolor!\tabcellcis}
\tabrefcite{hua2023incremental} 
&sEMG-based Gesture Classification: CIS (4 Ep.) 
&Rehearsal (experience replay) and Regularization (KD)
\\ \hline 

% \rowcolor{yellow}
\rowcolor{hybridcolor!\tabcellci}
\tabrefcite{chen2023generative} & 3 separate applications including (a) Optic disc: DIS (3 Ep.), (b) cardiac: DIS (3 Ep.), (c) prostate segmentation: DIS (6 Ep.) & Rehearsal (GAN based generative replay) + Regularization (KD from past segmentation model)  \\ \hline

% \rowcolor{yellow}
\rowcolor{hybridcolor!\tabcellcis}
\tabrefcite{sun2023few} 
&Multi-class classification on time-series signals: CIS (4-10 Ep.)& Rehearsal (experience replay)+ Regularization (KD) 
\\ \hline 

% \rowcolor{yellow}
\rowcolor{hybridcolor!\tabcellci}
\tabrefcite{li2024dual} & MR Cardiac segmentation: DIS (5 Ep.) & Rehearsal (privacy-aware generative replay with CGAN) + Architecture (ConvLSTM based domain-customised expansion block) 
\\ \hline

% \rowcolor{yellow}
\rowcolor{hybridcolor!\tabcellcis}
\tabrefcite{ceccon2024multi} 
&Chest X-ray disease classification: NIC (7 Ep.)& Pseudo-Labeling and memory buffer based replay
\\ \hline

% \rowcolor{yellow}
\rowcolor{hybridcolor!\tabcellci}
\tabrefcite{bayasi2024gc} 
&(a) Skin lesion classification: CIS (3 Ep.), DIS (4 Ep.), CIS+DIS (5 Ep.), (b) blood cell classification: CIS (4 Ep.), (c) Colon tissue classification: CIS (4 Ep.)& Architecture (fixed but partitioned network: culprit unit pruning mechanism) + Regularization (knowledge distillation from a generalized network to expert network to enhance generalizability)
\\ \hline

% \rowcolor{yellow}
\rowcolor{hybridcolor!\tabcellcis}
\tabrefcite{ye2024continual} &Multi-modal (medical report, MRI, X-ray, CT, histopathology) Representation learning: TIS (5 Ep.) & Rehearsal (K-means sampling based buffer creation for experience replay ) + Regularization (MSE based knowledge distillation)
\\ \hline 

\end{tabular}
}
\label{tab:hybridLiterature}
\end{table*}
%-----------------------------------------------
%%%%%%%%%%%%%%%%%%%%%%%%%%%%%%%%%%%%%%%%%%%%%%%%%%%5
\begin{table*}[!ht]
\caption{List of various comparative studies of CL techniques in literature}
\vspace{-1em}
\centering
\resizebox{.99\textwidth}{!}{
\begin{tabular}{ p{4cm} | p{7cm} | p{9cm} } 
% header			
\bottomrule
\rowcolor{gray!20}
\textbf{Reference (year)} 
& \textbf{ Application \& CL scenario} 
& \textbf{CL technique}
\\ \hline 

\rowcolor{hybridcolor!\tabcellcis}
\tabrefcite{lenga2020continual} 
& Chest X-ray classification: DIS (2 Ep.)
& Comparative study: joint training, EWC, LwF 
\\ \hline

\rowcolor{hybridcolor!\tabcellci}
\tabrefcite{morgado2021incremental} 
& Dermatological imaging modality classification: DIS 
& Comparative study: Regularization (EWC), rehearsal based (AGEM, experience replay) 
\\ \hline

\rowcolor{hybridcolor!\tabcellcis}
\tabrefcite{derakhshani2022lifelonger} 
& Disease classification: TIS, CIS, DIS (4 Ep.) 
& Regularization, memory-replay (comparative: EWC, MAS, LwF, iCaRL, EEIL) 
\\ \hline

\rowcolor{hybridcolor!\tabcellci}
\tabrefcite{ranem2022continual} 
& MRI-based binary hippocampus segmentation: DIS (3 Ep.) 
& Comparative study: transformers with EWC fisher matrix \tabdescite{kirkpatrick2017overcoming}, EWC with Riemannian walk \tabdescite{chaudhry2018riemannian}, Modeling the Background (knowledge distillation and a modified Cross Entropy Loss) \tabdescite{cermelli2020modeling}, Pseudo-labeling and Local Pod (multi-scale spatial distillation loss with pseudo labeling) \tabdescite{douillard2021plop}, Pooled Outputs Distillation \tabdescite{douillard2020podnet} 
\\ \hline

\rowcolor{hybridcolor!\tabcellcis}
\tabrefcite{kaustaban2022characterizing} 
&Tumor classification: DIS, CIS, TIS, IIL
& Regularization, replay (comparative study: EWC, LwF, CoPE, iCaRL, A-GEM) 
\\ \hline

\rowcolor{hybridcolor!\tabcellci}
\tabrefcite{bandi2023continual} 
& Cancer detection across organs such as breast, colon, and head-neck: DIS
& Comparative study of existing approaches: PackNet (architecture), EWC (regularization), GEM (rehearsal) 
\\ \hline 

% \rowcolor{yellow}
\rowcolor{hybridcolor!\tabcellcis}
\tabrefcite{verma2023privacy} 
& Disease classification on (a) Fundus: CIS (3 Ep.), (b) pathology images: CIS (3 Ep.)& Comparative study of existing buffer-free (privacy-aware) approaches in three categories: Rehearsal (GR, BIR), Regularization (EWC, SI, MAS, MUC-MAS, RWalk, OWM, GPM, LwF, LwM), Architecture (EFT)
\\ \hline

% \rowcolor{yellow}
\rowcolor{hybridcolor!\tabcellci}
\tabrefcite{kim2024continual} 
&Arrhythmia detection on ECG data: DIS (4 Ep.)& Comparative study of 3 existing regularization techniques (LwF, EWC, MAS)  
\\ \hline

% \rowcolor{yellow}
\rowcolor{hybridcolor!\tabcellcis}
\tabrefcite{ceccon2024fairness} 
&Chest X-ray disease classification: CIS (5 Ep.)&Comparative study of approaches in regularization (LwF, pseudo-label), rehearsal (replay), hybrid (rehearsal+regularization) categories for fairness evaluation. 
\\ \hline

\end{tabular}
}
\label{tab:comparativeLiterature}
\end{table*}
%-----------------------------------------------

%% file: content/sections/cl-technique/hybrid-techniques.tex
\subsection{Hybrid techniques}
\label{sec:hybrid-techniques}
Combining two or more individual CL strategies has recently gained interest owing to the enhanced performance gained by harnessing the merits of different CL strategies. \citet{ozdemir2018learn} mitigate catastrophic forgetting for incremental medical segmentation applications using a distillation loss inspired by LwF \cite{li2017learning}. Further, they extend the approach by incorporating memory-replay strategies and task-specific segmentation heads. Specifically, representative images were selected based on abstraction layer response and content distance in the last embedding layer. \citet{liu2022learning} also use knowledge distillation and a memory buffer to store the prototypical representation of different organ categories. \citet{akundi2022incremental} also keep class-specific classification head along with knowledge distillation to avoid forgetting for chest X-ray classification application where data arrives in class incremental fashion. A pseudo-rehearsal technique was combined with task-specific dense layers for pneumothorax classification \cite{ravishankar2019feature} by \citet{ravishankar2019feature}; however, it suffers from the linear increase of parameters and unbounded memory to store features for every domain.

\citet{li2020continual} introduce dual distillation as well as a fixed memory-based experience replay to continually learn the effective model. Specifically, they maintain an expert classifier, which is nothing but the previous model fine-tuned on new classes after replacing the old fully-connected layer with a new one. Then, a final classifier is distilled using the old classifier and the expert classifier. A fixed amount of old samples is also stored and replayed with current data while training the updated classifier. Then, the nearest-mean-of-exemplars (often calculated by averaging the feature vectors of the stored or selected subset of data in the feature space for each class) method is used for category prediction at the test time.

\citet{chee2023leveraging} utilize the dynamic expanding network, regularization, as well as data replay for continually learning new classes. A low-level feature extractor is shared across tasks, but a high-level feature extractor is especially learned for each subsequent task. Further, an alternate training procedure so as to learn new classes, i.e., newly added high-level feature extractor (by freezing others) and learning the old classes (freezing the new task feature extractor) is followed. A fixed memory per task is kept to replay samples with current samples. 

\citet{sadafi2023continual} use experience replay-based rehearsal and binary cross-entropy-based regularization terms for mitigating catastrophic forgetting. There is fixed memory and each class has an equal contribution to it. The sampling strategy for exemplar selection for a specific class is twofold: half of the samples are selected based on their distance from the mean, i.e., the closest samples to the class mean are selected, and another half of the samples for the class are selected based on model's uncertainty for each sample as given in epistemic uncertainty estimation \cite{mukhoti2021deep}.

\citet{zhang2023continual} exploit knowledge distillation as pseudo-labeling for old classes along with architecture-based CL strategy. In vanilla Swin UNETR architecture, they replace the conventional output layer responsible for segmentation with organ-specific light-weight segmentation heads, which thus enable independent predictions for any new or previously learned classes. Thus, there is a single encoder and decoder module and multiple organ-specific heads (few MLP layers) on the output of the encoder. In a similar fashion \citet{ji2023continual} keep organ-specific encoder and apply knowledge distillation. They also apply network architecture search-based pruning on decoders to maintain network complexity. But in comparison to \citet{zhang2023continual} strategy, the network complexity of maintaining multiple decoders is very high. 

\citet{zhang2023adapter} propose a rehearsal and architecture-based approach. In order to effectively extract discriminative features from a pre-trained feature extractor for different diseases, a learnable lightweight adapter is added between consecutive convolutional stages at each subsequent task. Furthermore, each task-specific classification head is also added at each step. Then, in each task, the model aims to find optimal parameters in the newly added task-specific adapters and classification head using the training data from the new task (new disease) and preserve a small subset for each previously learned disease.

%% file: content/sections/cl-technique/comparative-studies.tex
\subsection{Comparative studies}
\label{sec:comparative-studies}

There have been multiple comparative studies of different kinds of CL approaches on various applications like histopathology-based cancer detection \cite{bandi2023continual}, histopathology-based tumor detection \cite{kaustaban2022characterizing}, chest X-ray classification \cite{lenga2020continual}, dermatological image modality classification \cite{morgado2021incremental}, etc. \citet{lenga2020continual} show a comparative study of joint training and CL techniques such as EWC, and LwF to improve model adaptation to new chest X-ray domains arising from cross-sites while mitigating catastrophic forgetting effects. Upon model evaluation on two datasets from different sites (ChestX-ray14 and MIMIC-CXR) referring to domain shifts, the best results are achieved with joint training, but EWC and LwF offer practical solutions as the previous samples need not be stored with the CL approaches. In the digital pathology research field, \citet{kaustaban2022characterizing} systematically evaluate various CL methods, including regularization-based and rehearsal-based approaches on self-augmented domain shift on the H\&E dataset. The authors concluded that though regularization-based methods performed well for DIS and IIS, only the rehearsal-based method (iCaRL) is effective for CIS which is a more challenging scenario. Further, TIS may be even more challenging for digital pathology as compared to other image domains.

Further, with simulated datasets, i.e., MedMNIST (MNIST-like collections of biomedical images) various benchmark CL methods in rehearsal-based (iCaRL), regularization-based (EWC, MAS, LwF), and bias-correction method (EEIL) were evaluated under different CL scenarios \cite{derakhshani2022lifelonger}. For the majority of the experiments including CIS, TIS, and cross-DIS, the rehearsal-based approach, i.e., iCaRL shows the most promising results for disease classification. Further authors indicated that the existing CL methods which tend to do well in natural image applications (non-medical) may perform inadequately in medical disease classification applications due to their inherent complexity, such as the spatial locality of diseases. 
%%%%%Their study highlights the potential of rehearsal-based approaches in disease classification.
Similarly, in the case of a dermatological imaging modality classification problem under domain-shifted condition, \cite{morgado2021incremental} showed a comparison of EWC, averaged GEM, and experience replay. They concluded that MobileNetV2 with experience replay performs best among the others. \cite{gonzalez2023lifelong} also provided open-source implementation of five CL strategies including rehearsal, EWC, LwF, Riemannian Walk (RW), and Modeling the Background (MiB), on top of nnU-Net for various CL based applications like hippocampus, prostate, or cardiac segmentation.

%% file: content/sections/cl-supervision/index.tex
% %==================================
% \begin{table*}[!ht]
% \caption{Level of Supervision}
% \centering
% \begin{tabular}{|c|c|c|}
% \hline
% Task/Domain ID required? & \begin{tabular}[c]{@{}c@{}}Rigid Task/Domain boundary?\end{tabular} &Reference \\ \hline

% yes& yes& \cite{derakhshani2022lifelonger}\\ \hline
% no& yes &\cite{derakhshani2022lifelonger, gonzalez2022task}\\ \hline
% inferred& yes &   \\ \hline

% yes& no &\\ \hline
% no& no &\\ \hline
% inferred& no &\\ \hline

% \end{tabular}
%  \label{tab:supervision}
% \end{table*}
% %==================================
\begin{figure*}[!ht]
    \centering
    \includegraphics[scale=0.75]{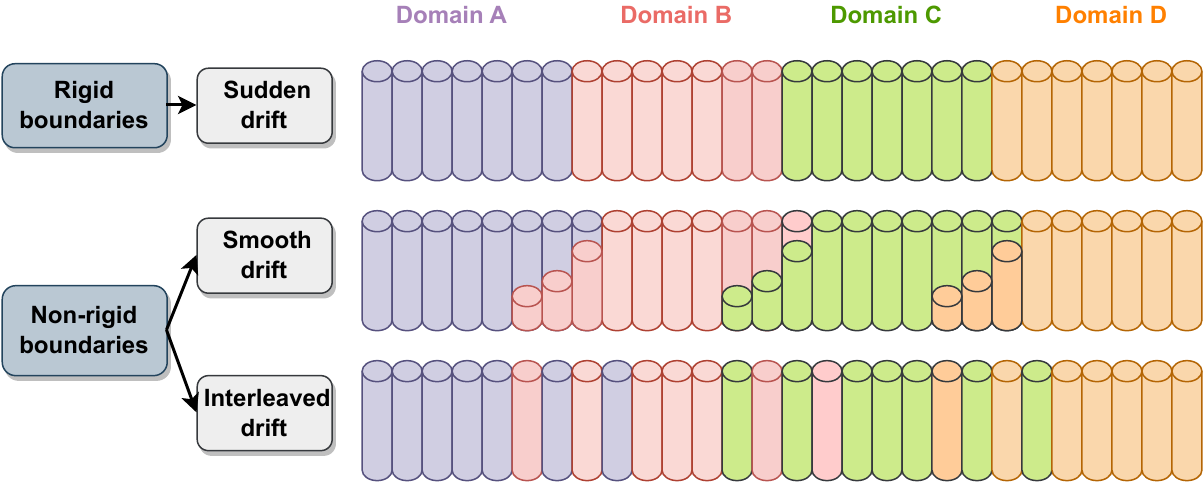}
    \caption{A demonstration for rigid and non-rigid task/domain boundaries for an example sequence of four domains in a CL pipeline}
    \label{fig:domainBoundaries}
\end{figure*}
\section{Level of supervision}
\label{sec:CLsupervision}

Here we provide a discussion on what level of supervision is considered and explored in literature while developing a CL method.

% Task or domain ID
\input{content/sections/cl-supervision/task-or-domain-id}

% Task boundaries
\input{content/sections/cl-supervision/task-boundaries}

% Sample annotation
\input{content/sections/cl-supervision/sample-annotation}

%% file: content/sections/cl-supervision/task-or-domain-id.tex
\subsection{Task or domain ID}
\label{sec:task-or-domain-id}
Most of the approaches specifically require information about task/domain ID, i.e., which task/domain the current samples belong to. This information is then used to select or activate the task-specific branch in the model for classification/segmentation/feature extraction. If task/domain ID is needed then the approach is regarded as (a) task/domain aware else (b) task/domain agnostic. However, a well-designed CL approach should not rely on a task/domain ID to perform prediction~\cite{aljundi2017expert,bayasi2021culprit}. 
Most of the approaches presume a-priori knowledge of domain IDs~\cite{derakhshani2022lifelonger,zhang2023continual} while some infer from the data~\cite{zhang2023adapter,bayasi2021culprit,aljundi2017expert}. A very few do not require this information~\cite{derakhshani2022lifelonger,gonzalez2022task,perkonigg2022continual}. %%%%%Table~\ref{tab:task_id_supervision} can be refereed to see the partition of literature that requires ID, infer, and the works that do not need it.

%% file: content/sections/cl-supervision/task-boundaries.tex
\subsection{Task boundaries}
\label{sec:task-boundaries}
Typically, a shift is regarded as a change of data source, i.e., datasets coming from different sites, acquisition environments, etc. Therefore, there are well-separated domain shifts or task boundaries. Thus, the necessary measures to adapt to the changed domain are triggered when there is a change in task/domain. Most of the works in the literature follow the assumption of well-separated task/domain boundaries. 

However, in real-life cases, the change in domain/task could be interleaved or smooth rather than an abrupt transition. Such settings come under the umbrella of `blurred boundary continual learning' \cite{wang2023comprehensive}. To better visualize the rigid and non-rigid boundary-based sequence of domains, a demonstration is provided via \refx{fig:domainBoundaries}. Recently a few works \cite{hofmanninger2020dynamic, srivastava2021continual,gonzalez2022task,perkonigg2022continual} have explored such settings. For example, \citet{gonzalez2022task} considers two kinds of non-rigid shifts in hippocampus segmentation datasets. In the first kind, termed `shifting source', they create slowly shifting distributions with 3 datasets (HarP, Dryad, and Decathlon) by interleaving samples from these datasets for segmentation application. %%%% as shown in Fig.~\refx{fig:gonjalez_example}. 
In another case termed `transformed', they create shift using TorchIO library \cite{perez2021torchio} which performs various intensity rescaling and affine transform in the Decathlon dataset to create five episodes. Another example of a non-rigid boundaries-based CL scenario is given by \citet{srivastava2021continual} for the 
chest X-ray classification problem. They consider 3 multi-site datasets (NIH Chest-X-rays14, PadChest, and CheXpert) as 3 domains; however, they curate a smooth transition from one dataset to another at the boundaries. %%% as depicted through Fig.~\refx{fig:Srivastava}
 %%%%%%Further, a tabularization of literature over rigid and non-rigid boundaries of task is provided in Table~\refx{tab:task_boundaries_supervision} for reference.

%% file: content/sections/cl-supervision/sample-annotation.tex
\subsection{Sample annotation}
\label{sec:sample-annotation}
Apart from information regarding the task IDs, annotation for the samples in tasks is also a major concern. Almost all the available CL strategies in the medical domain require sample annotation for learning the classification or segmentation task. It is undeniable that labeling is highly costly in the medical domain, especially in the histopathology field, and hence may hinder the rapid advancement of the field. In other domains, unsupervised CL strategies have been explored; for example, \citet{ashfahani2022unsupervised} use only a few labeled samples to associate clusters to classes and the learning is completely unsupervised.
%%talk about few shot/ self-supervised work...recent in miccai, active learning paper one too

%% file: content/sections/evaluation/index.tex
%===========================================
\begin{figure}[!ht]
\centering
\includegraphics[scale=0.65]{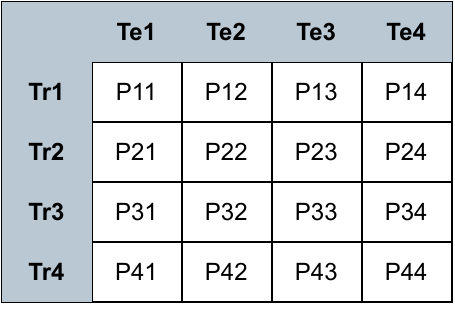}
\vspace{-0.7em}
\caption{Train-test performance matrix}
\label{fig:cl_conf_mat}
\end{figure}
%===========================================
%%%%%%%%%%%%%%%%%%%%%%%%%%%%%%%%%%%%%%%%%%%%%%%%%%

\input{content/tables/bwt-matrices}

\input{content/tables/fwt-matrices}

%%%%%%%%%%%%%%%%%%%%%%%%%%%%%%%%%%%%%%%%%%%%%%%
\section{Evaluation strategy and metrics}
\label{sec:evaluation}

%%% commented on Dec. 19, 2023.... will discuss it when asked in revision...\textit{Typically we have training and test pairs as $(Tr_i, Te_i)$ for each $i^{th}$ episodes. \textbf{Some works use only one held-out test set. CHECK CAREFULLY THE WORK BY ~\cite{karthik2022segmentation}} Very few works have used validation sets~\cite{kaustaban2022characterizing,perkonigg2022continual,hofmanninger2020dynamic,morgado2021incremental}.  \textbf{WRITE SOME MORE EXAMPLE AND DETAIL.} }

To evaluate a CL strategy, various aspects like the performance of the model on past, current, or future episodes, resource consumption~\cite{zhang2023continual}, memory size~\cite{hofmanninger2020dynamic}, model size growth, execution time, etc can be explored. CL frameworks involving dynamic architecture or memory also report time and memory analysis. For example, \citet{gonzalez2022task} maintains model history as mean and variance parameters of multivariate Gaussian, which dynamically grow in the presence of drift; hence, authors also provided training time analysis. If the framework does not involve dynamic architecture, primarily performance on data is reported. Further, the performance of data mainly involves evaluating and reporting the episode performance, stability, and plasticity.

In CL, there is a given sequence of episodes ($1,...,T$) to learn sequentially. Typically, we have well-defined training and test pairs as $(Tr_i, Te_i)$ for each $i^{th}$ episode. Upon sequential training, a \textbf{train-test performance matrix} $P\in \mathit{P}^{T\times T}$ is generated for the given sequence of episodes ($1,...,T$). An example matrix for $T=4$ is shown in \refx{fig:cl_conf_mat}. The cell $P_{t,i}$ is performance on test data of $i^{th}$ episode when the model training up to $t^{th}$ episodes is complete. Thus, after training up to $t^{th}$ episodes, the performance $P_{t,i}$ for {$i<t$, $i=t$, and $i>t$ correspond to performances on past, current, and future episodes}, respectively.

The performance metric is chosen depending on the application. For example, IoU, dice score/coefficient, structural similarity, Hausdorff distance, average symmetric surface distance, etc., are popular choices for segmentation, whereas accuracy, recall, F1-score, and AUC are usually computed for classification applications. Once the performance metric is chosen, we can compute metrics for measuring stability and plasticity on top of it. Various metrics have been proposed to quantify stability (\refx{tab:bwtMatrices}) and plasticity (\refx{tab:fwtMatrices}); however, all the metrics are derived from the same train-test performance matrix (\refx{fig:cl_conf_mat}) to measure the amount of forgetting or forward transfer. 

%%%%%%%%%%%%%%%%%%%%%%%%%%%%%%%%%%%%%%%%%%%%%%%%%%%%%%%%%%%%%%%%

% Stability: Backward Transfer
\input{content/sections/evaluation/stability}

\input{content/sections/evaluation/plasticity}

%%%%%%%%%%%%%%%%%%%%%%%%%%%%%%%%%%%%%%%%%%%%%%%%%%
%%%%\subsubsection{Task accuracy}

% Episode performance and others
\input{content/sections/evaluation/performance}

%% file: content/tables/bwt-matrices.tex
%-----------------bwtMatrices--------------------
\begin{table*}[!ht]
	\caption{Metrics for backward transfer. For ease of readability, terms in equations are color-coded to refer to the corresponding cells in the matrix (last column).}
    \vspace{-1em}
    \centering

	\begin{tabular}{cccc}
		% header
        \bottomrule
        \rowcolor{gray!20}
		\textbf{Reference}
		& \textbf{Eq.}
		& \textbf{Equation}
		& \textbf{Pictorial representation}
		\\ \hline
		
		\tabrefcite{lopez2017gradient} & Eq.(A)
		& $ \displaystyle
		\frac{1}{T-1} \sum_{i=1}^{T-1} \left( \tikzmarky[a1]{P_{T,i}} - \tikzmarkp[a2]{P_{i,i}} \right) 
		$
		& \adjustbox{valign=c}{\includegraphics[scale=0.4]{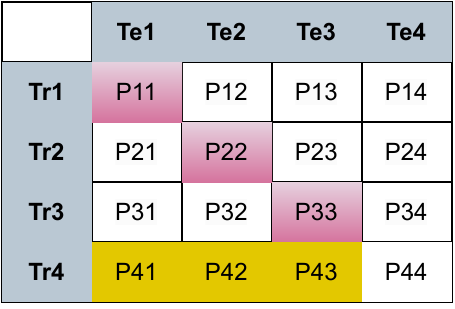}}
		\\  

        \rowcolor{gray!5}
		\tabrefcite{derakhshani2021kernel}& Eq.(B) 
		& $ \displaystyle
		\frac{1}{T-1} \sum_{i=1}^{T-1} \max_{j=1}^{T-1} \left( \tikzmarky[b1]{P_{j,i}} - \tikzmarkp[b2]{P_{T,i}} \right) 
		$
		& \adjustbox{valign=c}{\includegraphics[scale=0.4]{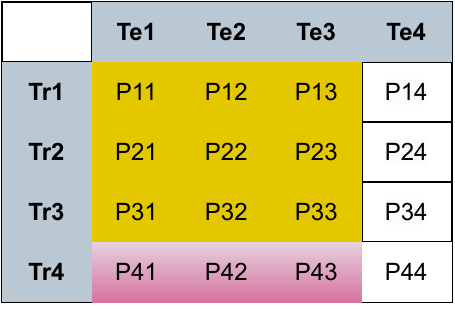}}
		\\ 
		
		\tabrefcite{chee2023leveraging} & Eq.(C) 
		& $ \displaystyle
		\frac{1}{T-1} \sum_{t=2}^{T} \left( \frac{1}{t-1} \sum_{i=1}^{t-1} \max_{j=1}^{t-1} \left( \tikzmarky[c1]{P_{j,i}} -\tikzmarkp[c2]{P_{t,i}} \right) \right)
		$
		& \adjustbox{valign=c}{\includegraphics[scale=0.4]{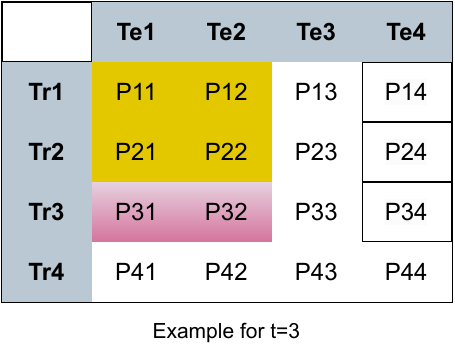}} 
		\\ 
		
		\rowcolor{gray!5}
		\tabrefcite{gonzalez2022task} & Eq.(D) 
		& $ \displaystyle
		\frac{1}{T-1} \sum_{i=1}^{T-1} \left( \frac{1}{\left | \left \{  t_j \right \}_{j>i} \right | } \sum_{j>i}^{} \left( \tikzmarky[d1]{P_{j,i}} - \tikzmarkp[d2]{P_{i,i}} \right) \right)
		$
		& \adjustbox{valign=c}{\includegraphics[scale=0.4]{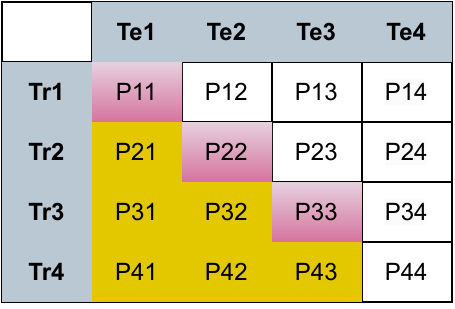}} 
		\\ 
		
		\tabrefcite{ozgun2020importance} & Eq.(E) 
		& $ \displaystyle
		\frac{2}{T(T-1)} \sum_{t=2}^{T} \sum_{i=1}^{t-1} \max \left( \tikzmarky[e1]{P_{t,i}}- \tikzmarkp[e2]{P_{i,i}}, 0 \right)
		$
		& \adjustbox{valign=c}{\includegraphics[scale=0.4]{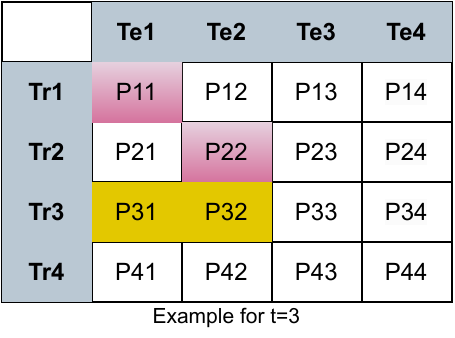}} 
		\\ \hline 
		
	\end{tabular}
 \label{tab:bwtMatrices}
	
\end{table*}
%--------------------------------------------

%% file: content/tables/fwt-matrices.tex
%-----------------fwtMatrices--------------------
\begin{table*}[!ht]
	\caption{Metrics for forward transfer. For ease of readability, terms in equations are color-coded to refer to the corresponding cells in the matrix (last column).}
    \vspace{-1em}
    \centering
	
	\begin{tabular}{cccc} 
		% header			
        \bottomrule
        \rowcolor{gray!20}
		\textbf{Reference} 
		& \textbf{Eq.} 
		& \textbf{Equation} 
		& \textbf{Pictorial representation}
		\\ \hline 
		
		\tabrefcite{lopez2017gradient} & Eq.(F) 
		& $ \displaystyle
		\frac{1}{T-1}\sum_{i=2}^{T} \left( \tikzmarky[f]{P_{i-1,i}} - \bar{b}_{i} \right)
		$ 
		& \adjustbox{valign=c}{\includegraphics[scale=0.4]{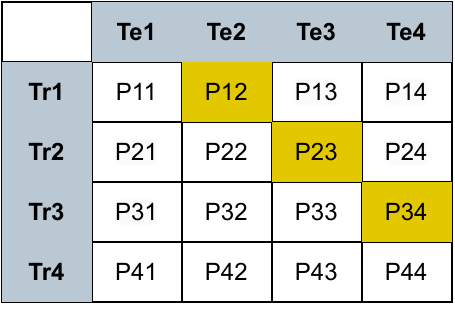}} 
		\\ 
		
		\rowcolor{gray!5}
		\tabrefcite{gonzalez2022task} & Eq.(G) 
		& $ \displaystyle
		\frac{1}{T-1}\sum_{i=2}^{T} \left( \frac{1}{\left |\left \{  t_j\right \}_{j\leq i} \right |}\sum_{j\leq i}^{} \left(P_{j,i}- P_{j-1,i} \right) \right)
		$
		&
		\\ 
		
		\tabrefcite{shu2022replay} & Eq.(H) 
		& $ \displaystyle 
		\frac{1}{T-1}\sum_{i=2}^{T} \left( \tikzmarky[h1]{P_{i,i}}-\tikzmarkp[h2]{P_{1,i}}\right) 
		$ 
		& \adjustbox{valign=c}{\includegraphics[scale=0.4]{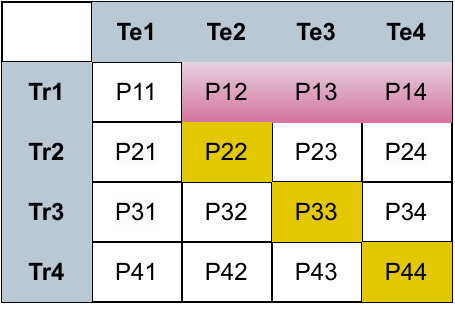}} 
		\\ 
		
		\rowcolor{gray!5}
		\tabrefcite{ozgun2020importance} & Eq.(I) 
		& $ \displaystyle
		\frac{2}{T(T-1)} \sum_{t=1}^{T-1} \sum_{i> t} \tikzmarky[i1]{P_{t,i}} 
		$ 
		& \adjustbox{valign=c}{\includegraphics[scale=0.4]{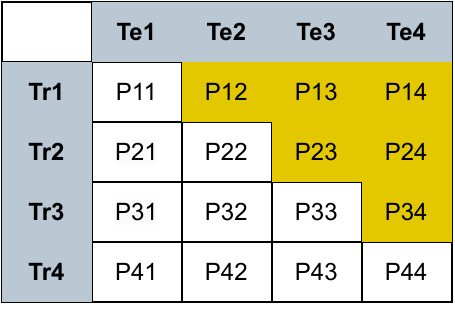}} 
		\\ 
		
		\tabrefcite{ozgun2020importance} & Eq.(J) 
		& $ \displaystyle
		 \frac{1}{(T-1)} \sum_{t=2}^{T} \left( \tikzmarky[j1]{P_{i,i}}-b_{i} \right) 
		$ 
		& \adjustbox{valign=c}{\includegraphics[scale=0.4]{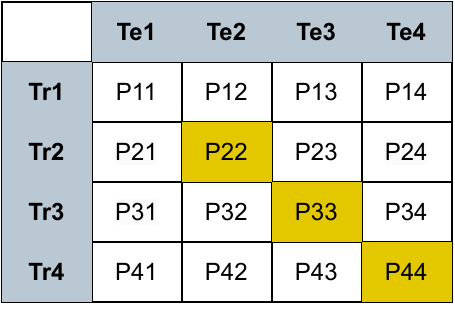}} 
		\\ \hline 
		
	\end{tabular}
	\label{tab:fwtMatrices}
	
\end{table*}
%-----------------------------------------------

%% file: content/sections/evaluation/stability.tex
\subsection{Stability: Backward transfer}
\label{sec:evaluation-stability}
It refers to measuring the catastrophic forgetting of a model upon learning new episodes. A model is regarded as stable if it gives a similar performance for an already learnt episode over time. However, upon learning the new episode naively, it is possible to forget the past learning and hence disrupt the performance of the previous episode. In contrast, a carefully designed CL strategy will cause less forgetting or improvement on past episodes upon learning the new episodes. A backward transfer (BWT) metric is used to measure the amount of forgetting. In other words, it is a way to measure stability, i.e., how well the model would retain the previously acquired knowledge to prevent catastrophic forgetting.

\tabrefcite{lopez2017gradient} define BWT as the influence of new learning $t_i$ task on a previous task. When learning a task $t_i$ increases the performance on a {previous} task, it is $+ve$ BWT, and if it causes deterioration, it results in $-ve$ BWT. Large $-ve$ BWT is also termed as catastrophic forgetting. The BWT metric given by \tabrefcite{lopez2017gradient} is very frequently adopted \cite{srivastava2021continual,hofmanninger2020dynamic} and it is computed after finishing all the episodes as shown in Eq.(A) in \refx{tab:bwtMatrices}. 
Average forgetting \cite{derakhshani2021kernel} is another popularly used metric \cite{derakhshani2022lifelonger,chen2022breast}. It is computed as the average of the difference in highest performance, and the final performance reached after training on all episodes is finished (Eq.(B) in \refx{tab:bwtMatrices}). \tabrefcite{chee2023leveraging} compute BWT after learning each subsequent task ($t^{th}$) rather than computing it at the last episode (Eq.(B) in \refx{tab:bwtMatrices}) as shown through Eq.(C) in \refx{tab:bwtMatrices}. \tabrefcite{gonzalez2022task} compute BWT as the change in performance after training with each subsequent task averaged over the number of tasks (Eq.(D) in \refx{tab:bwtMatrices}). Note that BWT is not defined for the last task. Another BWT metric contribution is by \tabrefcite{ozgun2020importance} as shown via Eq.(E) in \refx{tab:bwtMatrices}.

%% file: content/sections/evaluation/plasticity.tex
\subsection{Plasticity: Forward transfer}
\label{sec:evaluation-plasticity}
The plasticity of a model is reflected in its capability to accommodate more and more knowledge or exploit the knowledge learned so far to learn the new. In literature, this is frequently referred to as forward transfer (FWT). It is also said that FWT measures the ``zero-shot" learning capability of a model \cite{lopez2017gradient}. 
In contrast to the usage of BWT, not all works report FWT. 
%%%%Certain approaches might enforce rigorous regularization on the model, which can result in poor performance on new tasks, consequently causing a comparatively reduced level of forgetting, i.e., high $+ve$ BWT. 
Certain approaches might enforce rigorous regularization on the model to get a comparatively reduced level of forgetting (high $+ve$ BWT); however, this results in poor performance on new tasks. 
Therefore, reporting FWT along with BWT is equally important for unbiased evaluation of the CL model.

The very popular FWT metric \cite{srivastava2021continual,hofmanninger2020dynamic} shown via Eq.(F) in \refx{tab:fwtMatrices} is proposed by \citet{lopez2017gradient} that quantifies FWT as the impact of learning a task $t_i$ on a future task. It is the average of the difference in the performance on $t_i$ task before learning it and the performance given by a model with random weights ($\bar{b}_{i}$). Note that FWT is not defined for the first task. Further, \citet{gonzalez2022task} computes FWT as the change in performance in each stage before and up to $Tr_i$, averaged over the number of tasks (Eq.(G) in \refx{tab:fwtMatrices}).  Another metric for FWT was proposed by \citet{shu2022replay} as shown via Eq.(H) in \refx{tab:fwtMatrices}. \citet{ozgun2020importance} defines two separate forward transfer metrics, one for unseen (average of elements above diagonals of $P$) and another for seen episodes as reported via Eq.(I) and Eq.(J) in \refx{tab:fwtMatrices}, respectively. There, $b_{i}$ is the performance by a standalone model only trained on $i^{th}$ episode. Further, they report a ``transfer-learning metric" as the sum of diagonal elements in matrix $P$ as measures of plasticity, i.e., the ability to adapt to new tasks.

% \begin{equation}
% \frac{1}{T-1}\sum_{i=2}^{T} P_{i-1,i}- \bar{b}_{i}
%   \label{eq:fwt_lopez}
% \end{equation}  
% %------------------------------
% \begin{figure}
%     \centering
%     \includegraphics[scale=0.5]{DRAWN/forward_lopez.pdf}
%     \caption{forward\_lopez}
%     \label{fig:forward_lopez}
% \end{figure}
% %------------------------------
%
% \begin{equation}
%  \frac{1}{T-1}\sum_{i=2}^{T}\frac{1}{\left |\left \{  t_j\right \}_{j\leq i} \right |}\sum_{j\leq i}^{}\left [ P_{j,i}- P_{j-1,i} \right ]
% \label{eq:fwt_gonzalez}
% \end{equation}  
%
%

% \begin{equation}
% \frac{1}{T-1}\sum_{i=2}^{T} P_{i,i}- P_{1,i}
%   \label{eq:fwt_shu}
% \end{equation}  
% %------------------------------
% \begin{figure}
%     \centering
%     \includegraphics[scale=0.5]{DRAWN/forward_shu.pdf}
%     \caption{forward\_shu}
%     \label{fig:forward_shu}
% \end{figure}
% %------------------------------
%

% \begin{equation}
% \frac{2}{T(T-1)} \sum_{t=1}^{T-1} \sum_{i> t}P_{t,i}
%   \label{eq:fwt_unseen_li2022domain}
% \end{equation}  
%
% \begin{equation}
% \frac{1}{(T-1)} \sum_{t=2}^{T} P_{i,i}-b_{i}
%   \label{eq:fwt_seen_li2022domain}
% \end{equation} 
%------------------------------
% \begin{figure}
%     \centering
%     \includegraphics[scale=0.5]{DRAWN/forward_ozgun.pdf}
%     \caption{forward\_ozgun}
%     \label{fig:forward_ozgun}
% \end{figure}
% %------------------------------

%% file: content/sections/evaluation/performance.tex
\subsection{Episode performance and others}
\label{sec:evaluation-performance}
Average accuracy~\cite{derakhshani2021kernel} is computed as the model performance after training up to $t^{th}$ episode. At any time $t$, the average performance is the mean of accuracy values for episodes ($1,...,t$) as below:
%--------------------
\begin{equation}
 \frac{1}{t}\sum_{i=1}^{t} P_{t,i}
\label{eq:avg_acc}
\end{equation}
%--------------------
Thus, after processing all the episodes, we get total $T$ average accuracy values which can be plotted and visualized for different approaches~\cite{derakhshani2022lifelonger, kaustaban2022characterizing}. Further, researchers~\cite{derakhshani2022lifelonger} also compute mean over these values ($\frac{1}{T}\sum_{i=1}^{T} [\frac{1}{t}\sum_{i=1}^{t} P_{t,i}$]) to get a single scalar value which can be used to directly compare with other approaches. Another popular definition of average accuracy is to compute the above-mentioned metric only after finishing all the episodes~\cite{lopez2017gradient,chen2022breast,kaustaban2022characterizing} as below. 
%--------------------
\begin{equation}
 \frac{1}{T}\sum_{i=1}^{T} P_{T,i}
\label{eq:avg_acc_lopez}
\end{equation}
%--------------------
Another performance metric namely ``incremental-learning metric" is contributed which is the average of elements in lower-triangular matrix $P$~\cite{ozgun2020importance, li2022domain}.

%% file: content/tables/classification.tex
\begin{table*}[!ht]
\caption{Literature for CL based classification. (**) indicates non-imaging modality}
\vspace{-1em}
\centering
\resizebox{.99\textwidth}{!}{
\begin{tabular}{p{2cm}|p{2.3cm}|p{2.7cm}|p{1.3cm}|p{4cm}|p{2cm}|p{1.8cm}|p{2cm}|p{5cm}}
% header			
\bottomrule
\rowcolor{gray!20}
\textbf{Reference (year)} & \textbf{Imaging modality} & \textbf{Application} 
& \textbf{CL scenario} & \textbf{CL strategy} & \textbf{Model}
& \textbf{Learning type} & \textbf{Evaluation metric} & \textbf{Dataset}
\\ \hline 

% body
\rowcolor{classificationcolor!\tabcellci}
\tabrefcite{ravishankar2019feature} & X-ray, Ultrasound & (a) Cardiac view classification, (b) Pneumothorax identification
& DIS, TIS, IIS & Rehearsal + Architectural & CNN
& Supervised & Accuracy, BWT & Subset of ChestXRay \tabdescite{wang2017chestx}
\\ \hline

\rowcolor{classificationcolor!\tabcellcis}
\tabrefcite{hofmanninger2020dynamic} & CT & Classification of synthetic object in chest images
& DIS & Rehearsal & Res-Net50 \tabdescite{he2016deep}
& Supervised & Accuracy, BWT, FWT & Private
\\ \hline

\rowcolor{classificationcolor!\tabcellci}
\tabrefcite{lenga2020continual} & X-ray & Chest X-ray classification 
& DIS & Comparitive: EWC, LWF & DenseNet121 \tabdescite{huang2017densely}
& Supervised & AUC, BWT, FWT & ChestX-ray14, MIMIC-CXR
\\ \hline

\rowcolor{classificationcolor!\tabcellcis}
\tabrefcite{li2020continual} & Dermoscopic & Skin disease classification
& CIS & Regularization + Rehearsal & ResNet18, AlexNet, VGG19
& Supervised & Accuracy & Skin8, Skin40 from ISIC2019 \tabdescite{tschandl2018ham10000}, CIFAR100
\\ \hline

\rowcolor{classificationcolor!\tabcellci}
\tabrefcite{srivastava2021continual} & X-ray & Chest X-ray classification
& DIS & Rehearsal & ResNet101 
& Supervised & AUC, FWT, BWT & NIH Chest-X-rays14 \tabdescite{mcdermott2020chexpert++}, PadChest \tabdescite{bustos2020padchest}, CheXpert \tabdescite{wang2017chestx}
\\ \hline

\rowcolor{classificationcolor!\tabcellcis}
\tabrefcite{bayasi2021culprit} & Dermoscopic & Skin lesion classification
& DIS & Architectural (fixed but partitioned network) & ResNet-50
& Supervised & Accuracy & HAM10000 \tabdescite{tschandl2018ham10000}, Dermofit \tabdescite{ballerini2013color}, Derm7pt \tabdescite{kawahara2018seven}, MSK \tabdescite{codella2018skin}, PH2 \tabdescite{mendoncca2013ph}, UDA \tabdescite{codella2018skin} 
\\ \hline

\rowcolor{classificationcolor!\tabcellci}
\tabrefcite{morgado2021incremental} & Dermoscopic & Dermatological imaging modality classification
& DIS & Regularization, rehearsal & VGG-16, MobileNetV2 
& Supervised & Accuracy, precision, recall, F1-score, FWT, BWT & Private 
\\ \hline

\rowcolor{classificationcolor!\tabcellcis}
\tabrefcite{shu2022replay} & Fundus images & Eyes disease classification
& DIS & Rehearsal + Regularization & Resnet-18, 2-layer MLP \tabdescite{saha2021}
& Supervised & F1-score, BWT, FWT & ODIR, RIADD, REFUGE, iSee \tabdescite{fang2020attention}, EyeQ \tabdescite{fu2019evaluation}, PMNIST \tabdescite{rupesh2013compete} 
\\ \hline

\rowcolor{classificationcolor!\tabcellci}
\tabrefcite{akundi2022incremental} & X-ray & Chest X-ray classification
& CIS & Regularization & DenseNet121
& Supervised & AUC, BWT & CheXpert
\\ \hline

\rowcolor{classificationcolor!\tabcellcis}
\tabrefcite{kaustaban2022characterizing} & Histopathology & Tumor classification (breast cancer, colorectal cancer) 
& IIS, DIS, CIS, TIS & Regularization, Rehearsal (comparative:EWC, LwF, AGEM, CoPE, iCaRL) & ResNet-18 
& Supervised & Accuracy, FWT, BWT & CRC \tabdescite{kather2019predicting}, PatchCam \tabdescite{bejnordi2017diagnostic} 
\\ \hline

\rowcolor{classificationcolor!\tabcellci}
\tabrefcite{chen2022breast} & Histopathology & Breast cancer identification
& CIS & Regularization & Alexnet, Densenet201, Resnet152
& Supervised & Accuracy, forgetting & BreakHis
\\ \hline

\rowcolor{classificationcolor!\tabcellcis}
\tabrefcite{derakhshani2022lifelonger} & Multiple & Multiple disease classification
& TIS, CIS, cross-DIS & Regularization, Rehearsal (comparative: EWC, MAS, LwF, iCaRL, EEIL) & ResNet18
& Supervised & Accuracy, forgetting & MedMNIST \tabdescite{yang2023medmnist} 
\\ \hline

\rowcolor{classificationcolor!\tabcellci}
\tabrefcite{chee2023leveraging} & Histopathology, retinal, skin lesions & (a) Colorectal cancer detection, (b) diabetic retinopathy classification, (c) skin lesion classification 
& CIS & Rehearsal + Architecture + Regularization & ResNet18
& Supervised & Accuracy, forgetting & CCH5000 \tabdescite{kather2016multi}, EyePACS \tabdescite{kaggleDiabeticRetinopathy}, HAM10000 \tabdescite{tschandl2018ham10000} 
\\ \hline

\rowcolor{classificationcolor!\tabcellcis}
\tabrefcite{zhang2023adapter} & 
Multiple & Disease classification (multiple organ source) 
& CIS & Rehearsal + Architectural & ResNet18
& Supervised & MCR & Skin8, Path16, CIFAR100
\\ \hline

\rowcolor{classificationcolor!\tabcellci}
\tabrefcite{bandi2023continual} & Histopathology & Cancer detection cross organs (breast, colon, head-neck)
& DIS & Comparative: (Architecture: PackNet) (Regularization: EWC, GEM) & DenseNet
& Supervised & FROC, ROC, Cohen's Kappa & CAMELYON16, CAMELYON17, private for colon and head-neck 
\\ \hline

\rowcolor{classificationcolor!\tabcellcis}
\tabrefcite{bai2023revisiting} & VQA & Surgical visual-question localized-answering 
& CIS & Regularization & VisualBERT
& Supervised & Accuracy, IoU & EndoVis18, EndoVis17, M2CAI 
\\ \hline

\rowcolor{classificationcolor!\tabcellci}
\tabrefcite{sadafi2023continual} & 
Microscopic & WBC classification
& DIS, CIS, DIS + CIS & Rehearsal + Regularization & ResNeXt-50
& Supervised & Accuracy & Matek-19 \tabdescite{matek2019single}, INT-20, and Acevedo-20 \tabdescite{acevedo2020dataset} 
\\ \hline

\rowcolor{classificationcolor!\tabcellcis}
\tabrefcite{yang2023few} & Multiple & Disease classification 
& DIS+CIS & Rehearsal & ResNet18
& Few-shot & Accuracy, performance dropping rate & MedMNIST \tabdescite{yang2023medmnist} (PathMNIST, DermaMNIST, OrganAMNIST, RetinaMNIST, BreastMNIST, BloodMNIST) 
\\ \hline

\end{tabular}
} 
\label{tab:classificationCL}
\end{table*}
%---------------------------------- 
%%%%%%%%%%%%%%%%%%%%%%%%%%%%%%%%%%%%%%%%%%%%%%%%%%%%%%%%%%%%%%%%
\begin{table*}[!ht]
\ContinuedFloat
\caption{Literature for CL based classification. (**) indicates non-imaging modality (\textit{Continued})}
\vspace{-1em}
\centering
\resizebox{.99\textwidth}{!}{
\begin{tabular}{p{2cm}|p{2.3cm}|p{2.7cm}|p{1.3cm}|p{4cm}|p{2cm}|p{1.8cm}|p{2cm}|p{5cm}}
% header			
\bottomrule
\rowcolor{gray!20}
\textbf{Reference (year)} & \textbf{Imaging modality} & \textbf{Application} 
& \textbf{CL\hspace{1em} scenario} & \textbf{CL strategy} & \textbf{Model}
& \textbf{Learning type} & \textbf{Evaluation metric} & \textbf{Dataset}
\\ \hline

% \rowcolor{yellow}
\rowcolor{classificationcolor!\tabcellci}
\tabrefcite{byunconditional} & 2D fundus, dermoscopic & (a) Diabetic retinopathy severity classification, (b) skin lesion classification 
& DIS & Rehearsal & ViT-B/16
& supervised & AUC, backward transfer &fundus (Messidor-2, APTOS), skin lesion (BCN2000, PAD-UEFS-20, HAM10000)
\\ \hline 

% \rowcolor{yellow}
\rowcolor{classificationcolor!\tabcellcis}
\tabrefcite{xiao2023fs3dciot} 
&dermoscopic \& clinical images&Skin disease classification  & CIS & Rehearsal (Memory buffer optimization)  & ResNet-18
& Few-shot & accuracy, sensitivity, and specificity & private curated from various public sources (ISIC-2018, ISIC-2019, ISIC-2020, SD-198, SD-256,
PAD-OFED-20, Asan, Hally, etc.) 
\\ \hline 

% \rowcolor{yellow}
\rowcolor{classificationcolor!\tabcellci}
\tabrefcite{yang2023continual} 
&Dermoscopic, clinical&Skin disease classification  & CIS & Architectural  & ResNet101 + GMMs
& Supervised & MCR & Skin7~\tabdescite{codella2018skin}, Skin40~\tabdescite{sun2016benchmark}
\\ \hline

% \rowcolor{yellow}
\rowcolor{classificationcolor!\tabcellcis}
\tabrefcite{verma2023privacy} 
& (a) Fundus, (b) pathology &Disease classification   & CIS & Comparative (Rehearsal, Regularization, Architecture)  & ResNet50
& Supervised & MCR &(a) OCT~\tabdescite{kermany2018large}, 
(b) PathMNIST~\tabdescite{yang2023medmnist}
\\ \hline

% \rowcolor{yellow}
\rowcolor{classificationcolor!\tabcellci}
\tabrefcite{huang2023conslide} 
&WSI& Tumor subtype classification & CIS& Rehearsal  & ConvNeXt (image encoder) + transformer (classifier)  
& supervised &BWT, AUC, Masked AUC, Accuracy & 4 datasets (NSCLC, BRCA, RCC, ESCA) from TCGA\footnote{https://www.cancer.gov/ccg/research/genome-sequencing/tcga} project 
\\ \hline 

% \rowcolor{yellow}
\rowcolor{classificationcolor!\tabcellcis}
\tabrefcite{hua2023incremental} 
&sEMG**& Gesture classification & CIS & Rehearsal+ Regularization & CNN
& supervised &Accuracy, Remember & Ninapro DB2 \tabdescite{atzori2014electromyography}
\\ \hline 

% \rowcolor{yellow}
\rowcolor{classificationcolor!\tabcellci}
\tabrefcite{sun2023adaptive} 
&time-series signals**& Mortality identification, Sepsis identification & CIS& Rehearsal  &  LSTM
& online &BWT, FWT, AUC, Accuracy & COVID-19 \tabdescite{yan2020interpretable} , SEPSIS \tabdescite{seymour2017time}
\\ \hline 

% \rowcolor{yellow}
\rowcolor{classificationcolor!\tabcellcis}
\tabrefcite{sun2023few} 
&time-series signals**& Multi-class classification & CIS& Rehearsal + Regularization &1D-ConvNet
& few-shot &Accuracy, relative performance drop rate &Mit-BIH \tabdescite{goldberger2000physiobank}, FaceAll \tabdescite{dau2019ucr}, UWave \tabdescite{dau2019ucr}, Mit-BIH Long-Term ECG \tabdescite{goldberger2000physiobank}
\\ \hline 

% \rowcolor{yellow}
\rowcolor{classificationcolor!\tabcellci}
\tabrefcite{li2023doctor} 
&Physiological signals**&Disease classification  & DIS, CIS, TIS& Rehearsal   & MLP
& Supervised & Accuracy, F1 score, BWT & CovidDeep\tabdescite{hassantabar2021coviddeep}, DiabDeep\tabdescite{yin2019diabdeep}, MHDeep\tabdescite{hassantabar2022mhdeep}
\\ \hline

% \rowcolor{yellow}
\rowcolor{classificationcolor!\tabcellcis}
\tabrefcite{kim2024continual} 
&ECG**&Arrhythmia detection   & DIS & Comparative (Regularization: LwF, EWC, MAS)  & 1D-CNN+MLP
& Supervised & AUC & \tabdescite{zheng202012}, \tabdescite{wagner2020ptb}, \tabdescite{alday2020classification}, \tabdescite{liu2018open}
\\ \hline

% \rowcolor{yellow}
\rowcolor{classificationcolor!\tabcellci}
\tabrefcite{aslam2024cel} 
& Time series signal**&Disease classification    & DIS &Regularization (EWC)  & LSTM
& Supervised & Accuracy, forgetting & 
 Mpox~\tabdescite{mathieu2022mpox}, Influenza~\tabdescite{cdcFluViewInteractive}, and Measles~\tabdescite{europaSurveillanceAtlas}
\\ \hline

% \rowcolor{yellow}
\rowcolor{classificationcolor!\tabcellcis}
\tabrefcite{ceccon2024multi} 
&X-ray&Chest disease classification  & NCI & Rehearsal + Regularization  &  --
& Supervised & AUC, F1 score, forgetting & ChestX-ray14\tabdescite{wang2017chestx}, CheXpert\tabdescite{irvin2019chexpert}
\\ \hline

% \rowcolor{yellow}
\rowcolor{classificationcolor!\tabcellci}
\tabrefcite{verma2024confidence} 
& (a) Fundus, (b) histopathology&Disease classification   & TIS & Architectural  & ResNet50
& Supervised & Accuracy &(a) OCT~\tabdescite{kermany2018large}, 
(b) PathMNIST~\tabdescite{yang2023medmnist}
\\ \hline

% \rowcolor{yellow}
\rowcolor{classificationcolor!\tabcellcis}
\tabrefcite{bayasi2024gc} 
&(a) Dermoscopic, (b) microscopic, (c) histopathology&(a) Skin lesion classification, (b) blood cell classification, (c) Colon tissue classification  & CIS, DIS, CIS+DIS &Regularization + Architectural  & ResNet50
& Supervised & Avg. recall, forgetting, AUPRC& 
(a) HAM10000~\tabdescite{tschandl2018ham10000},
Dermofit~\tabdescite{ballerini2013color}, Derm7pt~\tabdescite{kawahara2018seven}, MSK~\tabdescite{codella2018skin},
UDA~\tabdescite{codella2018skin}, BCN~\tabdescite{combalia2019bcn20000}, PH2~\tabdescite{mendoncca2013ph} (b) PBS-HCB~\tabdescite{acevedo2020dataset}, (c) NCT-CRC-HE~\tabdescite{kather2019predicting}
\\ \hline

% \rowcolor{yellow}
\rowcolor{classificationcolor!\tabcellci}
\tabrefcite{ceccon2024fairness} 
&X-ray & Chest disease classification  &CIS  &Comparative (Regularization, Rehearsal, hybrid) & ResNet50
& Supervised & AUC, TPR gap& ChestX-ray14 \tabdescite{wang2017chestx}, CheXpert \tabdescite{irvin2019chexpert}
\\ \hline

% \rowcolor{yellow}
\rowcolor{classificationcolor!\tabcellcis}
\tabrefcite{thandiackal2024multi} 
& Histopathology&Tissue classification    & DIS &Rehearsal   & ResNet18
& Unsupervised & F1 score & 
K-19~\tabdescite{kather2016multi}, K-16~\tabdescite{kather2019predicting}, CRC-TP~\tabdescite{javed2020cellular} 
\\ \hline

% \rowcolor{yellow}
\rowcolor{classificationcolor!\tabcellci}
\tabrefcite{qazi2024dynammo} 
& (a) Histopathology, (b) dermoscopic&Disease classification    & CIS &Architectural   & ResNet18
& Supervised &Accuracy, FLOPS & 
Skin8, Path16
\\ \hline

% \rowcolor{yellow}
\rowcolor{classificationcolor!\tabcellcis}
\tabrefcite{bringas2024cladsi} 
& Time-series (motion-sensor)** &Alzheimer's disease stage identification  & IIS (2 Ep.,3 Ep.,4 Ep.)&Rehearsal   & 1D-CNN
& Supervised &Accuracy, F1-score, forgetting & 
private
\\ \hline

\end{tabular}
} 
\label{tab:classificationCL2}
\end{table*}
%----------------------------------   

%% file: content/tables/segmentation.tex
%----------------segmentationCL-------------------
\begin{table*}[!ht]
\caption{Literature for CL based segmentation}
\vspace{-1em}
\centering
\resizebox{.99\textwidth}{!}{
\begin{tabular}{p{2.5cm}|p{2cm}|p{2.6cm}|p{1.25cm}|p{2cm}|p{2cm}|p{1.8cm}|p{1.7cm}|p{7cm}}
% header			
\bottomrule
\rowcolor{gray!20}
\textbf{Reference (year)} & \textbf{Imaging modality} & \textbf{{Application}} 
& \textbf{CL scenario} & \textbf{CL strategy} & \textbf{Model}
& \textbf{Learning type} & \textbf{Evaluation metric} & \textbf{Dataset}
\\ \hline 

% body
\rowcolor{segmentationcolor!\tabcellci}
\tabrefcite{karani2018lifelong} & MRI & Brain 
& DIS & Architecture & U-Net 
& Supervised & DSC & HCP \tabdescite{van2013wu}, ADNI~\tabdescite{uscADNIAlzheimerapossDisease}, ABIDE \tabdescite{di2014autism} and IXI~\tabdescite{brainixidataset} 
\\ \hline

\rowcolor{segmentationcolor!\tabcellcis}
\tabrefcite{baweja2018towards} & MRI & Brain structure segmentation (normal, white matter lesions)
& TIS & Regularization: EWC & DeepMedic 3D convnet \tabdescite{kamnitsas2017efficient}
& Supervised & DSC coefficient & UK Biobank \tabdescite{miller2016multimodal} 
\\ \hline

\rowcolor{segmentationcolor!\tabcellci}
\tabrefcite{mcclure2018distributed} & SMRI &Brain structure segmentation (axial and sagittal) 
& DIS & Regularization: DWC & MeshNet \tabdescite{fedorov2017end} (CNN-based)
& Supervised & DSC & HCP \tabdescite{van2013wu}, NKI \tabdescite{nooner2012nki}, Buckner \tabdescite{biswal2010toward}, WU120 \tabdescite{power2017temporal} 
\\ \hline

\rowcolor{segmentationcolor!\tabcellcis}
\tabrefcite{ozdemir2018learn} & MRI &Segmentation of humerus and scapula 
& CIS & Regularization (LwF) + rehearsal & U-Net
& Supervised & DSC, SSD & Private 
\\ \hline

\rowcolor{segmentationcolor!\tabcellci}
\tabrefcite{ozdemir2019extending} & MRI & Knee segmentation
& CIS & Regularization & U-Net
& Supervised & DSC, MSD & SKI10 MICCAI Grand Challenge \tabdescite{heimann2010segmentation} 
\\ \hline

\rowcolor{segmentationcolor!\tabcellcis}
\tabrefcite{van2019towards} & 4 MR sequences: pre- and post-contrast T1w, T2w, \& T2w FLAIR & Brain tumor segmentation
& DIS & Regularization: EWC & 3D U-Net CNN
& Supervised & Dice score & BraTS 2018 \tabdescite{bakas2017advancing, menze2014multimodal}, private 
\\ \hline

%%not a cl paper
% \tabdescite{venkataramani2019towards} ***** & X-ray & Lung 
% & Domain- incremental (3 Ep.) & Memory- replay***
% & U-Net & Supervised & Dice coefficient 
% &Montgomery \tabdescite{jaeger2014two}, JSRT \tabdescite{shiraishi2000development}, Pneumoconiosis (private) \\ \hline

\rowcolor{segmentationcolor!\tabcellci}
\tabrefcite{ozgun2020importance} & MRI & Brain MRI segmentation
& DIS & Regularization: MAS & QuickNAT \tabdescite{roy2019quicknat} 
& Supervised & CL-DSC, REM, BWT+, TL, FWT & CANDI \tabdescite{kennedy2012candishare}, MALC \tabdescite{asman2013non}, and ADNI \tabdescite{jack2008alzheimer} 
\\ \hline

\rowcolor{segmentationcolor!\tabcellcis}
\tabrefcite{memmel2021adversarial} & MRI & Hippocampal segmentation
& DIS &  & U-Net
& Supervised & IoU, DSC & \tabdescite{simpson1902large}, \tabdescite{kulaga2015multi}, \tabdescite{boccardi2015training} 
\\ \hline

\rowcolor{segmentationcolor!\tabcellci}
\tabrefcite{perkonigg2021continual} & MRI & Brain MRI segmentation
& DIS & Rehearsal & 3D-ModelGenesis  \tabdescite{zhou2021models}
& Supervised & MAE, FWT, BWT & IXI~\tabdescite{brainixidataset}, OASIS3 \tabdescite{lamontagne2019oasis} 
\\ \hline

\rowcolor{segmentationcolor!\tabcellcis}
\tabrefcite{ranem2022continual} & MRI & Hippocampus segmentation
& DIS & Comparative (rehearsal, regularization, hybrid)& ViT U-Net
& Supervised & DSC, FWT, BWT & Decathlon \tabdescite{antonelli2022medical}, Drayd \tabdescite{denovellis2021hippocampal}, HarP \tabdescite{boccardi2015training} 
\\ \hline

\rowcolor{segmentationcolor!\tabcellci}
\tabrefcite{liu2022learning} & CT & Organ segmentation (liver, spleen, pancreas, right kidney left kidney) 
& CIS & Rehearsal + regularization & nnUNet 
& Supervised & DC, HD95 & \tabdescite{MultiAtl34online,simpson2019large}, KiTS \tabdescite{heller2019kits19} + private 
\\ \hline

\rowcolor{segmentationcolor!\tabcellcis}
\tabrefcite{karthik2022segmentation} & MRI & Multiple sclerosis lesions (brain) segmentation
& DIS & Rehearsal & 3D U-Net
& Supervised & DSC, BWT & \tabdescite{kerbrat2020multiple} 
\\ \hline

\rowcolor{segmentationcolor!\tabcellci}
\tabrefcite{gonzalez2022task} & MRI & Hippocampus segmentation
& DIS & Architectural & nnUNet \tabdescite{isensee2021nnu}
& Supervised & DSC, FWT, BWT & HarP \tabdescite{boccardi2015training}, Dryad \tabdescite{kulaga2015multi}, Decathlon \tabdescite{simpson2019large} 
\\ \hline

\rowcolor{segmentationcolor!\tabcellcis}
\tabrefcite{li2022domain} & CMR & Cardiac segmentation
& DIS & Rehearsal & U-Net
& Supervised & DSC, HD95, BWT, FWT & M\&Ms \tabdescite{campello2021multi} 
\\ \hline

\rowcolor{segmentationcolor!\tabcellci}
\tabrefcite{zhang2023s, zhang2021comprehensive} & MRI & (a) Prostate segmentation, (b) optic cup and disc segmentation
& DIS & Regularization & U-Net
& Supervised & DSC, ASSD &
RUNMC \tabdescite{bloch2015nci},
BMC \tabdescite{bloch2015nci}, 
HCRUDB \tabdescite{lemaitre2015computer}, 
UCL \tabdescite{litjens2014evaluation},
BIDMC \tabdescite{litjens2014evaluation}, 
HK \tabdescite{litjens2014evaluation}, \tabdescite{sivaswamy2015comprehensive,fumero2011rim,orlando2020refuge} 
\\ \hline

\rowcolor{segmentationcolor!\tabcellcis}
\tabrefcite{zhang2023continual} & CT & (a) Abdomen multi-organ segmentation, (b) abdomen to liver tumor segmentation 
& CIS & Architectural & Swin UNETR \tabdescite{hatamizadeh2021swin}
& Supervised & Average DSC & BTCV \tabdescite{landman2015miccai}, LiTS \tabdescite{bilic2023liver}, JHH \tabdescite{xia2022felix} (private) 
\\ \hline

\rowcolor{segmentationcolor!\tabcellci}
\tabrefcite{ji2023continual} & 3D CT & Whole-body organ segmentation
& CIS & Architectural & nnUNet
& Supervised & DSC HD95 & TotalSegmentator \tabdescite{wasserthal2023totalsegmentator}, ChestOrgan, HNOrgan, EsoOrgan (3 private) 
\\ \hline

\rowcolor{segmentationcolor!\tabcellcis}
\tabrefcite{bera2023memory} & MRI & Prostate segmentation, hippocampus segmentation, spleen segmentation 
& TIS, DIS & Rehearsal & Residual UNet
& Supervised & DSC, average forgetting, BWT, accuracy & (Prostate158 \tabdescite{adams2022prostate158}, NCI-ISBI \tabdescite{NCIISBI291online}, Promise12 \tabdescite{litjens2014evaluation}, Decathlon  \tabdescite{antonelli2022medical}), (Drayd \tabdescite{denovellis2021hippocampal}, HarP \tabdescite{boccardi2015training}), (Spleen dataset in Decathlon \tabdescite{NCIISBI291online}) 
\\ \hline

\rowcolor{segmentationcolor!\tabcellci}
\tabrefcite{zhu2023uncertainty} & CT, MRI & (a) Abdomen segmentation, (b) muscles segmentation, (c) prostate segmentation 
& DIS & Regularization & U-Net
& Supervised & DSC, ASSD & Prostate \tabdescite{liu2020shape,bloch2015nci,lemaitre2015computer}, abdomen \tabdescite{MultiAtl34online,kavur2021chaos}, muscles \tabdescite{zhu2021deep} 
\\ \hline

\rowcolor{segmentationcolor!\tabcellcis}
\tabrefcite{liu2023incremental} & MRI & Brain tumor segmentation
& CIS + domain shift & Regularization & ResNet-based 2D nnU-Net 
& Supervised & DSC, HD95 & BraTS2013 \tabdescite{menze2014multimodal}, TCIA \tabdescite{clark2013cancer}, CBICA \tabdescite{bakas2018identifying} 
\\ \hline

\rowcolor{segmentationcolor!\tabcellci}
\tabrefcite{wang2023rethinking} & Endoscopic images & (a) Endoscopy segmentation, (b) surgical instrument segmentation 
& CIS & Rehearsal + regularization & ResNet101
& Supervised & mIoU & EDD2020 \tabdescite{ali2021deep,ali2020endoscopy}, EndoVis18 \tabdescite{allan20202018}, EndoVis17 \tabdescite{allan20192017} 
\\ \hline 
\end{tabular}
}
\label{tab:segmentationCL}
\end{table*}
%--------------------------------------------------
%%%%%%%%%%%%%%%%%%%%%%%%%%%%%%%%%%%%%%%%%%%%%%%%%%%%%%%%%%

\begin{table*}[!ht]
\ContinuedFloat
\caption{Literature for CL based segmentation (\textit{Continued})}
\vspace{-1em}
\centering
\resizebox{.99\textwidth}{!}{
\begin{tabular}{p{2.5cm}|p{2cm}|p{2.6cm}|p{1.25cm}|p{2cm}|p{2cm}|p{1.8cm}|p{1.7cm}|p{7cm}}
% header			
\bottomrule
\rowcolor{gray!20}
\textbf{Reference (year)} & \textbf{Imaging modality} & \textbf{{Application}} 
& \textbf{CL scenario} & \textbf{CL strategy} & \textbf{Model}
& \textbf{Learning type} & \textbf{Evaluation metric} & \textbf{Dataset}
\\ \hline 

% \rowcolor{yellow}
\rowcolor{segmentationcolor!\tabcellci}
\tabrefcite{wei2023representative} & MRI &Brain tumor segmentation
& IIS & Rehearsal  & UNet
& Supervised & DSC & LGG Segmentation \tabdescite{buda2019association} 
\\ \hline

% \rowcolor{yellow}
\rowcolor{segmentationcolor!\tabcellcis}
\tabrefcite{chen2023generative} & Fundus, MRI &(a) Optic disc segmentation, (b) cardiac segmentation, (c) prostate segmentation
& DIS& Rehearsal + Regularization (KD) & UNet
& Unsupervised & DSC & (a) REFUGE~\tabdescite{orlando2020refuge}, IDRiD~\tabdescite{porwal2018indian}, RIM-ONE DL~\tabdescite{batista2020rim}, (b) M\&Ms challenge datasets~\tabdescite{campello2021multi}, (c) RUNMC~\tabdescite{liu2020shape},
NCI-ISBI13~\tabdescite{bloch2015nci}, I2CVB~\tabdescite{lemaitre2015computer},  PROMISE12~\tabdescite{litjens2014evaluation}
\\ \hline

% \rowcolor{yellow}
\rowcolor{segmentationcolor!\tabcellci}
\tabrefcite{li2024dual} & MRI &Cardiac segmentation
& DIS & Rehearsal + Architecture & Res-UNet
& Supervised & DSC, BWT, FWT & ACDC~\tabdescite{bernard2018deep}, M\&M~\tabdescite{campello2021multi} 
\\ \hline

% \rowcolor{yellow}
\rowcolor{segmentationcolor!\tabcellcis}
\tabrefcite{zhu2024boosting} & MRI &(a) Prostate segmentation (b) Cardiac
segmentation& DIS & Regularization & UNet
& Supervised &HD95, DSC, BWT & (a) RUNMC~\tabdescite{bloch2015nci}, BMC~\tabdescite{bloch2015nci}, I2CVB~\tabdescite{lemaitre2015computer}, UCL~\tabdescite{litjens2014evaluation}, BIDMC~\tabdescite{litjens2014evaluation}, HK~\tabdescite{litjens2014evaluation}, (b) M\&M~\tabdescite{campello2021multi} 
\\ \hline

\end{tabular}
}
% \label{tab:segmentationCL2}
\end{table*}
%--------------------------------------------------

%% file: content/tables/class-seg-litreature.tex
%----------------segmentationPluClassificationCL-------------------
\begin{table*}[!ht]
\caption{CL Literature for classification + segmentation {and other application}}
\vspace{-1em}
\centering
\resizebox{.99\textwidth}{!}{
\begin{tabular}{p{2cm}|p{2cm}|p{2.7cm}|p{2cm}|p{2cm}|p{1.7cm}|p{1.8cm}|p{2cm}|p{5.5cm}}
% header			
\bottomrule
\rowcolor{gray!20}
\textbf{Reference (year)} & \textbf{Imaging modality} & \textbf{{Application}} 
& \textbf{CL scenario} & \textbf{CL strategy} & \textbf{Model}
& \textbf{Learning type} & \textbf{Evaluation matrices} & \textbf{Dataset}
\\ \hline 

% body
% \rowcolor{yellow}
\rowcolor{bothcolor!\tabcellci}
\tabrefcite{zhang2019continually} &Longitudinal MRI & Alzheimer’s
disease progression modelling (regression)
& TIS (7 Ep.) & Regularization & MLP
& Supervised &wCC, PCC, rMSE& ADNI-1 \tabdescite{jack2008alzheimer},
\\ \hline 

\rowcolor{bothcolor!\tabcellcis}
\tabrefcite{perkonigg2022continual} & CMR, CT, MRI & Cardiac segmentation, lung nodule detection, brain age estimation
& TIS & Rehearsal & 2D-UNet, Faster R-CNN, ResNet-50
& Supervised & Dice score, AP, MAE & Cardiac \tabdescite{campello2021multi},  LIDC \tabdescite{setio2017validation} + LNDb challenge \tabdescite{pedrosa2019lndb}, IXI \tabdescite{brainixidataset} + OASIS-3 \tabdescite{lamontagne2019oasis}
\\ \hline 

% \rowcolor{yellow}
\rowcolor{bothcolor!\tabcellci}
\tabrefcite{wu2024modal} &(PD, T1, T2) weighted MRI, X-ray & Super-resolution
& TIS & Regularization & HAN \tabdescite{niu2020single}
& Supervised &SSIM, PSNR, BWT, forgetting, Intransigence (plasticity) & IXI \tabdescite{brainixidataset}, Chest X-ray \tabdescite{wang2017chestx}
\\ \hline 

% \rowcolor{yellow}
\rowcolor{bothcolor!\tabcellcis}
\tabrefcite{ye2024continual} &Multi-modality (medical report, MRI, X-ray, CT, histopathology) & Representation learning
& TIS (5 Ep.)& Rehearsal + Regularization & Transformer
& Self-supervised &DSC, HD95, AUC, accuracy, F1-score &
MIMIC-CXR 2.0.0 \tabdescite{johnson2019mimic},
ADNI-1+ADNI-2+ADNI-GO \tabdescite{jack2008alzheimer},
DeepLesion \tabdescite{yan2018deeplesion},
TCGA
\\ \hline

% \rowcolor{yellow}
\rowcolor{bothcolor!\tabcellci}
\tabrefcite{zhu2024lifelong} &Histopathology & WSI retrieval
& CIS (4 Ep.)& Rehearsal & TransMIL \tabdescite{shao2021transmil}
& Supervised &KRC, SRC, slide-level mean AP &
TCGA\\ \hline

\end{tabular}
}
\label{tab:segmentationPluClassificationCL}
\end{table*}
%----------------------------------------------------------------

%% file: content/sections/discussion/index.tex
%%%%%%%Challenges and future trends
%%https://www.thelancet.com/journals/landig/article/PIIS2589-7500(21)00076-5/fulltext
%%%%\st{CL methods were also computationally efficient, taking only about 28\% of the runtime as joint training. Though patient data evolve quickly nowadays, FDA has not approved algorithms based on CL [28] and extensive research is needed to establish regulations for safely incorporating CL in clinical settings.}
%https://pubs.rsna.org/doi/epdf/10.1148/radiol.2020200038
%https://www.thelancet.com/journals/landig/article/PIIS2589-7500(20)30102-3/fulltext
%%https://link.springer.com/chapter/10.1007/978-3-030-87234-2_16
%%%%%%%%%%%%%%%%%%%%%%%%%%%%%%%%%%%%%%%%%%%%%%%%%%%%%%%%%%%%%%%%%%%%%%%%%%
\section{Discussion: Towards the future}
\label{sec:futureDirection}
Our exploration of the CL methods for medical imaging reveals a mix of challenges and strategies. The literature reflects a struggle with traditional paradigms, notably the need for explicit task/domain identification. While some methods demand prior knowledge of domain IDs, others explore task/domain-agnostic approaches that aim to eliminate this reliance \cite{aljundi2017expert, bayasi2021culprit}. The evolving notion of task/domain boundaries adds another layer of complexity, with real-world scenarios often defying the assumption of well-separated shifts. Non-rigid transitions and blurred boundaries, as explored by \citet{gonzalez2022task,wang2023comprehensive}, pose challenges to existing CL approaches.
Sample annotation, a common requirement for CL in medical imaging, introduces resource challenges, especially in histopathology. While supervised strategies dominate, \citet{ashfahani2022unsupervised} showcase the feasibility of unsupervised CL, using minimal labeled samples. The evaluation of CL strategies involves a diverse set of metrics, including stability (backward transfer - BWT) and plasticity (forward transfer - FWT). These metrics, encompassing aspects like average forgetting and incremental learning, facilitate nuanced comparisons.
In the classification domain, applications like chest X-rays and skin disease classification employ diverse CL strategies. Latent replay mitigates catastrophic forgetting but increases resource consumption, while knowledge distillation balances stability and plasticity \cite{li2020continual,srivastava2021continual}. Segmentation tasks, spanning brain sclerosis lesion segmentation to cardiac segmentation, rely on regularization techniques like EWC and knowledge distillation \cite{karthik2022segmentation,li2022domain}. The simultaneous handling of classification and segmentation tasks, evident in studies on dermoscopic image-based skin disease classification \cite{li2020continual} and abdominal organ segmentation \cite{liu2022learning}, introduces a trade-off between resource efficiency and overall model performance.
The synthesis of literature and comparative analyses unveils the current state of CL in medical imaging. However, challenges persist, and future research should explore unsupervised CL, address labeling costs, and refine evaluation metrics. As the field evolves, understanding the dynamic interplay between medical imaging tasks and CL methodologies will be crucial for driving innovation. In \refx{tab:classificationCL,tab:segmentationCL,tab:segmentationPluClassificationCL}, we present the overview of the selected publications in the domain of classification, segmentation, and both combined, respectively. We highlight the domain/application, dataset information, imaging modalities, CL scenarios, and strategies, including the learning process and methodology adopted/proposed by different studies. {From the last year (2023), we also witness some exploration in non-imaging medical applications including disease classification, gesture classification, arrhythmia detection, etc. Further, apart from classification and segmentation, applications like super-resolution~\mbox{\cite{wu2024modal}}, WSI retrieval~\mbox{\cite{zhu2024lifelong}}, representation learning~\mbox{\cite{ye2024continual}}, etc., also benefit from CL.}
Based on our critical analysis of existing literature, we discuss some of the open challenges and thus possible research directions for CL in the medical field.

%%%%%%%%%%%%%%%%%%%%%%%%%%%%%%%%%%%%%%%%%%%%%%%%%%%%%%%%%%%%%%%%%%%%%%%%%%
\textbf{Non-imaging or other applications}
{Recently, there have also been explorations in wearable medical sensor data. For example, \mbox{\cite{hua2023incremental}} explore experience replay (various sample selection strategies were explored) and knowledge distillation (on old data) based CL for incrementally learning gesture classes with sEMG data. Similarly, \mbox{\cite{sun2023adaptive}} and \mbox{\cite{sun2023continuous}} incrementally learn classes from time series signals (ECG, categorical data) for applications including time-series health monitoring, mortality identification, sepsis identification, gesture classification, etc. Further, there have been studies to handle domain shifts due to multi-site data \mbox{\cite{kim2024continual}}, patient-level splits \mbox{\cite{li2023doctor}}, data distributional shifts (change in mean and standard deviation) \mbox{\cite{aslam2024cel}} in ECG, physiological signal, or other time-series data for disease classification and arrhythmia detection.}

{Largely CL has been explored for mainstream applications such as classification and segmentation, but some other application, such as image reconstruction, registration, translation, generation, anomaly detection, etc., can also benefit from CL techniques. Recently, \mbox{\citet{wu2024modal}} explored CL for the image super-resolution task across different imaging modalities including PD, T1, and T2 weighted MRI, and X-ray is learned in a continual manner with a single super-resolution model. Notably, to deal with forgetting, authors use the idea of constraining gradient updates according to the importance of parameters as well as adding a distillation loss as a regularization. }
%%%%%%%%%%%%%%%%%%%%%%%%%%%%%%%%%%%%%%%%%%%%%%%%%%%%%%%%%%%%%%%%%%%%%%%%%%

\textbf{Availability of benchmark datasets}: Contrary to other natural imaging fields, there are no standard benchmark datasets for CL in the medical field. Therefore, a fair comparison of different research advancements made in the field is challenging. MedMNIST is an attempt to provide an MNIST-like dataset for the medical field, however, it is limited by its non-complex nature of various sub-datasets and not mainly developed for CL. Hence, it does not offer various drift conditions that a CL technique could be explored to evaluate and quantify.

%%%%%%%%%%%%%%%%%%%%%%%%%%%%%%%%%%%%%%%%%%%%%%%%%%%%%%%%%%%%%%%%%%%%%%%%%%
\textbf{Intense labor requirements}: Most of the CL approaches are supervised in nature and thus demand large annotated datasets for sequential training. This creates a bottleneck in model development especially in the histopathology domain as it is the most expensive in terms of annotation. To cope with a limited labeling budget in problems like cardiac segmentation, lung nodule detection, and brain age estimation, \citet{perkonigg2022continual,perkonigg2021continual} explored an active learning strategy within the CL framework. The model indicates important samples and then only those samples are annotated by the experts. On the other hand, there are fully unsupervised CL approaches \cite{ke2022unsupervised,pratama2021unsupervised,ashfahani2022unsupervised} or self-supervised CL approaches \cite{liu2023class} in other domains that should be explored here to tackle the annotation scarcity problem.

%%%%%%%%%%%%%%%%%%%%%%%%%%%%%%%%%%%%%%%%%%%%%%%%%%%%%%%%%%%%%%%%%%%%%%%%%%
%%%%%\textbf{Plasticity vs. stability dilemma}

%%%%%%%%%%%%%%%%%%%%%%%%%%%%%%%%%%%%%%%%%%%%%%%%%%%%%%%%%%%%%%%%%%%%%%%%%%

\textbf{Unexplored distributional changes}: In the medical realm, drifts arising from covariance shifts are mainly explored that too by incorporating datasets from different centers or acquisition protocols for the same task. Thus, the type of drift is always ``sudden drift'' and easier to handle. 
On the other hand, there are many realistic unexplored drift settings. For example, drift within the center is close to a realistic scenario as it gives rise to smooth drift scenarios \cite{perkonigg2022continual}. 
%%%%%%Further, some drifts occur temporarily for example, the cell tissue may degrade and cause the segmentation problem harder over time as observed in Bichlmayer et al.~\cite{bichlmayer20223d}.
Further, some drifts occur temporarily; for example, \citet{bichlmayer20223d} observe one novel kind of domain shift in kidney-CAM-model data \cite{bichlmayer20223d}, even if they follow a fixed acquisition protocol. The tissue structure in histopathology data of mouse kidneys shows degradation over time, which causes the segmentation problem to be harder over time. 
Another very important kind of unexplored drift is ``concept shift" or ``concept drift", which refers to a situation where the shift causes a change in the relationship of the input variable with its target value \cite{lu2018learning,kumari2022anomaly,bhatt2022experimental,kumari2021situational}.
% \textcolor{red}{
For example, in the context of diabetic patient identification, consider a scenario where individuals aged 30-50 are deemed likely to be diabetic patients. However, if, for unforeseen reasons or based on lifestyle and food habits, there is a shift in the severity level and the age group to 20-50, it would be considered a concept drift for diabetic patient identification. This implies a fundamental change in the underlying patterns and characteristics associated with diabetic patients, thus influencing the performance and reliability of machine learning models. 

% \textcolor{blue}{
%For example, in a tumor classification problem the definition of the tumor class may change due to unforeseen reasons, i.e., the non-tumor can become a tumor and the model needs to learn this concept shift. 
%%%%%%%%%%%%%%%%%%%%%%%%%%%%%%%%%%%%%%%%%%%%%%%%%%%%%%%%%%%%%%%%%%%%%%%%%%%%%%

\textbf{Scalability for large-Scale medical datasets:}
Scalability emerges as a critical challenge in the context of CL for medical imaging, particularly as datasets continue to grow in size and complexity \cite{gonzalez2020wrong}. Efficient strategies are needed to handle the scale of large medical datasets, considering both computational resources and model complexity. Research efforts, as explored by \citet{zhang2023continual}, should focus on optimizing model architectures, memory management, and training algorithms to ensure scalability without compromising performance.

%%%%%%%%%%%%%%%%%%%%%%%%%%%%%%%%%%%%%%%%%%%%%%%%%%%%%%%%%%%%%%%%%%%%
\textbf{Interpretable and explainable models:}
The interpretability and explainability of CL models constitute another open challenge in the medical imaging domain \cite{rymarczyk2023icicle}. As highlighted in \citet{derakhshani2022lifelonger}, models used in clinical settings should not only achieve high performance but also provide meaningful insights into decision-making processes. Future research directions should prioritize the development of CL methodologies that can offer explanations for their predictions, fostering trust and acceptance in the clinical community.
%%%%%%%%%%%%%%%%%%%%%%%%%%%%%%%%%%%%%%%%%%%%%%%%%%%%%%%%%%%%%%%%%%%%

\textbf{Ethical considerations and patient privacy:}
Ethical considerations and patient privacy concerns are paramount in CL for medical imaging. The use of patient data for model training raises ethical questions regarding consent, data anonymization, and potential biases. Future research, in alignment with ethical guidelines discussed by \citet{chee2023leveraging}, should prioritize the development of frameworks that ensure responsible data usage, privacy preservation, and ethical guidelines. Striking a balance between innovation and the protection of patient rights is imperative for the sustainable progress of CL in medical imaging. {\mbox{\citet{verma2023privacy}} stress over exemplar-free CL approaches for colon pathology and retina data-based disease classification. They present a comparative study of 12 existing approaches in all three major categories offering privacy-aware solutions. Specifically, they use LwF \mbox{\cite{radio3}}, LwM \mbox{\citet{dhar2019learning}}, EWC \mbox{\cite{kirkpatrick2017overcoming}}, SI \mbox{\cite{zenke2017continual}}, MAS \mbox{\cite{aljundi2017expert}}, MUC-MAS \mbox{\cite{liu2020more}}, RWalk \mbox{\cite{chaudhry2018riemannian}}, OWM \mbox{\cite{zeng2019continual}}, and GPM \mbox{\cite{saha2021}} in regularization category, GR \mbox{\cite{shin2017continual}} and BIR \mbox{\cite{van2020brain}} in generative replay category, and EFT \mbox{\cite{verma2021efficient}} in architecture-based category. }

{\textbf{Limitations with Gigapixel WSIs:} 
The high resolution of whole slide images (WSIs), often 50,000 x 50,000 pixels, presents significant computational challenges for deep learning model design. The variability in WSI imaging technology and staining protocols affects model performance on new data, requiring continual adaptation of WSI analysis methods \mbox{\cite{LAI2024100474}}. Directly applying standard CL approaches to hierarchical WSI models can lead to severe knowledge forgetting of previously seen datasets. Additionally, WSIs are gigapixel images with only slide-level labels, making storage and computation for rehearsal-based CL impractical due to limited memory. \mbox{\citet{huang2023conslide}} proposed a CL method, "ConSlide", to progressively update a hierarchical WSI analysis architecture using sequentially acquired heterogeneous WSI datasets. In this approach, a representative set of past datasets is stored and periodically reorganized and replayed during model updates using an asynchronous updating mechanism. \mbox{\citet{thandiackal2024multi}} proposed a rehearsal-based domain incremental scenario with unsupervised learning for tissue classification in WSIs patches. They utilize the generative feature-driven image replay in conjunction with a dual-purpose discriminator. Further, \mbox{\citet{zhu2024lifelong}} enhance reservoir sampling for WSI retrieval by incorporating distance consistency-based rehearsal.}

%%%%%%%%%%%%%%%%%%%%%%%%%%%%%%%%%%%%%%%%%%%%%%%%%%%%%%%%%%%%%%%%%%%%

\textbf{Continual learning in diffusion models:}
The integration of CL into diffusion models \cite{kazerouni2023diffusion} represents a promising avenue for advancing the capabilities of these models in dynamic and evolving environments. Diffusion models, widely employed in diverse fields, including image processing and medical imaging, often face challenges when confronted with changing data distributions. CL techniques can play a pivotal role in allowing these models to adapt and accumulate knowledge over time, ensuring sustained performance in the face of evolving datasets and tasks. By incorporating CL, diffusion models can enhance their adaptability and generalizability, making them more resilient to variations in data characteristics and distribution shifts \cite{gao2023ddgr}. This approach can significantly contribute to the robustness and effectiveness of diffusion models in real-world applications.

%%%%%%%%%%%%%%%%%%%%%%%%%%%%%%%%%%%%%%%%%%%%%%%%%%%%%%%%%%%%%%%%%%%%
\textbf{Continual learning in implicit neural representations:}
The exploration of CL within implicit neural representations \cite{molaei2023implicit} opens up new frontiers in leveraging the power of neural networks for dynamic learning scenarios.
{Implicit neural representations refer to a class of models where functions, such as geometric shapes or scenes, are represented as continuous signals rather than discrete entities \mbox{\cite{sitzmann2020implicit}}. These representations are commonly used in generative modeling and function approximation due to their ability to capture complex, high-dimensional data in a compact form \mbox{\cite{molaei2023implicit}}. Implicit neural representations can benefit immensely from CL techniques to adapt to new data and tasks seamlessly.} 
As the data landscape evolves, implicit neural representations face challenges related to retaining past knowledge and efficiently incorporating new information \cite{po2023instant,yan2021continual}. CL offers a solution to these challenges by enabling models to update their representations while preserving previously acquired knowledge. This not only enhances the model's ability to handle changing data distributions but also supports the development of intelligent systems that can learn and evolve over time. CL in implicit neural representations is instrumental in creating adaptable and intelligent models that can navigate the complexities of dynamic environments.

%% file: content/sections/conclusion/index.tex
\section{Conclusion}
\label{sec:conclusions}

It has been proved and accepted by the research community that traditional machine learning models are ill-suited to handle the dynamic nature of data, and CL offers a promising solution. The systematic review in this manuscript provides a comprehensive overview of the state-of-the-art research in the field of CL in medical image analysis. We have explored various aspects of this evolving topic, including the challenges posed by changing data distributions, hardware, imaging protocols, data sources, tasks, and concept shifts in clinical practice and the need for models to adapt seamlessly. Through a meticulous analysis of the existing literature, we have examined the CL scenarios, strategies, level of supervision, experimental setup, evaluation schemes, and metrics employed to deal with the drifting nature of medical image data. Furthermore, diverse applications of CL in medical image analysis are discussed, ranging from disease classification and detection to intricate tasks of image segmentation. Each application area presents unique challenges and opportunities for research and development, and our review has shed light on the progress made in these domains. Additionally, a thorough collection and analysis of various evaluation matrices for forward and backward transfer facilitate robust evaluation and benchmarking of approaches.

In conclusion, this systematic review provides valuable insights into the current state of CL adaptation in medical image processing tasks, unexplored challenges, and promising future research directions. 
As medical image analysis technology continues to advance, the development of CL models will be crucial for improving diagnostic accuracy, patient care, and overall healthcare outcomes. 

%%%%%In conclusion, CL in medical image analysis is a rapidly evolving research direction with significant potential to revolutionize healthcare in terms of the development of robust, sustainable, and reusable models. 
%%%%%%Our survey highlights the significance of CL in medical image, especially in the context of changing hardware, image protocols, data sources, end tasks, and concept shifts.  

%%%%%%%%%%%%%%%%%%%%%%%%%%%%%%%%%%%%%%%%%%%%%%%%%%%%%%%%%%%%%%%%%%%%%%%%%%%%
\section*{Acknowledgments}
This work was supported by the Federal Ministry of Education and Research (BMBF) of Germany under the grant no. FKZ 01IS21067A-C and the German Research Foundation (Deutsche Forschungsgemeinschaft, DFG) under the grant no. 445703531.

\section*{Declaration of Competing Interest}
The authors declare that they have no known competing financial interests or personal relationships that could have appeared to influence the work reported in this paper.